\documentclass{aa}

\usepackage{txfonts}
\usepackage{color, colortbl}
\usepackage[colorlinks, citecolor=blue, linkcolor=blue, urlcolor=blue]{hyperref}

\usepackage{graphicx}
\usepackage{txfonts}
\usepackage{pdflscape}	
\usepackage{color, colortbl}
\usepackage{dutchcal}

\usepackage{newtxtext,newtxmath}

\usepackage[T1]{fontenc}
\usepackage{ae,aecompl}

\usepackage{pdflscape}	

\usepackage{listings}
\usepackage{pdflscape}
\usepackage{longtable}
\usepackage{wasysym}
\usepackage{multirow}

\usepackage{graphicx}	
\usepackage{amsmath}	
\usepackage{amssymb}	
\usepackage{multirow}
\usepackage{tablefootnote}

\newcommand{\gaia}{{\it Gaia}}
	
\makeatletter
\newcommand*{\rom}[1]{\expandafter\@slowromancap\romannumeral #1@}
\makeatother

\usepackage{soul}
\usepackage{blindtext}
\usepackage{hyperref}
\begin{document}

\title{The Galactic Bulge exploration V.: \\ The secular spherical and X-shaped Milky Way bulge\thanks{Based on observations collected at the European Southern Observatory under ESO programme 093.B-0473.}}

\author{Z.~Prudil\inst{1}, V.~P.~Debattista\inst{2}, L.~Beraldo~e~Silva\inst{3,4}, S.~R.~Anderson\inst{2}, S.~Gough-Kelly\inst{2}, A.~Kunder\inst{5}, M.~Rejkuba\inst{1}, O.~Gerhard\inst{6}, R.F.G.~Wyse\inst{7}, A.~J.~Koch-Hansen\inst{8}, R.~M.~Rich\inst{9}, A. Savino\inst{10}}

\institute{
European Southern Observatory, Karl-Schwarzschild-Strasse 2, 85748 Garching bei M\"{u}nchen, Germany; \email{Zdenek.Prudil@eso.org}
\and Jeremiah Horrocks Institute, University of Central Lancashire, Preston PR1 2HE, UK
\and Department of Astronomy \& Steward Observatory, University of Arizona, Tucson, AZ, 85721, USA
\and{Observatório Nacional, Rio de Janeiro - RJ, 20921-400, Brasil}
\and Saint Martin's University, 5000 Abbey Way SE, Lacey, WA, 98503 
\and Max-Planck Institut f\"ur extraterrestrische Physik, Giessenbachstra\ss e, 85748 Garching, Germany
\and Department of Physics and Astronomy, The Johns Hopkins University, Baltimore, MD 21218, USA
\and Astronomisches Rechen-Institut, Zentrum f{\"u}r Astronomie der Universit{\"a}t Heidelberg, M{\"o}nchhofstr. 12-14, D-69120 Heidelberg, Germany
\and Department of Physics and Astronomy, UCLA, 430 Portola Plaza, Box 951547, Los Angeles, CA 90095-1547, USA
\and Department of Astronomy, University of California, Berkeley, Berkeley, CA, 94720, USA}

\date{\today}

\abstract
{In this work, we derive systemic velocities for $8456$ RR~Lyrae stars, the largest dataset of these variables in the Galactic bulge to date. In combination with \textit{Gaia} proper motions, we compute their orbits using an analytical gravitational potential akin to that of the Milky Way (MW), identifying interlopers from other MW structures, which amount to $22$ percent of the total sample. Our analysis reveals that most interlopers are associated with the halo, with the remainder linked to the Galactic disk. We confirm the previously reported lag in the rotation curve of bulge RR~Lyrae stars regardless of the removal of interlopers. Metal-rich RR~Lyrae stars' rotation patterns are consistent with that of non-variable metal-rich giants, following the MW bar, while metal-poor stars exhibit slower rotation. The analysis of orbital parameter space is used to distinguish bulge stars that, in the bar reference frame, have prograde orbits from those in retrograde orbits. We classify the prograde stars into orbital families and estimate the chaoticity (in the form of frequency drift, $\log\Delta\Omega$) of their orbits. RR~Lyrae stars with banana-like orbits have a bimodal distance distribution, similar to the distance distribution seen in the metal-rich red clump stars. The fraction of stars with banana-like orbits decreases linearly with metallicity, as does the fraction of stars on prograde orbits (in the bar reference frame). The retrograde moving stars (in the bar reference frame) form a centrally concentrated nearly spherical distribution. Analyzing an $N$-body+SPH simulation, we find that some stellar particles in the central parts oscillate between retrograde and prograde orbits and only a minority stays prograde over a long period of time. Based on the simulation, the ratio between prograde and retrograde stellar particles seems to stabilize within a couple of gigayears after bar formation. The non-chaoticity of retrograde orbits and their high numbers can explain some of the spatial and kinematical features of the MW bulge that have been often associated with a classical bulge.}

\titlerunning{Galactic bulge in 6D using RR~Lyrae stars}
\authorrunning{Prudil et al.}
\maketitle

\section{Introduction}

The Galactic bulge is a key component of the Milky Way (MW), representing a dense collection of stars in the inner Galaxy. Understanding its formation and evolution provides critical insights into the history of our Galaxy. The central regions of galaxies alongside the associated bars have been extensively studied through theoretical models, including cosmological and $N$-body simulations which have significantly advanced our understanding of these structures \citep[e.g,][]{Athanassoula2003,Debattista2004,Athanassoula2005,Debattista2017,Fragkoudi2018,Fragkoudi2020,Tahmasebzadeh2022}. 

Photometrically, the Galactic center shows an inclined elongated, barred shape \citep{Weiland1994,Dwek1995,Stanek1997} that we nowadays refer to as a box-/peanut-shape \citep[B/P-shape, e.g.,][]{McWilliam2010,Nataf2010,Wegg2013}. Such bulges are mainly produced either by resonance trapping/crossing \citep[e.g.,][]{Combes1981,Combes1990,Quillen2002,Quillen2014,Sellwood2020,Anderson2024} or by the buckling instability \citep[e.g.,][]{Raha1991,Merritt1994,Debattista2006,Xiang2021}. Bars, in general, drive the redistribution of angular momentum within galaxies, from the central regions (e.g., bulge) to the outer parts (disk and halo). Exchange of angular momentum occurs mainly near resonances, particularly near the inner Lindblad resonance \citep[ILR, e.g.,][]{LyndenBell1972,Tremaine1984,Athanassoula2003} and the outer Lindblad resonance (OLR). As a bar loses angular momentum, it slows down \citep[its pattern speed decreases, e.g.,][]{Chiba2021,Chiba2022} and grows stronger \citep{Debattista1998}. Another evolutionary path leading to the formation of bulges is through the hierarchical galaxy formation via the accretion of satellites, which leads to so-called classical bulges \citep[e.g., the reviews of][]{Wyse1997,Kormendy2004}. Such bulges exhibit a spheroidal spatial distribution and random motion which contrasts starkly with the cylindrical rotation of B/P-shaped bulges.

The stellar content of the Galactic bulge has been extensively studied through spatial, chemical, and kinematical probes. For nearly three decades, we have known that the Galactic bulge has an elongated, barred shape \citep{Stanek1994,Stanek1997}. The Galactic bar, typically traced by red clump stars with ages ranging from $1$-$10$\,Gyr \citep{Ortolani1995,Salaris2002,Clarkson2008,Bensby2013,Renzini2018,Joyce2023}, forms an angle of approximately $25$ degrees with our line-of-sight towards the Galactic center \citep[see, e.g.,][]{Babusiaux2005,Wegg2013,Simion2017,Leung2023,Vislosky2024}. The chemistry of bulge giants shows a broad metallicity distribution varying across Galactic longitude and latitude \citep{Rich1988,McWilliam1994,Gonzalez2011GIBS,Ness2013III,Rojas-Arriagada2014,Zoccali2017,Rojas-Arriagada2020}.

The early studies of bulge kinematics focused on measuring the dispersion and rotation curve of non-variable giants (mainly red clump stars and M giants) and found them to be consistent with cylindrical rotation \citep[e.g.,][]{Minniti1992,Beaulieu2000}. Large-scale spectroscopic studies targeting thousands of giants across different Galactic longitudes and latitudes showed that the Galactic bulge exhibits cylindrical rotation across a variety of Galactic latitudes \citep[up to $\left| \mathcal{b} \right| < 10$\,deg,][]{Rich2007,Howard2008,Howard2009,Kunder2012,Ness2013IV,Zoccali2017,Rojas-Arriagada2020,Lucey2021COMBSII,Wylie2021}. The rotation curve of the Galactic bulge is observed to differ for metal-rich and metal-poor giants. Metal-poor stars ([Fe/H]$<-0.5$\,dex) in the Galactic bulge have been reported to exhibit slower rotation than their metal-rich counterparts \citep{Zoccali2008,Ness2012,Ness2016,Arentsen2020PIGSI}. 

RR~Lyrae stars are an important complementary tracer of the Galactic bulge stellar distribution with respect to that of red clump stars. They were first discovered in the central regions of the MW almost a century ago \citep{vanGent1932,vanGent1933,Baade1946}. RR~Lyrae stars represent the old population of classical pulsators with ages generally assumed to be above $10$\,Gyr \citep{Catelan2009,Savino2020} and are located on the horizontal giant branch. They are often used as standard candles (due to the link between pulsation periods and luminosity) to study the MW structure and dynamics and help disentangle the MW formation history, mainly in the halo \citep[e.g.,][]{Erkal2019,Wegg2019,Fabrizio2021,Prudil2021Orphan,Prudil2022}, and in the Galactic bulge and disk \citep[e.g.,][]{Layden1996,Dekany2013,Pietrukowicz2015,Dorazi2024}. 

RR~Lyrae variables are divided into three categories based on their pulsation mode: RRab (fundamental), RRc (first-overtone), and RRd (double-mode), from the most to least numerous, respectively. They have been employed to study the Galactic bulge in numerous studies focusing on their photometry and spatial distribution \citep[e.g.,][]{Minniti1998,Alcock1998,Pietrukowicz2012,Dekany2013,Pietrukowicz2015,Semczuk2022}. There are significantly fewer spectroscopic kinematic studies of bulge RR~Lyrae stars, mostly confined to only a handful of variables in Baade's Window \citep[e.g.,][]{Butler1976,Gratton1986,Walker1991}. Only recently, thanks to the development of multi-object spectrographs, have larger surveys of bulge RR~Lyrae pulsators become possible, such as the Bulge Radial Velocity Assay for RR~Lyrae stars survey \citep[henceforth referred to as BRAVA-RR,][]{Kunder2016,Kunder2020}. Additionally, astrometric surveys like \textit{Gaia} \citep{Gaia2016,GaiaDR32023} provide precise proper motions that allow the measurement of transverse velocities for many RR~Lyrae stars, facilitating kinematic analyses \citep{Du2020}.

Comparison is often made between RR~Lyrae stars and non-variable giants (both metal-rich and metal-poor). RR~Lyrae variables show similar kinematical distributions to metal-poor giants \citep[e.g.,][]{Ness2013III,Arentsen2020PIGSI,Kunder2020,OlivaresCarvajal2024} while exhibiting slower rotation than metal-rich giants. The contrast between metal-poor and metal-rich giants is further magnified in their spatial distributions \citep[e.g.,][]{Zoccali2017,Lim2021,Johnson2022}, where metal-poor giants exhibit more spherical (non-barred) spatial properties, whereas metal-rich giants trace an elongated (barred) spatial distribution\footnote{We note that this depends on the definition of metal-poor population, which differs in aforementioned studies. The split is likely between $-1.0$ to $-0.5$\,dex in metallicity.}. A similar distribution to metal-poor giants is also observed for RR~Lyrae stars, for which we have more precise distance estimates than for non-variable giants. This led to a debate about whether the MW bulge is a composite bulge containing the B/P-shaped bulge as well as a classical bulge component. The recent conundrum \citep[highlighted by works of][]{Dekany2013,Pietrukowicz2015} regarding whether RR~Lyrae stars follow the Galactic bar and extinction variation has been addressed in our previous work \citep{Prudil20253D}. In that study, we showed that metal-rich ([Fe/H]$_{\rm phot} > -1.0$\,dex) RR~Lyrae pulsators display an elongated spatial distribution ($\iota = 18 \pm 5$\,deg) tracing the bar and nearly matching its angle \citep[as traced by, e.g.,][]{Wegg2013}, while metal-poor RR~Lyrae stars do not follow the bar (they show little to no tilt with the Galactic bar). The complexity is further enhanced by the interlopers from other MW structures. Thus, any analysis of the Galactic bulge requires a careful assessment of the impact of the interloping stars from the Galactic disk and halo.

The structure and dynamics of the Galactic bulge from the viewpoint of RR~Lyrae variables have been the aim of this series of papers \citep{Prudil2023,Prudil2024GBEXII,Kunder2024,Prudil20253D}. This study focuses on RR~Lyrae stars with full 6D information in phase space. We use this dataset to obtain orbital properties and compare our results with non-variable giants and a high-resolution $N$-body+Smooth Particle Hydrodynamics (SPH) simulation. Our manuscript is structured as follows. Section~\ref{sec:DataSets} describes the spectroscopic dataset we combined with our photometric and astrometric sample. Section~\ref{sec:Chemodyn6D} outlines our procedure for estimating orbits, obtaining a rotation curve, and removing interlopers. Section~\ref{sec:OrbitalParam} discusses orbital parameters of RR~Lyrae and non-variable giant datasets. Section~\ref{sec:RotMetalPoor} examines the orbital frequencies and parameters of our dataset. Section~\ref{sec:RetroOrbitsFam} assesses the periodicity of orbits and orbital families. Section~\ref{sec:simulationEND} compares our data with $N$-body+SPH simulations. In Section~\ref{sec:Discussion} we discuss the implication of our results. The last section, Section~\ref{sec:Summary}, summarizes our conclusions.

\section{Spectroscopic data available for the Galactic bulge RR~Lyrae stars} \label{sec:DataSets}

In this section, we describe the data used and sources of spectroscopic information for our data set. In the initial steps, we use photometric metallicities and distances of RR~Lyrae stars toward the Galactic bulge based on large-scale photometric studies from our previous work \citep{Prudil20253D}. For the vast majority ($96$ percent) of our final dynamical data set, the distances were obtained using the combination of $J$ and $K_{\rm s}$ passbands from the the Vista Variables in the V\'ia L\'actea survey \citep[VVV,][]{Minniti2010}, meaning their distances have relative uncertainties around six percent of the distance \citep{Prudil20253D}, and importantly, the dependence of the RR~Lyrae's absolute magnitude on metallicity is not significant. Proper motions were obtained from the third data release of the \textit{Gaia} catalog \citep[DR3,][]{Gaia2016,GaiaDR32023}. Similar to our previous study, we employ some of \textit{Gaia}'s quality flags and use them as criteria for the quality of proper motions and distances (to omit possible blended objects), specifically:
\begin{equation} \label{eq:Condition1}
\texttt{ipd\_frac\_multi\_peak} < 5, \hspace{0.5cm} \text{and} \hspace{0.5cm} \text{RUWE} < 1.4 \\,
\end{equation}
where the \texttt{ipd\_frac\_multi\_peak} represents the percentage of detection of a double peak in the PSF during \textit{Gaia} image processing, and RUWE is the re-normalized unit weight error. For further details see Section~2 in \citet{Prudil20253D}. 

In this work, we include an additional dimension in the form of systemic velocities\footnote{The systemic velocity of an RR~Lyrae star is a component of the line-of-sight velocity. It refers to the velocity of the star’s center of mass along the line-of-sight, as it moves through space. This velocity is corrected for the pulsation motion of the stellar atmosphere during the pulsation cycle.} for a subset of RR~Lyrae stars from our previous study. The determination of line-of-sight velocities and the estimation of systemic velocities for individual RR~Lyrae stars used in this study can be found in Appendix~\ref{sec:App:VlosVsys}. Below, we describe individual sources of spectra used in this study.

First, we utilize the data observed by the BRAVA-RR survey \citep{Kunder2016,Kunder2020}. The observations for the BRAVA-RR survey were conducted at the Anglo-Australian Telescope (AAT) using the AAOmega multifiber spectrograph. The BRAVA-RR survey provides spectra for $\approx 3100$ RRab RR~Lyrae stars (almost $12\,000$ co-added spectra in total) toward the Galactic bulge with a wavelength range centered around the Calcium Triplet ($8300$\,\AA~to $8800$\,\AA) with a resolution of approximately $10\,000$. The number of exposures for each RR~Lyrae star observed by BRAVA-RR ranges from $1$ to $17$, and the average signal-to-noise is $18$.

Our second source provides spectra collected by the Apache Point Observatory Galactic Evolution Experiment \citep[APOGEE survey,][]{Majewski2017,Blanton2017,Wilson2019APO} for RR~Lyrae stars in the direction of the Galactic bulge. APOGEE is an all-sky infrared spectroscopic survey of our Galaxy with observations conducted at the Las Campanas Observatory (LCO) and the Apache Point Observatory (APO). The APOGEE survey is established on two fiber-fed ($300$ fibers) multi-object spectrographs covering the photometric $H$-band ($1.51$--$1.70$\,$\mu$m) with spectral resolution above $20\,000$. For RR~Lyrae stars toward the Galactic bulge, APOGEE provides more than $10\,000$ spectra for over $6\,000$ RRab+RRc RR~Lyrae stars (using the $0.5$\,arcsec crossmatch with the OGLE survey) with an average signal-to-noise ratio equal to $9$, and each variable has between $1-12$ exposures. 

The third source is the ESO archive for program ID 093.B-0473. This program targeted RRc and RRab-type variables toward the Galactic bulge using the FLAMES with GIRAFFE spectrograph on the VLT (HR10 setup, with resolution $R \sim 21500$). We used reduced data from the ESO data archive. In total, this program collected $434$ spectra for $87$ RR~Lyrae stars ($65$ RRab and $22$ RRc, using $1.0$\,arcsec crossmatch), where each variable has between $4-5$ exposures. The signal-to-noise ratio of individual spectra varies between $10-40$ with a center at $\sim20$. The spectral range covers a region mostly populated by iron (\ion{Fe}{i} and \ion{Fe}{ii}) and Calcium lines (from $5300$~\AA~to $5600$~\AA).

In total, we obtained spectra for nearly $9000$ RR~Lyrae stars toward the Galactic bulge. Unfortunately, not all stars had suitable spectra for systemic velocity determination. Also, proper motions were not available for the entire spectroscopic data set. In the end, a sample of $8456$ RR~Lyrae stars could be used for dynamical analysis, which constitutes the largest sample of Galactic bulge RR~Lyrae stars with full 6D parameters assembled so far. In addition, we provide systemic velocities for the RRc pulsators in the Galactic bulge for the first time. Previous studies using APOGEE data, such as \citet{OlivaresCarvajal2024} and \citet{Han2024}, excluded first-overtone pulsators from their analyses, as they could not reliably determine their systemic velocities. In our case, we developed tools to estimate systemic velocities for RRc stars in the APOGEE survey \citep[see][]{Prudil2024GBEXII}, allowing us to enhance our dataset by including them. We do not expect any significant differences in the kinematical distribution between RRc and RRab stars.

\begin{figure}
\includegraphics[width=\columnwidth]{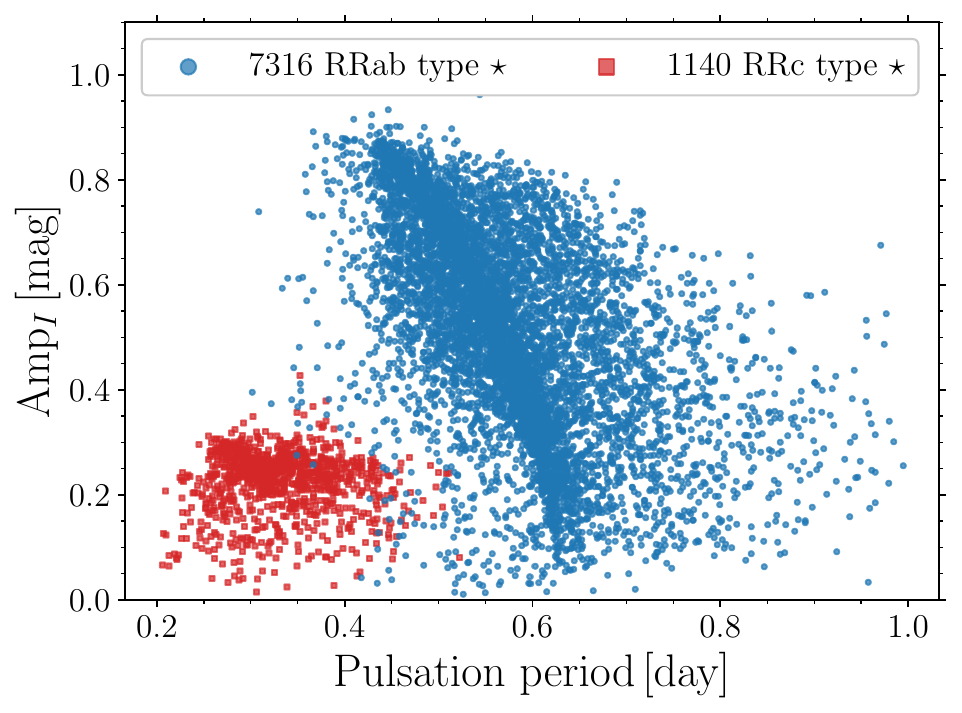}
\caption{The period amplitude diagram for RRab+RRc RR~Lyrae stars (RRab blue points and RRc red squares) in our final dynamical data set based on pulsation properties from the OGLE survey.}
\label{fig:MapPA}
\end{figure}

The obtained systemic velocities for RRab and RRc variables are included in Table~\ref{tab:FullVsys}. We will refer to this sample as our final dynamical data set. In Figure~\ref{fig:MapPA}, we present the period-amplitude diagram for our final dynamical data set. 

\begin{table}[h]
\caption{Table of calculated systemic velocities of our RR~Lyrae dataset.}
\label{tab:FullVsys}
\begin{tabular}{lll}
\hline \hline
ID & $v_{\rm sys}$ & $\sigma_{v_{\rm sys}}$ \\  
 -- & [km\,s$^{-1}$] & [km\,s$^{-1}$] \\ \hline
OGLE-BLG-RRLYR-00342 & $45$ & $18$ \\
OGLE-BLG-RRLYR-00343 & $57$ & $4$ \\
OGLE-BLG-RRLYR-00345 & $276$ & $22$ \\
OGLE-BLG-RRLYR-00351 & $34$ & $9$ \\
$\dots$ & $\dots$ & $\dots$ \\ \hline
\end{tabular}
\tablefoot{
The three columns contain; the OGLE identifier, systemic velocity and its uncertainty. The entire table is available at the CDS.}
\end{table}

\section{RR~Lyrae stars toward the Galactic bulge} \label{sec:Chemodyn6D}

The combination of positions (coordinates and distances) and motion (proper motions and systemic velocities) permits calculating and examining the orbits of stars in our final dynamical data set. Similar analyses have been done in the past for the Galactic bulge's metal-poor population to search for signs of rotation and a bar-like velocity dispersion \citep[e.g.,][]{Prudil2019Kin,Kunder2020,Arentsen2020PIGSI,Lucey2021COMBSII,Lucey2022COMBSIII,Arentsen2023,OlivaresCarvajal2024}. A common trait among these studies is the use of an analytical MW potential within which the dynamical data set gets evolved, and properties of individual objects assessed. To this end, we use publicly available code for Galactic dynamics, \texttt{AGAMA}\footnote{\url{http://agama.software/}} \citep{Vasiliev2019Agama}. From \texttt{AGAMA}, we use an analytical non-axisymmetric approximation of the MW and Galactic bulge model from \citet{Portail2017Pattern} provided by \citet{Sormani2022} and \citet{Hunter2024}. The potential is composed of the Plummer potential for the supermassive black hole, for the nuclear star cluster a flattened axisymmetric spheroidal potential \citep{Chatzopoulos2015}, and two-component flattened spheroidal potential for the nuclear stellar disk \citep{Sormani2020}. In addition, it contains an analytic bar density \citep[model based on data,][]{Portail2017Pattern,Sormani2022}, two stellar disks (thin and thick), and lastly, an Einasto dark matter potential \citep{Einasto1965}. For each star, we integrate its orbit for $5$\,Gyr with $100000$ time steps, and record eccentricities ($e$), maximum excursions from the Galactic plane ($z_{\rm max}$), and apocentric and pericentric distances ($r_{\rm apo}$ and $r_{\rm per}$). The integration time was selected based on the study of \citet{BeraldoSilva2023}, to capture precisely the orbital frequencies and chaotic orbits. We also calculated the angular momentum in the $z$-direction, $L_{z}$, and Jacobi integral, $E_{\rm J}$. Lastly, we rotated the analytical potential with the pattern speed equal to $\Omega_{\rm P} = 37.5$\,km\,s$^{-1}$\,kpc$^{-1}$ \citep[e.g.,][]{Sormani2015,Sanders2019Pata,Clarke2022}.

To further analyze the orbits, we also estimated fundamental frequencies using the \texttt{naif}\footnote{Available at \url{https://naif.readthedocs.io/en/latest/index.html}.} module \citep{BeraldoSilva2023}. We used complex time series as input for the frequencies in both inertial ($\Omega_{x}^{\rm ine}, \Omega_{y}^{\rm ine}, \Omega_{z}^{\rm ine}, \Omega_{R}^{\rm ine}, \Omega_{\phi}^{\rm ine}$) and bar reference frames ($\Omega_{x}^{\rm bar}, \Omega_{y}^{\rm bar}, \Omega_{z}^{\rm bar}, \Omega_{R}^{\rm bar}, \Omega_{\phi}^{\rm bar}$). We estimated frequencies in Cartesian and cylindrical coordinates for both reference frames to identify resonances and assess bar-supporting orbits. The fundamental frequencies are identified in \texttt{naif} as the leading frequencies in the spectra for each coordinate. We selected the complex time series to ensure the correct determination of the sign of $\Omega_{\phi}$. Following \citet{Papaphilippou1996,Papaphilippou1998} and \citet{BeraldoSilva2023} we define complex time series as:
\begin{gather}
f_{R} = R + i v_{R} \\
f_\varphi = \sqrt{2\left|L_{z}\right|}\left(\cos\varphi + i\sin\varphi \right)  \\
f_{z} = z + i v_{z} \hspace{2cm},
\end{gather}
where $R$, $\varphi$ and $z$ cylindrical coordinates, $L_{z}$ is the angular mometum, and $v_{z}$ with $v_{R}$ are the vertical and radial velocities.

In contrast, using a real time series would result in a symmetric power spectrum around zero frequency, and we would need to use the $L_{z}$ to identify the correct sign of $\Omega_{\phi}$. The complex combination overcomes this limitation by breaking the symmetry and enabling the correct identification of the sign.

Lastly, using the systemic velocity ($v_{\rm sys}$) we estimated the velocity in the Galactic Standard of Rest ($v_{\rm GSR}$), sometimes called Galactocentric radial velocity ($v_{\rm GV}$), for each RR~Lyrae star, using the following expression:
\begin{equation}\label{eq:vgv}
\begin{split} 
v_{\rm GV} = & v_{\rm sys} + 220\cdot\left ( \text{sin}(\mathcal{l})\text{cos}(\mathcal{b}) \right ) + 17.1\cdot ( \text{sin}(\mathcal{b})\text{sin}(22) \\ & + \text{cos}(\mathcal{b})\text{cos}(22)\text{cos}(\mathcal{l}-58) ) \hspace{0.5cm}.
\end{split}
\end{equation}
To encompass the uncertainty distribution of its dynamical properties, we varied each star's spatial and kinematical properties within its errors. We assumed that errors followed Gaussian distributions, integrated the orbit for each realization ($100$ in total), and recorded the center (assuming median) and percentiles ($15.9$ and $84.1$\,th) of the resulting distribution for each orbital parameter listed above.

\subsection{Interlopers and bulge stars} \label{subsec:SepBulgeOutside}

\begin{figure*}
\includegraphics[width=2\columnwidth]{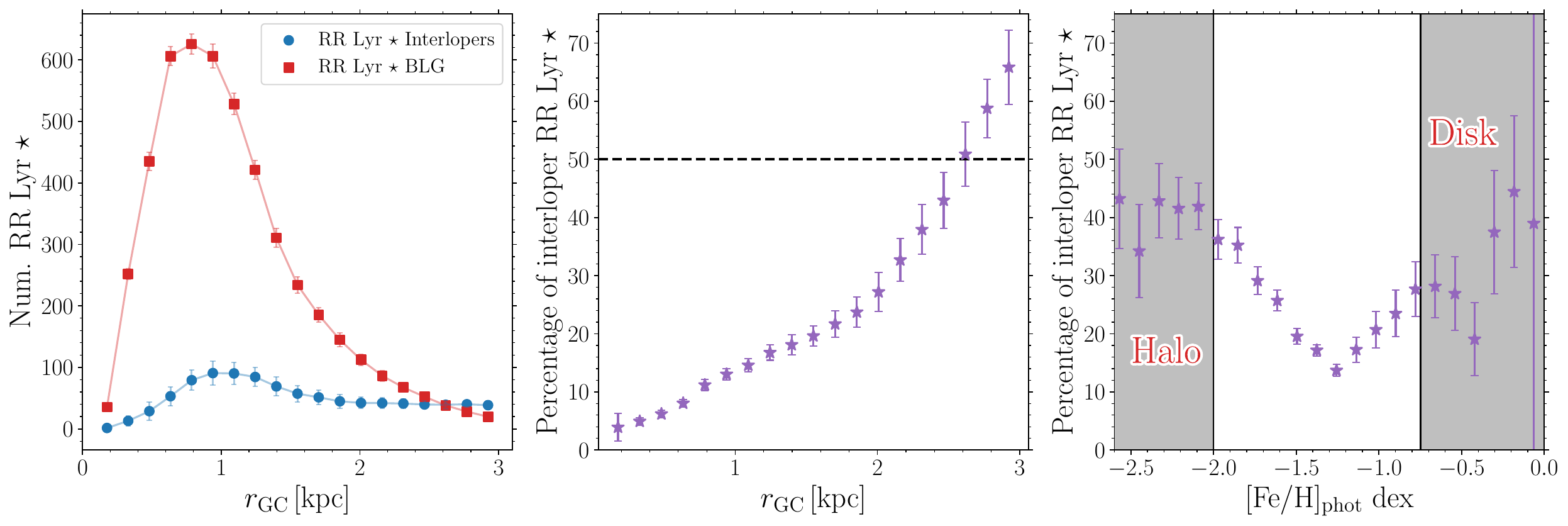}
\caption{Left: the distribution of Galactic bulge (red squares, $r_{\rm apo} < 3.5$\,kpc) and interloping (blue circles, $r_{\rm apo} > 3.5$\,kpc) RR~Lyrae as a function of Galactocentric radius. Middle: the fraction of interloping RR~Lyrae as a function of Galactocentric radius. Right: the fraction of interloping RR~Lyrae in bins of photometric metallicity. The highlighted regions mark the boundaries in metallicity for the Galactic halo and disk.}
\label{fig:InterlopersRFEH}
\end{figure*}

Interloper stars, which pass through the Galactic bulge, make a substantial contribution to the observed stellar population in this region \citep[$25$ to $50$ percent][]{Kunder2020,Lucey2021COMBSII,Lucey2022COMBSIII}. These stars complicate our understanding of the rotational dynamics of the inner MW. Earlier studies used various methods to reduce the impact of stars from other MW structures, primarily focusing on separation in the apocentric distance and the extent of deviation from the Galactic plane to filter out halo and disk interlopers \citep[see, e.g.,][]{Kunder2020,Lucey2021COMBSII,Lucey2022COMBSIII}. The distinction between bulge stars and interlopers is typically based on apocentric distance: stars on orbits that are confined within a Galactocentric distance of $3.5$\,kpc are labeled as bulge stars, while those beyond are labeled as interlopers. We note that our definition for bulge and interloper stars connects to their spatial distribution and not to a formation scenario. In our study, we applied the same criteria for apocentric distance to distinguish bulge RR~Lyrae stars from other RR~Lyrae stars that are not part of the bulge. This decision was partially motivated by \citet{Lucey2023} who found orbits that support the MW's bar extend vertically up to $3.5$\,kpc and by the extent of the structure of prograde (in the bar frame) RR~Lyrae stars, depicted in Figure~\ref{fig:ApocenZmaxRot}. We note that some other studies used stars with apocenter distances greater than $3.5$\,kpc, \citep[e.g.,][]{OlivaresCarvajal2024}; while here we are interested more in the purity of our bulge sample, and such a cut encompasses the majority of the bulge while limiting contamination from halo and disk stars. Approximately $22$ percent of RR~Lyrae in our final kinematical data have $r_{\rm apo} > 3.5$\,kpc, and we refer to these variables as interlopers. 

We examine the physical and spatial characteristics of our final dynamical data set, focusing on how these characteristics correlate with whether a star is more likely part of the bulge or an interloper. Figure~\ref{fig:InterlopersRFEH} illustrates the relationship between the Galactocentric spherical radius ($r_{\rm GC}$) and metallicity with the proportion of interlopers. As expected, the number of interlopers is highest at larger radii compared to the bulge RR~Lyrae population. Moreover, the number of RR~Lyrae stars not associated with the bulge increases with the distance from the Galactic center, surpassing the number of bulge RR~Lyrae stars beyond a radius of $2.5$\,kpc. A similar conclusion was found in the study of \citet{Arentsen2023}, where, despite their somewhat different spatial properties (in comparison with our dataset), they noticed a dependence of interloper fraction on the $r_{\rm GC}$ with the number of interloping stars increasing with increasing $r_{\rm GC}$.

In the metallicity analysis of interloper stars in the Galactic bulge, we observe three key contributors: stars from the Galactic disk, the outer bulge, and the halo. The disk's RR~Lyrae stars are known for their high metallicity \citep[e.g.,][]{Layden1996,Zinn2020,Prudil2020Disk,Iorio2021}, whereas the halo's RR~Lyrae stars predominantly occupy the lower metallicity range \citep{Fabrizio2019,Fabrizio2021}. The variables related to the outer bulge overlap both disk and halo interlopers in metallicity space, which makes distraction and clear separation difficult. This distinction is critical in understanding how each of these MW structures influences the bulge's interloper population. Based on the metallicity \citep[using photometric metallicities estimated in][]{Prudil20253D} boundaries shown in the right-hand panel of Figure~\ref{fig:InterlopersRFEH} right-hand panel (marked by the grey shaded regions), we find that approximately $25$ percent of the interlopers are likely disk RR~Lyrae stars, with the remaining exhibiting halo-like (remaining $75$\,percent of interlopers) characteristics in their orbits and metallicity. We note that from the identified $25$ percent of the disk interlopers, the majority of stars lie close to the Galactic plane, and only two percent have $z_{\rm max} > 3$\,kpc. These disk RR~Lyrae stars, mostly have $r_{\rm apo}$ between $3.5$\,kpc and $6$\,kpc (over $70$\,percent) and come from the outer bar region. 

In the Appendix, we include Figure~\ref{fig:HaloFreq}, which shows the distribution of orbital frequencies ($\Omega_{\phi}^{\rm ine}$ and $\Omega_{\phi}^{\rm bar}$) in the inertial and bar reference frames for our RR~Lyrae dataset. In this Figure, we divide the sample into bulge, interlopers, and likely halo RR~Lyrae variables (identified by $z_{\rm max} > 5$\,kpc). In the inertial frame, we do not observe a clear preference for prograde or retrograde rotation among the halo RR~Lyrae stars. In the bar frame, however, these pulsators lag behind the rotation of the bar and therefore appear retrograde with respect to the Galactic bar.

\subsection{Rotation pattern of RR~Lyrae variables across different metallicities} \label{subsec:Rot6DFEH}

\begin{figure*}
\includegraphics[width=2\columnwidth]{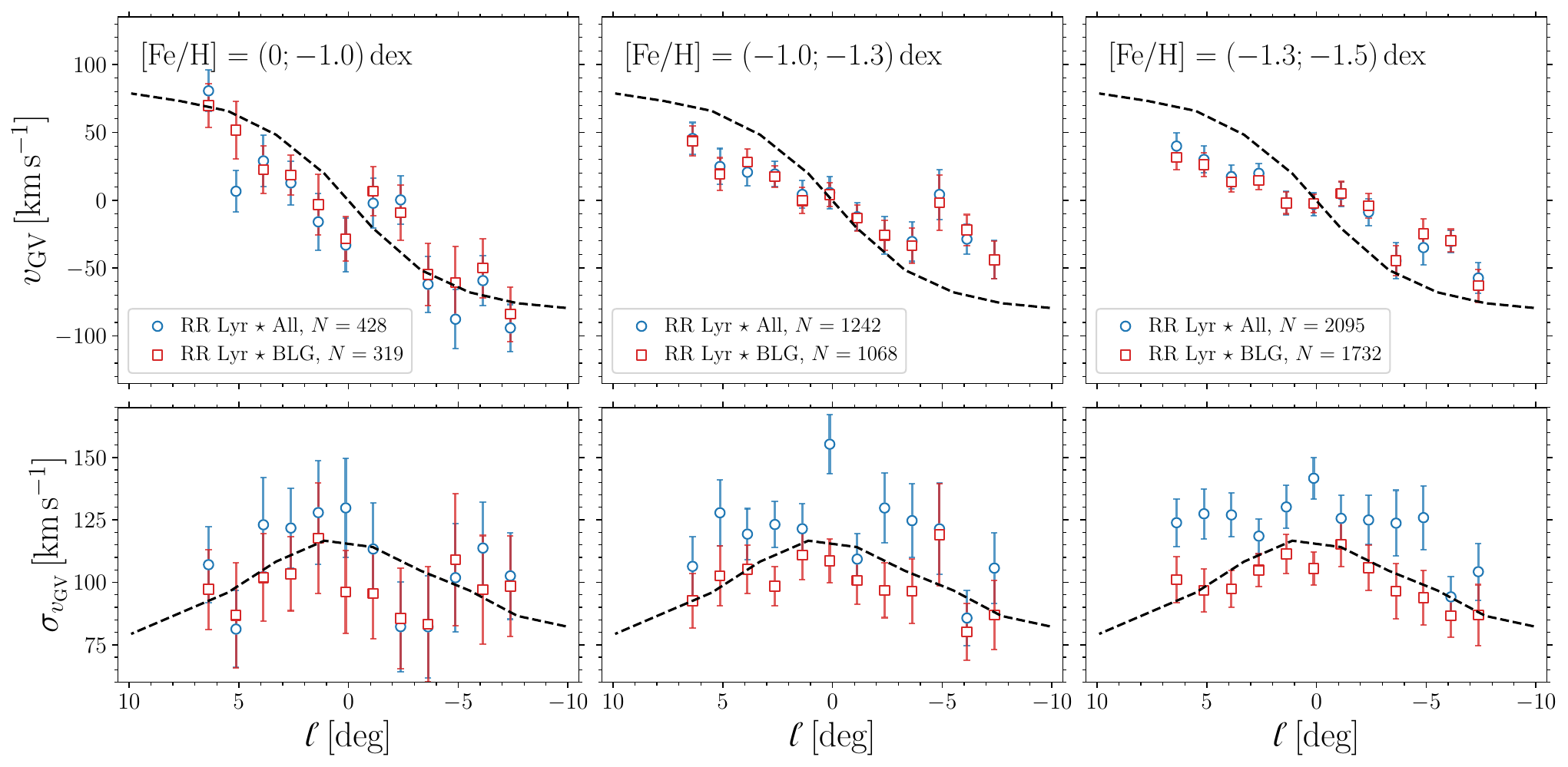}
\includegraphics[width=2\columnwidth]{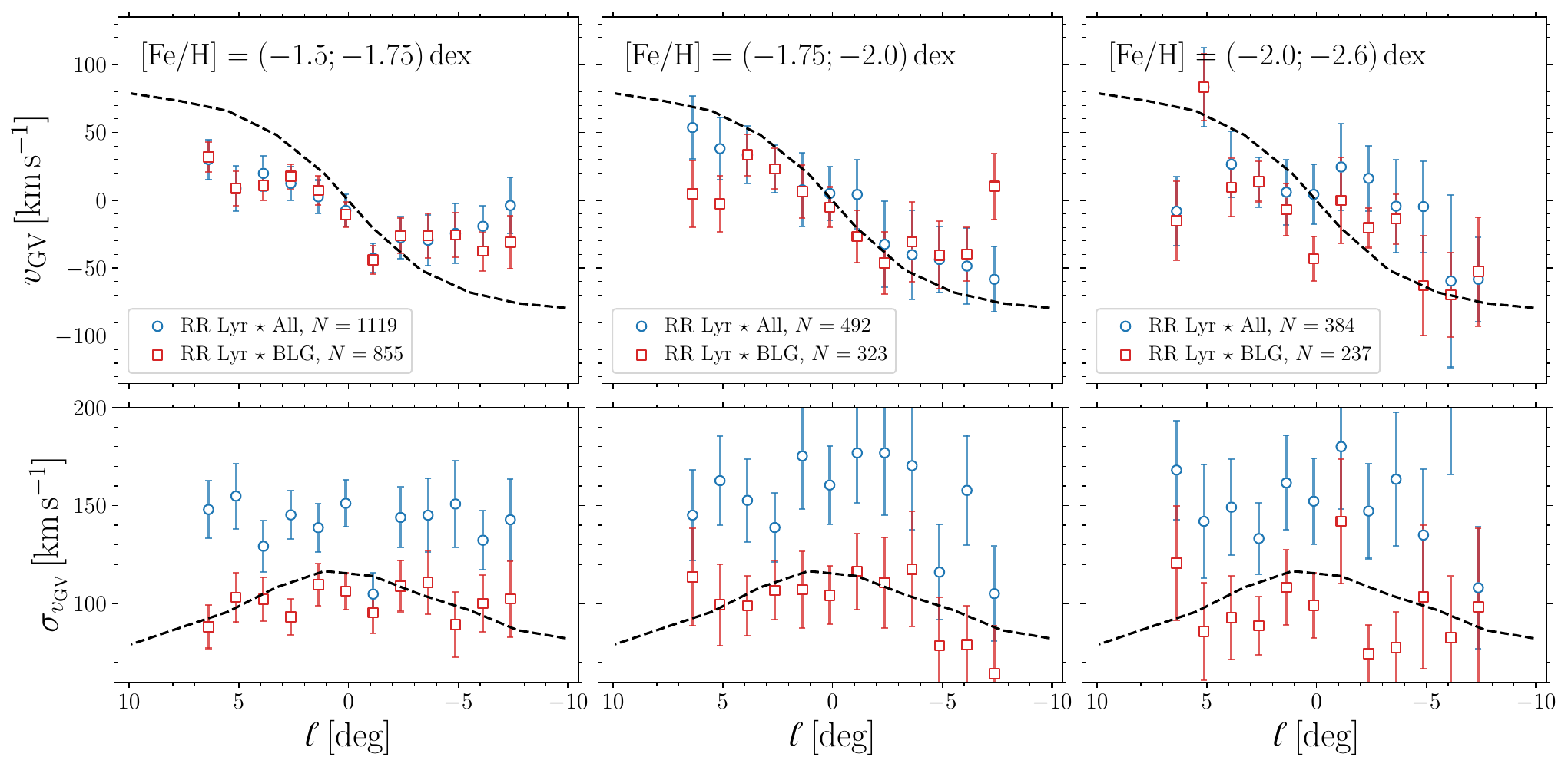}
\caption{The distribution of average $v_{\rm GV}$ and its dispersion $\sigma_{v_{\rm GV}}$ across the Galactic longitude bins and for different photometric metallicity cuts. The blue points represent the final dynamical data set, and the red squares stand for RR~Lyrae variables with $r_{\rm apo} < 3.5$\,kpc (labeled BLG in the legend). The dashed lines trace the $v_{\rm GV}$ and $\sigma_{v_{\rm GV}}$ from the bar model of \citet{Shen2010} for $\mathcal{b} = 4$\,deg. For this Figure we use equally spaced bins between $\mathcal{\mathcal{l}}=(-8.0;8.25)$\,deg with a step of $1.25$\,deg. The asymetric distribution in $\mathcal{l}$ was driven by our RR~Lyrae dataset.}
\label{fig:VGVRotDisp-FEH}
\end{figure*}

Here we explore the dynamical parameter space by looking at the distribution of average $v_{\rm GV}$ and its dispersion across the Galactic longitude bins and for different metallicities. We compare our observational dataset with the bulge model by \citet{Shen2010} which is based on the results from the BRAVA spectroscopic survey \citep{Rich2007,Howard2008,Howard2009}. In Figure~\ref{fig:VGVRotDisp-FEH}, we display the orbital properties for our entire data set (blue points) and for bulge RR~Lyrae stars with $r_{\rm apo}<3.5$\,kpc (red squares). The selection based on the apocentric distance to remove interlopers by construction decreases the dispersion in $v_{\rm GV}$. In the most metal-rich bin of our final dynamical data set (top left-hand corner of Figure~\ref{fig:VGVRotDisp-FEH}), we see clear signs of rotation that follow the bar model of \citet{Shen2010}. The dispersion $\sigma_{v_{\rm GV}}$ shows nearly a flat distribution after removing interlopers. This is in agreement with our previous work \citep{Prudil20253D} in which we showed that spatial and transverse properties of metal-rich RR~Lyrae ([Fe/H]$_{\rm phot} >-1.0$\,dex) follow the patterns characteristic of the Galactic bar. This is also in agreement with previous studies focused on metal-rich non-variable giants \citep[e.g.,][]{Kunder2012,Ness2013IV}. Here it is important to emphasize that the [Fe/H]$_{\rm phot}$ for the metal-rich bin ([Fe/H]$_{\rm phot} >-1.0$\,dex) is heavely skewed toward [Fe/H]$_{\rm phot} \approx -1.0$\,dex, with median metallicity (and $15.9$ and $84.1$ percentiles) of this bin equal to [Fe/H]$_{\rm phot} = -0.75_{-0.18}^{+0.32}$\,dex. In addition, removing the interlopers (both halo and disk in this case) partially enhances the rotation pattern in the $\mathcal{\mathcal{l}}$ vs. $v_{\rm GV}$ slightly more pronounced. 

Considering the metal-poor ([Fe/H]$_{\rm phot}$~$=(-1.5; -1.0)$\,dex) population, we find a lag in the average $v_{\rm GV}$ behind the rotation regardless of whether the halo interlopers are removed or not. The dispersion in $v_{\rm GV}$, as in the metal-rich case, has only a mild difference at $\mathcal{\mathcal{l}}=0$\,deg between the full and bulge (no interlopers) RR~Lyrae dataset. In this metallicity bin, the halo interlopers do not seem to affect the rotation pattern (their percentage is low, $\sim14$ to $20$ percent). Bins with metallicities below [Fe/H]$_{\rm phot}$~$<-1.5$\,dex start showing a significant contribution from halo interlopers, particularly in the $\sigma_{v_{\rm GV}}$ space where the total sample exhibits higher values for velocity dispersion. We again see a weak rotation that slowly decreases as we move toward lower metallicities. The rotation curve flattens at the metal-poor end of our distribution (RR~Lyrae variables with [Fe/H]$_{\rm phot} < -2.0$\,dex). 

Here, it is important to stress that the fraction of interlopers differs in each metallicity bin (as seen in Figure~\ref{fig:InterlopersRFEH}). We have mostly disk interlopers on the metal-rich side. As we move toward the metal-poor stars, the contribution decreases until we reach [Fe/H]$_{\rm phot}$~$=-1.5$\,dex. Then, the number of stars with apocentric distance above $3.5$\,kpc (mainly halo RR~Lyrae variables) increases again and reaches $40$ percent at the lowest metallicity bin in our sample. Therefore, for RR~Lyrae stars, the interlopers affect the rotation pattern only slightly, and their influence is more prominent in the velocity dispersion (particularly the removal of halo interlopers leads to its decrease), especially at the metal-poor end. The presence of interlopers in some cases leads to an increase in the velocity for the rotation curve, which was previously reported by \citet{Kunder2020,Lucey2021COMBSII,Arentsen2023}. The interlopers are not responsible for the disappearance or lag in rotation. Given that the metal-rich RR~Lyrae follow the Galactic bar, both spatially and kinematically \citep[][and Figure~\ref{fig:VGVRotDisp-FEH} here]{Prudil20253D}, other factors must explain the variation in the dynamics of RR~Lyrae and other metal-poor populations in the bulge, which we attempt to uncover next.
 
\section{Orbital properties of bulge stars} \label{sec:OrbitalParam}

In what follows, we aim to resolve the reason for the differences in the bar-supporting kinematics between the relatively old ($\approx12$\,Gyr) RR~Lyrae population\footnote{We note that recent studies, particularly by \citet{Iorio2021}, \citet{Bobrick2024} and \citet{Zhang2025} suggested that metal-rich RR~Lyrae stars are the result of binary interactions and are thus younger objects ($1 - 9$\,Gyr). We address the potential age of metal-rich RR~Lyrae stars from the kinematical point of view in Section~\ref{subsec:MetalRichKinAge}.} and the relatively younger ($< 10$\,Gyr) non-variable giants population. To achieve our goals, we use available kinematical data for non-variable giants as well as an $N$-body+SPH simulation together with our RR~Lyrae sample.

\subsection{Simulation} \label{subsec:Sim}

In this study, we use a high-resolution star-forming $N$-body+SPH simulation of a galaxy that evolves in isolation. It has been thoroughly described in \citet[][where it is referred to as HG1]{Cole2014,Gardner2014,Debattista2017,GoughKelly2022,Fernandez2025} where the simulation was extensively used as a comparison with the MW. The initial setup of this simulation starts with a hot gas corona inside a dark matter halo without any stars. The continued star formation is triggered as the gas corona cools down and settles into a disk, forming a disk galaxy. The model is evolved using the \texttt{GASOLINE} \citep{Wadsley2004} code for $10$\,Gyr. At the end of the evolution, it consists of $5\times10^{6}$ dark matter particles, and approximately $2.6\times10^{6}$ gas and $1.1\times10^{7}$ stellar particles. The model forms a bar around $3$\,Gyr that at $10$\,Gyr has a length of $3$\,kpc. In this work, we use snapshots at $1$, $2$, $4$, $5$, $6$, $7$, $8$, $9$ and $10$\,Gyr of this simulation.

We explore the orbital parameters and frequencies of the stellar particles located in the central regions. In particular, we adopt the approach of \citet{BeraldoSilva2023}, and use \texttt{PYNBODY}\footnote{\url{https://pynbody.github.io/pynbody/}} \citep{Pontzen2013} to center and align the simulation snapshots. For each snapshot, we measure the pattern speed, $\Omega_{\rm P}$, using the module\footnote{\url{https://github.com/WalterDehnen/patternSpeed}} provided by \citet{Dehnen2023} that estimates pattern speed from individual snapshots. We measure the bar angle (using the inertia tensor approach) in the simulation and align the bar with the x-axis. Using the \texttt{AGAMA} software, we also create a composite potential for each snapshot by describing stellar and gas particles using a triaxial cylspline potential and dark matter halo using an axisymmetric multipole potential. In these potentials, we integrate orbits of randomly selected stellar particles (of all ages) that are located within $r_{\rm GC} = 2$\,kpc. The boundary on spherical radius was motivated by the spatial scaling factor of $1.7$ \citep[e.g.,][]{GoughKelly2022} often used to match this simulation with the MW's spatial properties. We emphasize that at this step we did not scale the simulation to match the MW observed properties when computing the orbits \citep[as done in, e.g.,][]{GoughKelly2022}. 

\subsection{Data for non-variable giants} \label{subsec:nonG}

To compare and further explore our results for RR~Lyrae stars, we use a data set with a younger (on average) population provided by the APOGEE DR17 \texttt{StarHorse} value-added catalog \citep{Santiago2016,Queiroz2018,Queiroz2020}. To ensure \texttt{StarHorse} catalog compatibility with our RR~Lyrae sample, we mirror their spatial distribution (in Galactic coordinates and distances) with our RR~Lyrae dataset and enforce the following set of conditions on the APOGEE DR17 \texttt{StarHorse} catalog:
\begin{gather}
1.0 < \text{Heliocentric distance} < 20~\text{kpc} \\
0.0 < \text{log}\,g < 3.0~\text{dex} \hspace{0.5cm} ,
\end{gather}
These conditions ensure that we select preferentially giants (mostly red giants) with distances centered approximately at the Galactic bulge and spatially matching our RR~Lyrae dataset. In addition, we remove stars with the following flags in the APOGEE data: \textsc{PERSIST\_HIGH} and \textsc{LOW\_SNR}. Since we aim to obtain orbital properties for stars in the \texttt{StarHorse} catalog, we selected only objects with \textit{Gaia} proper motions and APOGEE line-of-sight velocities. Lastly, similar to our RR~Lyrae sample, we applied the criterion from Eq.~\ref{eq:Condition1}. This yielded a total of $9736$ red giants with spatial, kinematic, and chemical information. 

\begin{figure}

\includegraphics[width=\columnwidth]{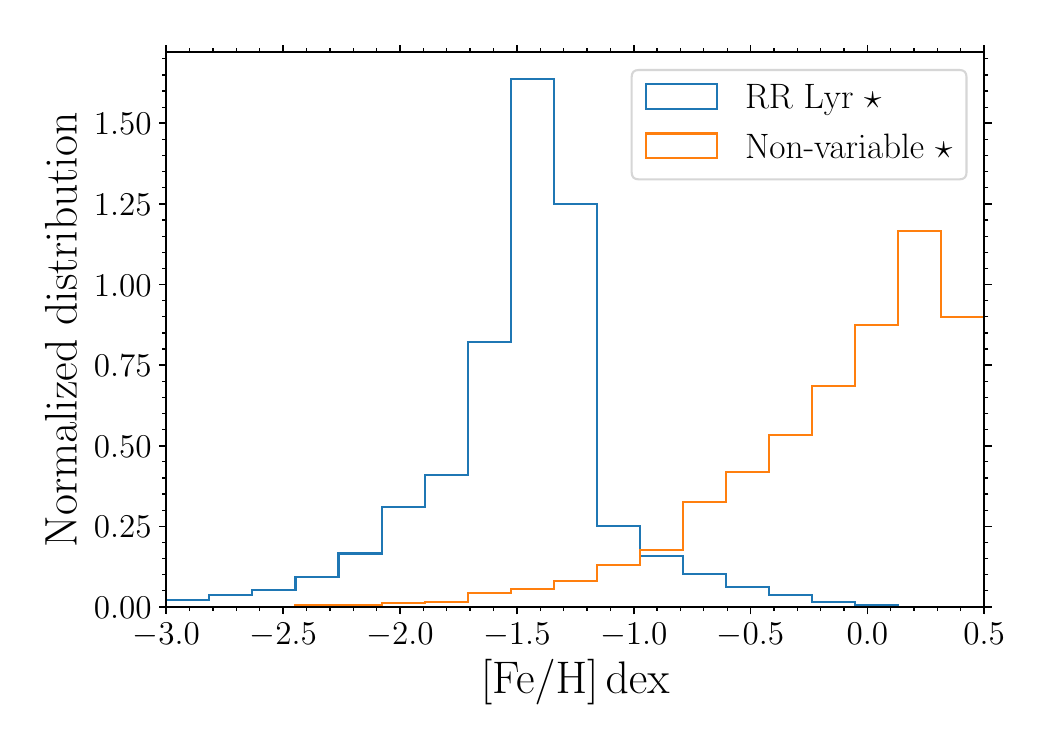}
\includegraphics[width=\columnwidth]{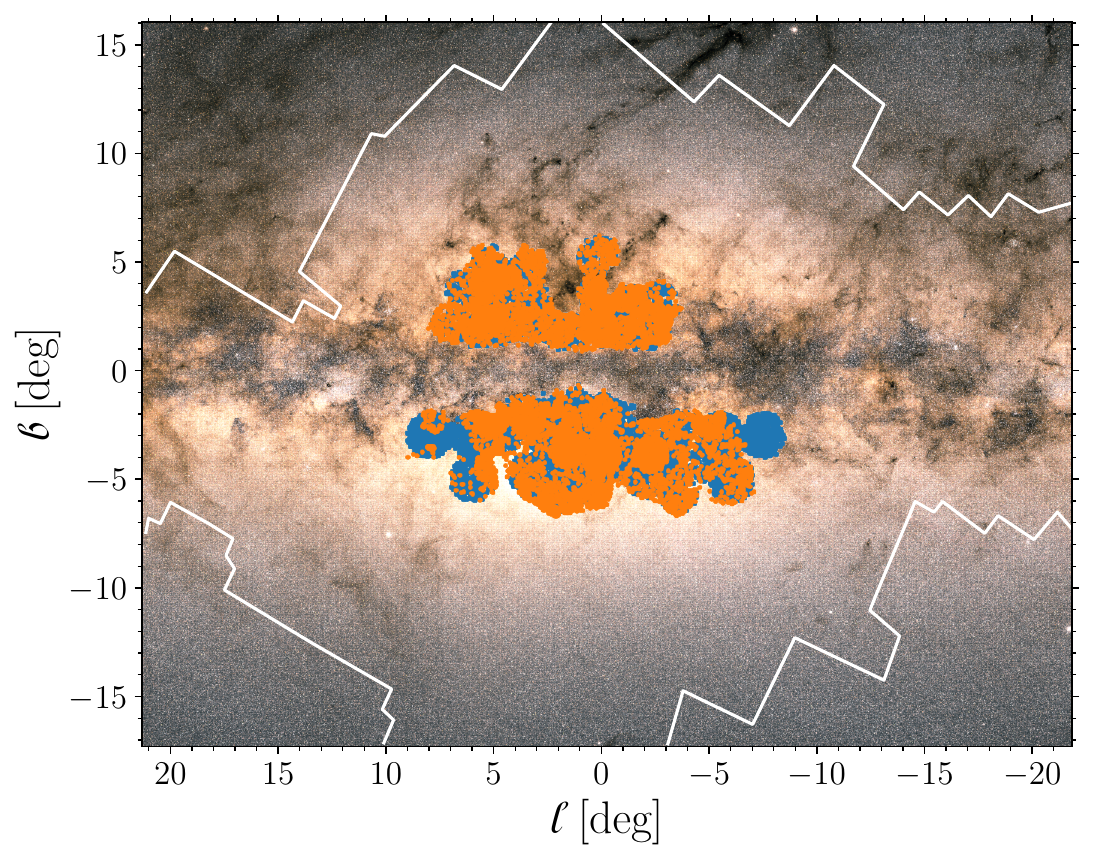}
\includegraphics[width=\columnwidth]{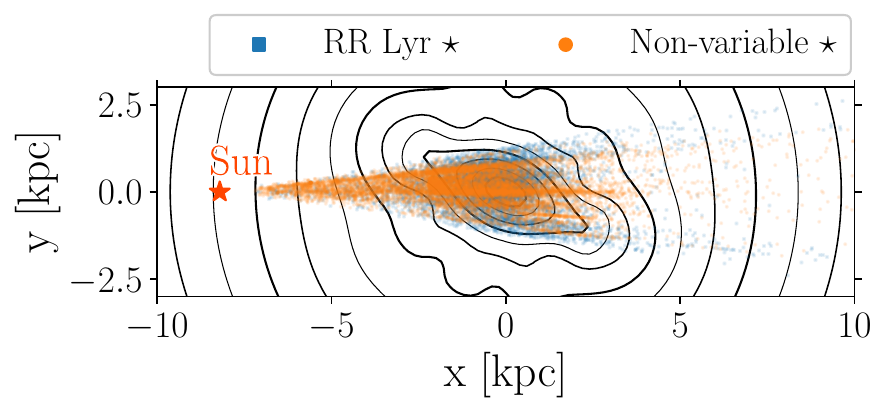}
\caption{The metallicity (top panel) and spatial distribution in Galactic coordinates (middle panel) and Cartesian coordinates (bottom panel) for RR~Lyrae (blue squares) and non-variable giants (orange points) data sets. The solid white lines in the top panel depict the approximate footprint of the OGLE survey. In the bottom panel, the black lines indicate the surface density contours (of the analytical potential) for a stellar component rotated by the bar angle ($25$ degrees). The underlying image in the top panel is \gaia's all-sky star density map. \textit{Image credit: ESA/Gaia/DPAC. Images released under CC BY-SA 3.0 IGO}.}
\label{fig:SpatialDistInPotential}
\end{figure}

As an additional and mostly independent source of non-variable stars, we used red clump stars identified in the Blanco DECam Bulge Survey \citep[BDBS,][]{Rich2020,Johnson2022}. This data set contains photometric metallicities and distances estimated based on the $ugrizY$ passbands. Similar to the pure APOGEE red giant data set described above, we used the same footprint (in Galactic coordinates and distances) on the red clump. We then crossmatched BDBS red clump stars with the \textit{Gaia} catalog to obtain proper motions. To derive line-of-sight velocities for BDBS giants, we combined several publicly available surveys, including \textit{Gaia} \citep{Katz2023}, GALactic Archaeology with HERMES \citep[GALAH,][]{Buder2021}, APOGEE, and GIRAFFE Inner Bulge Survey \citep{Zoccali2014,Gonzalez2015}. This resulted in $1751$ red clump stars with full spatial and kinematic information. Despite our best efforts to identify sources of line-of-sight velocity for the red clump dataset, their spatial distribution does not perfectly match the footprint of RR~Lyrae and red giant stars. However, we chose to keep them in our analysis as they represent an additional stellar evolutionary stage.

From here on, we will call the combination of the two samples the non-variable giant data set ($11\,487$ objects in total). Its metallicity and spatial distribution in Galactic and Cartesian coordinates are depicted in Figure~\ref{fig:SpatialDistInPotential}. This Figure shows that our RR~Lyrae and non-variable giants data sets match reasonably well. We note that the non-variable data do not cover the full extent of our RR~Lyrae sample in the region below the Galactic plane, where some of the BRAVA-RR fields reach broader Galactic longitudes. 

Similar to our RR~Lyrae data set, we used the analytical model of \citet{Portail2017Pattern,Sormani2022} and \citet{Hunter2024} for the MW implemented in \texttt{AGAMA} (see description in Section~\ref{sec:Chemodyn6D}) to obtain the orbital properties and frequencies \citep[using the \texttt{naif} software of][]{BeraldoSilva2023} of our non-variable giant sample. We used the same integration time of $5$\,Gyr, as in the case of RR~Lyrae stars. 

\subsection{Bifurcation in orbital parameters and recovering the rotation pattern for RR~Lyrae stars} \label{subsec:Bif}

\begin{figure*}
\includegraphics[width=2\columnwidth]{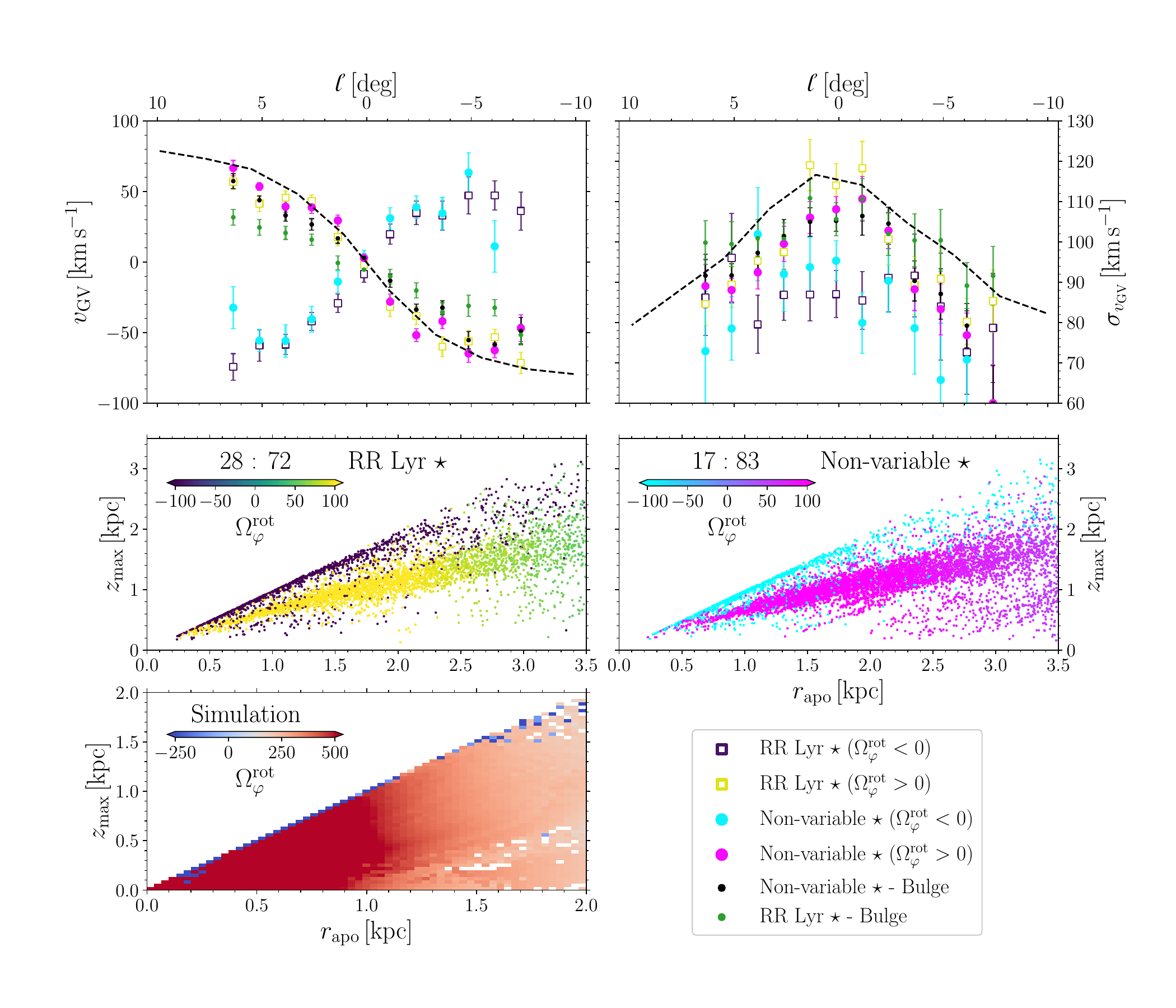}
\caption{The kinematic and orbital properties for RR~Lyrae variables, non-variable giants, and stellar particles from the simulation. The top panels show the distribution of average $v_{\rm GV}$ and its dispersion $\sigma_{v_{\rm GV}}$ across the Galactic longitude bins. The middle and bottom panels display the distribution of orbital properties $z_{\rm max}$ vs. $r_{\rm apo}$ for RR~Lyrae (middle left-hand panel), non-variable giants (middle right-hand panel), and model HG1 (bottom left-hand panel, at $10$\,Gyr). The color coding in all panels is based on $\Omega_{\varphi}^{\rm rot}$. The black dashed lines in the top panels represent the rotation and dispersion curve based on the bar model \citet{Shen2010} for $\mathcal{b} = 4$\,deg.}
\label{fig:ApocenZmaxRot}
\end{figure*}

Here, we combine and examine the data from the simulation (see Section~\ref{subsec:Sim}) with the RR~Lyrae and non-variable giants datasets. In Figure~\ref{fig:ApocenZmaxRot} we show the orbital properties for RR~Lyrae stars, non-variable giants, and the simulation. We chose the angular frequency in the bar's rest frame as an indicator for prograde (positive sign in bar reference frame) and retrograde (negative sign in bar reference frame) stellar motion with respect to the bar. We tested this approach using the parameter, $\lambda$, estimated for each time step of the orbit, defined as:
\begin{equation} \label{eq:SignCheck}
\lambda = L_{z} - \Omega_{\text{P}}R_{\text{cyl}}^{2} \\,
\end{equation}
where $L_{z}^{\text{ine}}$ and $R_{\text{cyl}}$ are the angular momentum and cylindrical radius in the inertial frame. For each star we estimate the median value of $\lambda$ and compare its sign with the sign of $\Omega_{\varphi}^{\rm rot}$ which we define as:
\begin{equation}
\Omega_{\varphi}^{\rm rot} = \Omega_{\varphi}^{\rm ine} - \Omega_{\rm P} \\.
\end{equation}
We find, in $91$ percent of the cases matching signs, while in the remaining cases, $\lambda$ oscillates around zero. We note that, by definition, $\Omega_{\varphi}^{\rm rot}$ is the same as $\Omega_{\varphi}^{\rm bar}$. However, in this work, we consider these two quantities to be different. Due to the presence of stars with chaotic orbits (see Section~\ref{sec:RetroOrbitsFam} for details), some stars exhibit different signs for $\Omega_{\varphi}^{\rm rot}$ and $\Omega_{\varphi}^{\rm bar}$. Subsequently, the fraction of stars with matching signs relative to $\lambda$ (as defined in Eq.~\ref{eq:SignCheck}) differs for these two quantities, with $\Omega_{\varphi}^{\rm rot}$ showing better agreement with $\lambda$ than $\Omega_{\varphi}^{\rm bar}$.

In Figure~\ref{fig:ApocenZmaxRot} we see that stars that fall on the identity line (forming one of the two substructures) between $z_{\rm max}$ and $r_{\rm apo}$ show almost exclusively negative $\Omega_{\varphi}^{\rm rot}$ and thus retrograde motion with respect to the Galactic bar. These stars, as expected (based on their negative $\Omega_{\varphi}^{\rm rot}$), to not contribute to the rotation and dispersion trends depicted in the upper panels but rather exhibit counter rotation and a flat dispersion profile. The second substructure, which we will refer to as bulk has positive $\Omega_{\varphi}^{\rm rot}$ values and clearly follows the rotation and dispersion patterns. From this figure we see that the retrograde orbits are significantly more vertically extended than the prograde orbits. Here and in what follows we use $\Omega_{\varphi}^{\rm rot}$ as a means to identify and separate prograde and retrograde stars in the Galactic bulge. We opted to use $\Omega_{\varphi}^{\rm rot}$ since regular (see Section~\ref{sec:RetroOrbitsFam} on regularity) orbits conserve frequencies in barred potentials unlike $L_{z}$ which is not conserved in non-axisymmetric potentials.

The top panels in Figure~\ref{fig:ApocenZmaxRot} shows how the positive and negative $\Omega_{\varphi}^{\rm rot}$ manifests in a rotation curve. The non-variable giants' sample shows a negligible difference in the rotation pattern between the entire sample (small black dots) and only those with positive $\Omega_{\varphi}^{\rm rot}$ (pink circles). On the other hand, for RR~Lyrae stars, the difference is quite significant. RR~Lyrae variables with positive $\Omega_{\varphi}^{\rm rot}$ (yellow squares) expectedly follow a bar rotation pattern while the bulge RR~Lyrae data set is lagging (green points). The reason behind the significant (for RR~Lyrae pulsators) and negligible (for non-variable giants) difference is in the ratio of number of stars with positive versus negative $\Omega_{\varphi}^{\rm rot}$. Another way of looking at results presented in Figure~\ref{fig:ApocenZmaxRot} is that the angular momentum distribution of the RR~Lyrae stars is much less tilted towards the prograde $L_{z}$.

In the selected bulge fields, for the non-variable giants populations, we have a fraction of $83$\,\% stars with prograde kinematics (see the middle right-hand panel of Figure~\ref{fig:ApocenZmaxRot}), while for RR~Lyrae stars, this probability is $\approx72$\,\%. If we randomly draw stars from our RR~Lyrae data set with the simple condition that at least $83$\,\% of them should have positive $\Omega_{\varphi}^{\rm rot}$ we nearly recover the same rotation pattern as for non-variable giants. In the bottom panel of Figure~\ref{fig:ApocenZmaxRot}, we present orbital properties colored by $\Omega_{\varphi}^{\rm rot}$ for our simulation (snapshot at $10$\,Gyr). We see that we partially recover the substructures found in the observational data. We notice that the stellar particles with negative $\Omega_{\varphi}^{\rm rot}$ clump on the identity line, and stellar particles with positive $\Omega_{\varphi}^{\rm rot}$ fall below it. The fraction of particles with negative $\Omega_{\varphi}^{\rm rot}$ is $14$\,\%. We note that in this case, we did not impose any selection criteria on the positions and orbital parameters of the stellar particles. 

Lastly, in addition to analyzing the stars confined to the Galactic bulge, we also looked at the foreground and background areas of the Galactic bulge. The discussion of objects that fall in these regions is included in Appendix \ref{sec:ForeBack}.

\section{Implications for the RR~Lyrae distribution from orbital frequencies} \label{sec:RotMetalPoor}

In this section we examine in depth the orbital frequencies and use them to distinguish between different spatial and kinematic populations of RR~Lyrae stars within the MW barred potential.

\subsection{Metal-poor RR~Lyrae stars} \label{subsec:metPoor}

Metal-poor giant stars ([Fe/H]~$< -2.0$\,dex) are found within the Galactic bulge \citep[e.g.,][]{Koch2016,Arentsen2020PIGSI,Rix2022,Arentsen2023} and a significant fraction (in general over $50$ percent) remains confined to the Galactic bulge. These objects exhibit a slight rotation in $v_{\phi}$ \citep[see Figure~10 in][]{Arentsen2023}. However, neither the metal-poor variable nor non-variable giants seem to follow the Galactic bar spatially and exhibit much slower rotation (if any) compared to the metal-rich population. This can be explained by the high fraction of interlopers ($40$\,percent of the total sample) and retrograde stars ($19$\,percent of the total sample) on the metal-poor end of the RR~Lyrae metallicity distribution ([Fe/H]$_{\rm phot} < -2.0$\,dex), resulting in only $41$ percent of total metal-poor RR~Lyrae stars being associated with the Galactic bar.

\begin{figure}
\includegraphics[width=\columnwidth]{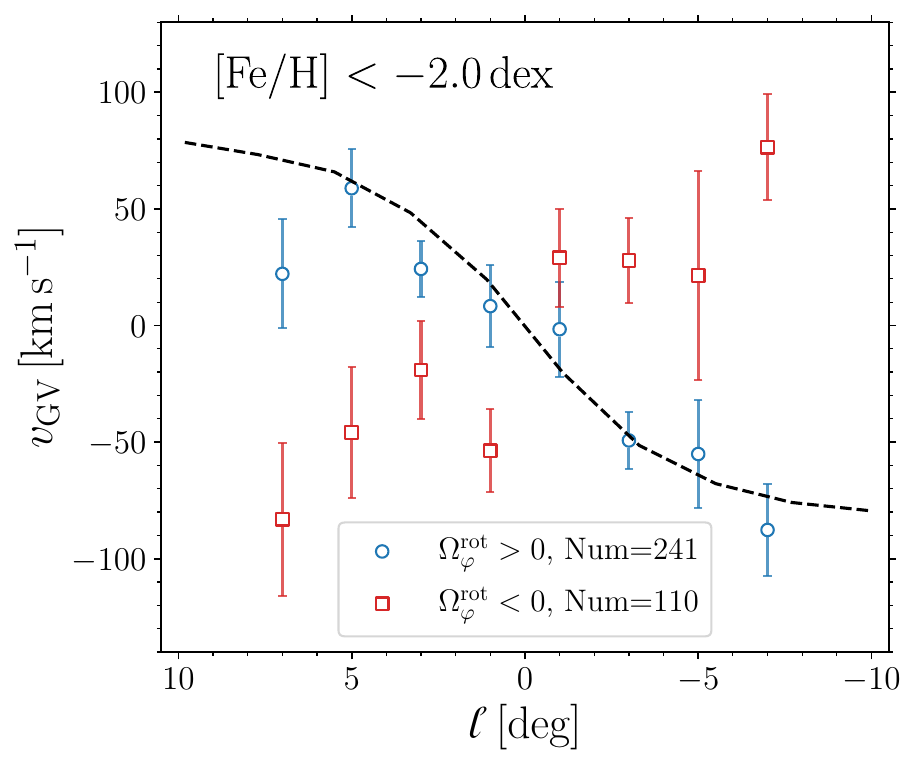}
\caption{The distribution average $v_{\rm GV}$ across Galactic longitude for metal-poor ([Fe/H] $< -2.0$\,dex) RR~Lyrae variables. The blue circles and red squares represent prograde and retrograde RR~Lyrae stars, respectively. Here we take bins in Galactic longitude between between $-8.0$ to $8.0$\,deg in steps of $2$\,deg. The dashed line represents the rotation curve for the bar model from \citet{Shen2010} at $\mathcal{b} = 4$\,deg.}
\label{fig:MetalPoorRot}
\end{figure}

We demonstrated previously (see the bottom right-hand panels of Figure~\ref{fig:VGVRotDisp-FEH}) that our dataset of RR~Lyrae stars shows little to no sign of rotation for metal-poor variables ($-2.0 > \textrm{[Fe/H]}_{\rm phot} > -2.6$\,dex). In Figure~\ref{fig:MetalPoorRot} we focus on RR~Lyrae pulsators with $\textrm{[Fe/H]}_{\rm phot}< -2.0$\,dex that remain on their orbits within the Galactic bulge (defined as $r_{\rm apo} < 3.5$\,kpc). In the metal-poor subset discussed here, retrograde RR~Lyrae stars represent approximately $30$ percent of the bulge metal-poor population. The prograde metal-poor RR~Lyrae variables in our dataset do not exhibit any preferred location within Galactic coordinates; they are evenly distributed across the analyzed footprint, and presumably well mixed with the dataset presented here. 

We note that the observed inconsistency in photometric metallicities for long-period RRab RR~Lyrae stars \citep[as described in][]{Hajdu2018,Jurcsik2021,Jurcsik2023} is likely present in our dataset. However, for the following reasons, it does not significantly affect the results outlined above. First, as mentioned in Section~\ref{sec:DataSets}, the vast majority of our RR~Lyrae dataset have distances determined using near-infrared passbands; therefore, the effect on distance is minimal \citep[see Appendix~B in][]{Prudil20253D}. Second, the ratio of the number of prograde and retrograde RR~Lyrae stars remains the same even if we shift our condition on metallicity towards more metal-poor stars\footnote{This is due to the reported offset of approximately $0.1$\,dex \citep{Jurcsik2021}.}. Thus the rotation and spatial distribution are not affected by the choice of metallicity limit when selecting the two populations.

\subsection{RR~Lyrae stars and B/P bulge} \label{subsec:BananasOrb}

\begin{figure}
\includegraphics[width=\columnwidth]{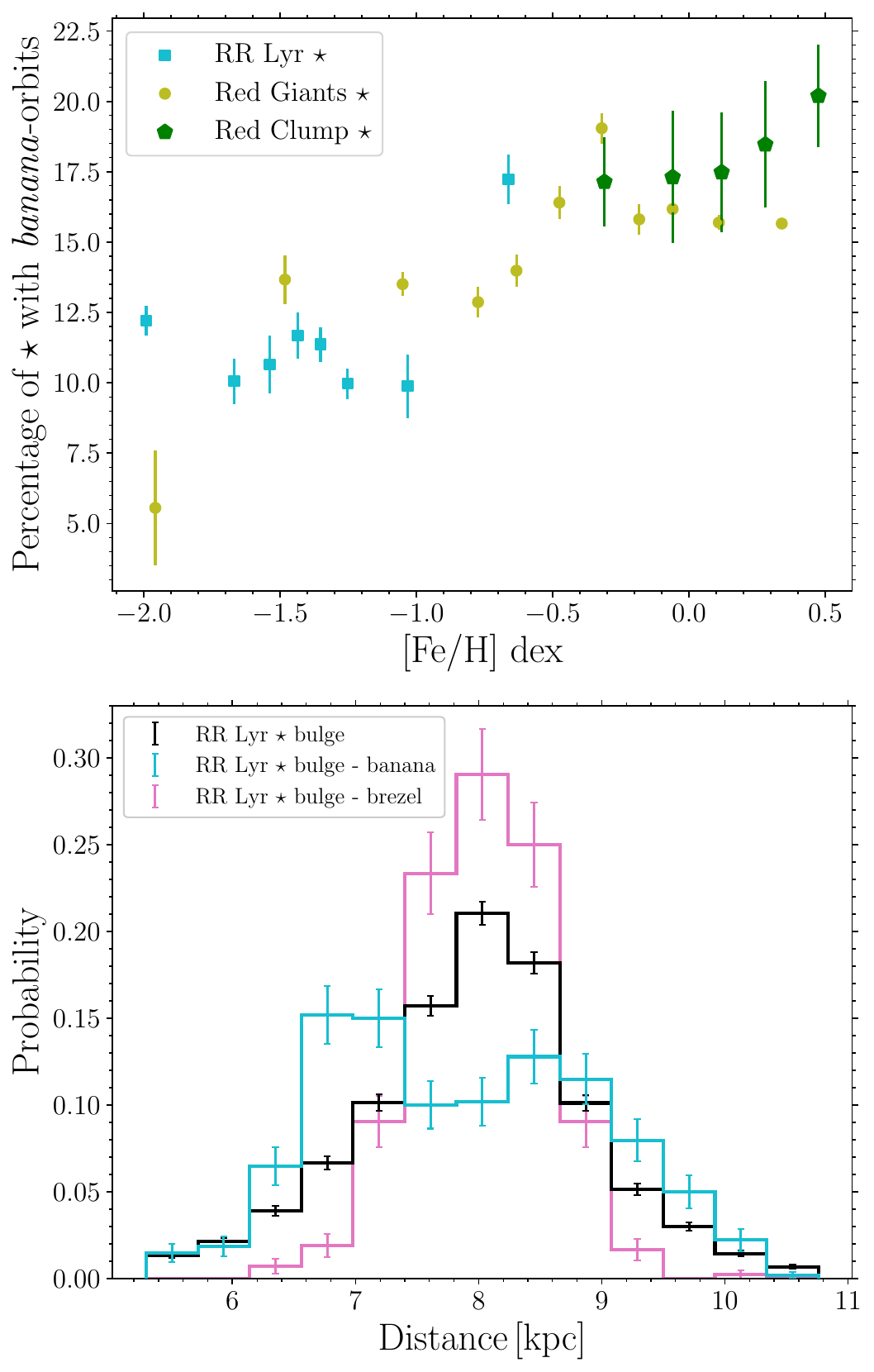}
\caption{The top panel shows the fraction of banana-orbits on metallicity for APOGEE red giants (yellow circles), BDBS red clumps (green pentagons), and RR~Lyrae (blue squares) data sets. The bottom panel shows the distance distribution of the RR~Lyrae data set. The black histogram denotes the distances of the bulge RR~Lyrae while the blue histogram shows the double-peaked distribution for RR~Lyrae stars with banana orbits. The pink histogram shows the RR~Lyrae with brezel-orbits which also has a single peak distribution. The uncertainties on individual bins (top panel) were estimated by varying the [Fe/H] by its errors and estimating the fraction of banana orbits in each iteration. For the bottom panel, the uncertainties on each bin represent the Poisson noise.}
\label{fig:RRlbanana}
\end{figure}

The bimodality in the distance distribution of the bulge red clump stars is well-known \citep{McWilliam2010,Nataf2010,Wegg2013,Lim2021} and is associated with the X-shape bulge. It is also understood that metal-poor ([Fe/H]$<-0.5$\,dex) red clump stars do not follow a bimodal distance distribution \citep[e.g.,][]{Ness2012,Rojas-Arriagada2014}. In simulations, the presence of a B/P-shaped bulge is often associated with banana-like orbits \citep[][]{Pfenniger1991,Patsis2002,Martinez-Valpuesta2006}, together with other orbital families \citep[e.g.,][]{Portail2015,Abbott2017}.

In this subsection, we use the calculated orbital and spatial properties to identify RR~Lyrae stars that support the B/P-shaped bulge. To identify stars on banana-orbits we used the following condition \citep[similar to the condition in][]{Portail2015}: 
\begin{equation} \label{eq:condBanana}
1.95 < \Omega_{\rm z}^{\rm bar} / \Omega_{\rm x}^{\rm bar} < 2.05 
\end{equation}
where $\Omega_{\rm z}^{\rm bar}$ and $\Omega_{\rm x}^{\rm bar}$ are orbital frequencies in the bar reference frame. We also follow \citet{Portail2015} in identifying brezel-orbits as
\begin{equation} \label{eq:BrezelOrb}
1.6 < \Omega_{\rm z}^{\rm bar} / \Omega_{\rm x}^{\rm bar} < 1.70 \\.
\end{equation}
In Figure~\ref{fig:RRlbanana} we show the results of our analysis. The top panel shows how the fraction of stars on banana-orbits changes with metallicity. We see a monotonic increase in the occurrence of banana orbits with metallicity with a plateau at the metal-rich tail. For the metal-rich red clump ($\textrm{[Fe/H]} > -0.5$\,dex) stars in banana orbits represent around $16$ to $20$ percent of the total number of orbits, while for metal-poor RR~Lyrae stars ([Fe/H]$_{\rm phot} < -1.0$\,dex) they constitute only $\approx 12$ percent of orbits. A likely explanation of this linear trend is kinematic fractionation \citep{Debattista2017,Fragkoudi2018}.

The bottom panel of Figure~\ref{fig:RRlbanana} shows the heliocentric distance distribution of our RR~Lyrae bulge kinematical dataset. As expected, from previous studies, RR~Lyrae stars confined to the Galactic bulge (black histogram) do not exhibit a double-peaked distance distribution. The same can be said when we look at the brezel orbits, which display similar single peak distance distributions. When we consider the RR~Lyrae stars on banana orbits however we recover a double-peaked distribution similar to the one seen for metal-rich red clump stars. One of the reasons why we generally do not see a double-peaked RR~Lyrae distribution (whole dataset) is the lower fraction of variables on banana orbits compared to the metal-rich non-variables. As shown in Section~\ref{subsec:MetalOmega} the number of retrograde orbits increases with decreasing metallicity and such orbits can dilute the potential B/P-shape of the Galactic bulge when traced with RR~Lyrae stars. 

\subsection{Metallicity, age, and $r_{\rm GC}$ dependence of $\Omega_{\varphi}^{\rm rot}$} \label{subsec:MetalOmega}

This subsection examines the chemical, age, and spatial properties of the retrograde stars and stellar particles used in this work. As seen in Figure~\ref{fig:ApocenZmaxRot}, the non-variable giants and RR~Lyrae data sets exhibit different fractions of stars with prograde and retrograde motion with respect to the MW bar. Here, we further explore this difference; in particular, we focus on the difference in ratios of stars with positive and negative $\Omega_{\varphi}^{\rm rot}$. It is important to emphasize here we use only stars (from the observed data) and stellar particles (rom the simulation) that stay confined within the Galactic bulge, i.e., with $r_{\rm apo} < 3.5$\,kpc, as in Sections~\ref{subsec:SepBulgeOutside} and~\ref{subsec:Sim}. This means that for the simulation we first selected stars within $r_{\rm GC} = 2$\,kpc, then obtained their orbits, and finally applied the condition on apocentric distance. The simulation, after forming the bar ($\approx 3$\,Gyr), forms an extended nuclear stellar disc that complicates our analysis due to its kinematic properties. Thus, in addition to the criterion on $r_{\rm apo}$, we decided to also remove stellar particles with $z_{\rm max} < 0.3$\,kpc to avoid including stellar particles associated with this structure. Likewise, our observational data set does not cover the MW nuclear stellar disc. The minimum in Galactic latitude for our dataset $\left| \mathcal{b} \right | = 2.0$\,deg, and the cut for simulation in $z_{\rm max} < 0.3$\,kpc translates to a similar restriction in Galactic latitude.

\begin{figure}
\includegraphics[width=\columnwidth]{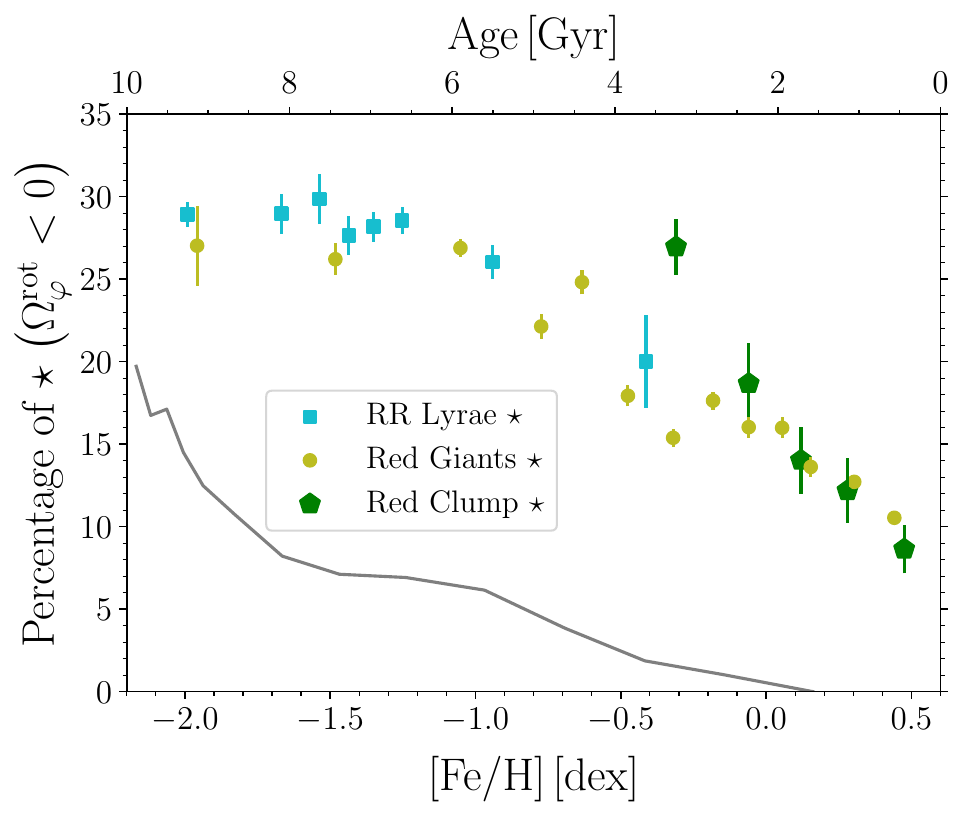}
\caption{The chemical and orbital properties for RR~Lyrae variables, BDBS Red Clump stars, and APOGEE Red Giants data sets together with results from the simulation. The Figure shows the dependence of the percentage of retrograde stars as a function of their [Fe/H]. The light blue squares represent metallicity bins for the RR~Lyrae sample. The green pentagons and yellow circles represent BDBS stars and APOGEE Red Giants, respectively. The grey line denotes a fraction of retrograde (negative $\Omega_{\varphi}^{\rm rot}$) stellar particles in the simulation as a function of age (upper axis).}
\label{fig:OmegaPhiChemistry}
\end{figure}

Figure~\ref{fig:OmegaPhiChemistry} shows a gradual increase in the number of retrograde stars with respect to the bar as the metallicity of a star decreases. Examining the metal-poor end ([Fe/H]~$<-1.0$\,dex) of our observational sample we notice a weak hint of a plateau around [Fe/H]~$\approx-1.5$\,dex. The increase in stars with prograde orbits with metallicity agrees well with previous results on giants toward the Galactic bulge \citep[see, e.g.,][]{Arentsen2020PIGSI,Lucey2021COMBSII} and RR~Lyrae stars \citep{Kunder2020}. We recover a similar trend (see grey line) by performing the binning process for our simulation but in age. There is an increase in bar counter-rotating particles (in the bar's reference frame) in the older stellar particles as compared to the younger particles (upper axis).

The gradual transition in the fraction of prograde stars with metallicity shown in Figure~\ref{fig:OmegaPhiChemistry} contrasts with the sharper transition seen in Figure~11 of \citet{Marchetti2024}, where a distinct change in kinematic alignment with the bar is observed across metallicity. Their analysis shows a somewhat sharper transition in metallicity space between a clear and diminished dipole pattern, using the correlation between Galactic proper motions as a proxy for rotation. Several factors could explain this discrepancy, including interlopers from other MW substructures, the broad selection function in our sample, or differences in how bar-supporting orbits are defined. Additionally, while the number of red clump stars does decrease significantly below [Fe/H]~$-0.5$,dex, the dominance of metal-rich stars in bar-like orbits, evident in the presence of a double red clump and strong dipole signature, suggests we might expect a sharper kinematic break. The smoother trend in our data could be due to the inclusion of overlapping bulge, bar, and inner disk populations that are not cleanly separated in our orbit-based selection. Notably, in our Figure~\ref{fig:OmegaPhiChemistry}, the most metal-poor bin ($\approx -0.35$\,dex) for red clump stars \citep[that come from BDBS, just like data from][study]{Marchetti2024} is significantly higher than the rest of the RR~Lyrae and red giant dataset at similar metallicity.

\begin{figure}
\includegraphics[width=\columnwidth]{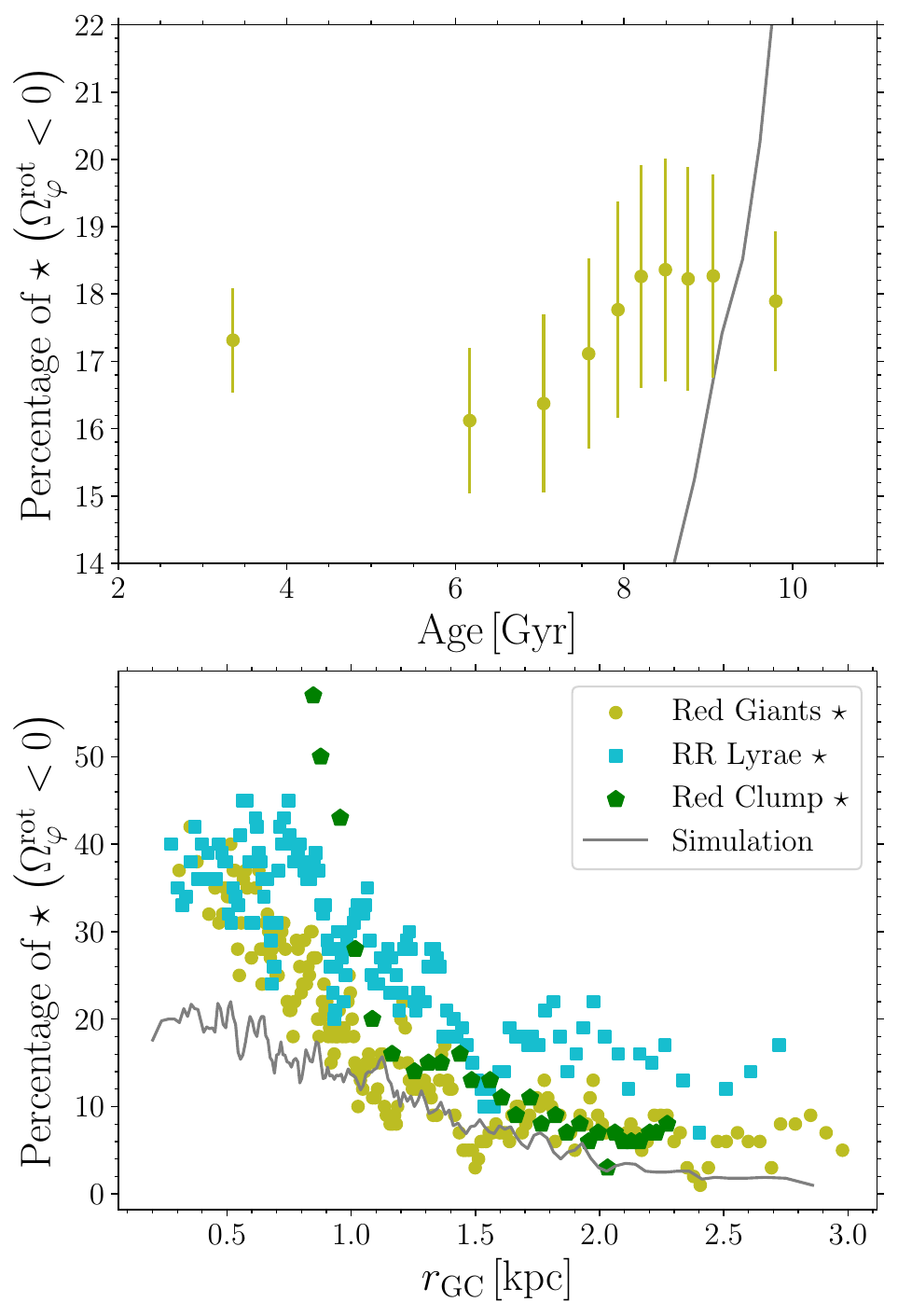}
\caption{The percentage of stars with negative $\Omega_{\varphi}^{\rm rot}$ as a function of age (top panel), and Galactocentric spherical radius (bottom panel). The grey solid line represents the trend obtained from the simulation.}
\label{fig:OmegaPhiRgc}
\end{figure}

In Figure~\ref{fig:OmegaPhiRgc}, we show the dependence of $\Omega_{\varphi}^{\rm rot}$ on stellar age and $r_{\rm GC}$. The stellar ages were taken for our APOGEE Red Giant dataset, from the \texttt{astroNN} value-added catalog \citep{Mackereth2019}. We again consider only stars on orbits that are confined to the Galactic bulge. As expected, based on results depicted in Figure~\ref{fig:OmegaPhiChemistry}, we see an increase of retrograde stars as a function of age (from ages six to ten Gyr). The simulation reproduces the increase as a function of time, but not the fraction of retrograde stars. In the bottom panel of Figure~\ref{fig:OmegaPhiRgc} we observe that the fraction of stars on retrograde orbits increases as we move closer to the Galactic center. At radii $r_{\rm GC} \approx 0.5$\,kpc, the fraction reaches almost $40$\% for RR~Lyrae variables and $35$\% for APOGEE red giants. The observed increase at smaller $r_{\rm GC}$ likely explains the findings in \citet{Kunder2020} and \citet{OlivaresCarvajal2024}, where the more centrally concentrated population of RR~Lyrae stars does not trace the Galactic bar. Our results indicate that the number of stars with prograde bar-supporting orbits decreases as we move toward the Galactic center. The same trend is also obtained from the simulation, where we observe an increase in retrograde stars for smaller $r_{\rm GC}$. We note that for our observational dataset, we used a running boxcar with size $100$ and step $30$ to create Figure~\ref{fig:OmegaPhiRgc}. In the case of the simulation, we used a box with a size of $1000$ and a step equal to $350$. We also scaled the spatial coordinates for each stellar particle by a factor of $1.7$ \citep{GoughKelly2022} to calculate $r_{\rm GC}$. This scaling was done to facilitate comparison between the simulation and the observational data since the bar of the model is smaller than that in the MW. The percent of $\Omega_{\varphi}^{\rm rot}$ is considerably higher at $r_{\rm GC} < 0.5$ in the observations than in the simulation. This could potentially indicate that the simulation is not exactly the MW (e.g., the difference in age of the bar in the simulation and in the MW) and may also point toward an additional old, spheroidal classical bulge population at small Galactocentric radii \citep[as reported in some of the previous studies, see][]{Rix2022,Belokurov2022}.

\begin{figure}
\includegraphics[width=\columnwidth]{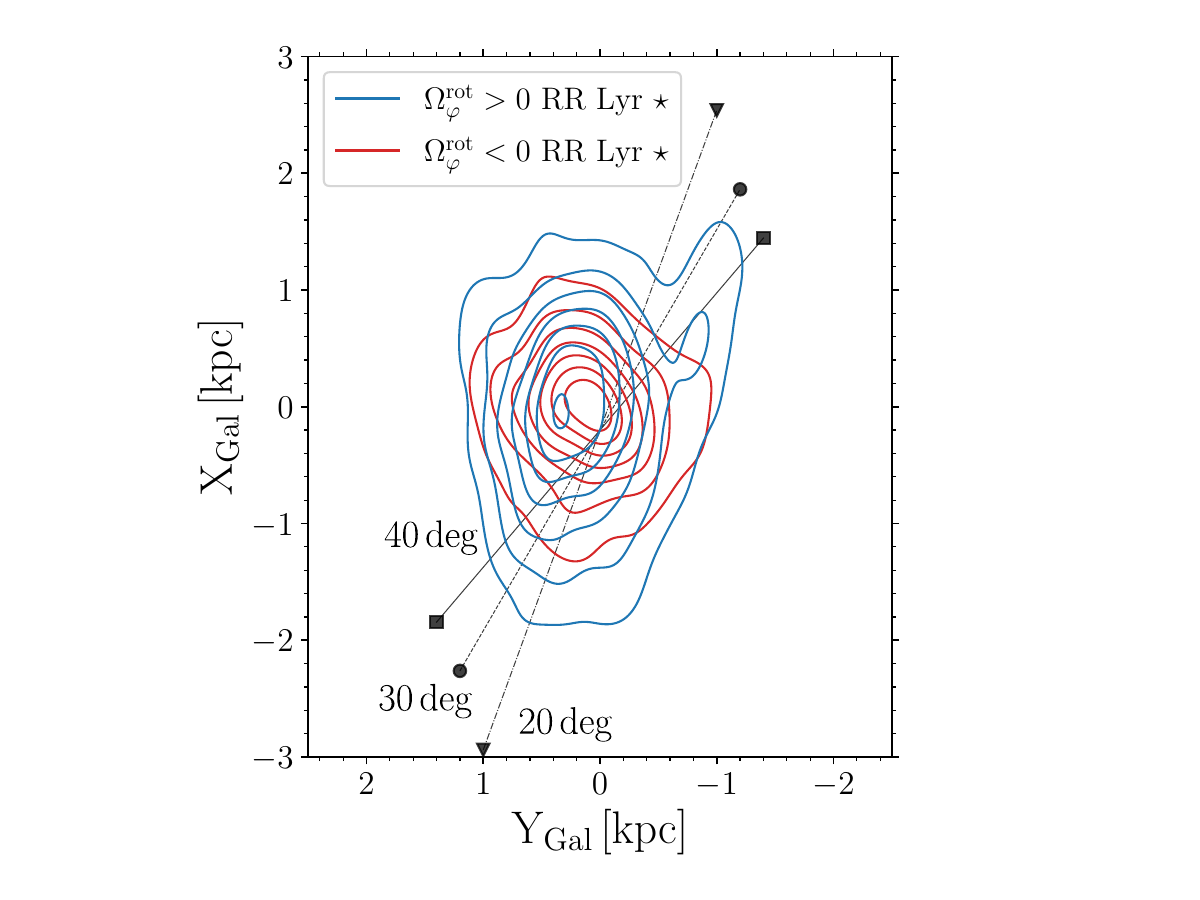}
\caption{The spatial distribution depicted using the kernel density estimates of the Cartesian coordinates for our RR~Lyrae bulge dataset. Blue and red contours show the distributions of prograde and retrograde RR~Lyrae stars. The black solid, dashed and dotted lines represent bar angles of $40$, $30$, and $20$ degrees, respectively. The Sun in this perspective would be at Y$_{\rm Gal} = 0$\,kpc and X$_{\rm Gal} = -8.2$\,kpc.}
\label{fig:SpatBar}
\end{figure}

As shown in Figures~\ref{fig:ApocenZmaxRot},~\ref{fig:MetalPoorRot},~\ref{fig:RRlbanana}, and~\ref{fig:OmegaPhiChemistry} orbital frequencies can separate stars supporting the bar from those that do not support the bar. They can also reveal some previously hidden features in the spatial properties which we explore further in Figure~\ref{fig:SpatBar} where we depict the spatial distribution in the Cartesian coordinates using kernel density estimates for our RR~Lyrae bulge dataset. We split our variable sample into prograde and retrograde based on the sign of the $(\Omega_{\varphi}^{\rm rot})$. We see that the retrograde RR~Lyrae population is more centrally concentrated with a somewhat more circular distribution while prograde RR~Lyrae stars have a more elongated shape. To quantify this difference, we measured the bar angle for both the prograde and retrograde RR~Lyrae groups. We applied the inertia tensor method, following our previous work \citep[see Section~6 in][]{Prudil20253D}, and selected only stars located within $0.2 < |Z| < 0.5$\,kpc. For the prograde and retrograde populations, we obtained bar angles of $\iota = 16 \pm 3$\,deg and $\iota = -5 \pm 5$\,deg, respectively. We note that a smaller value of $R_{\rm cyl}^{\rm lim} = 1.2$\,kpc was used to reduce potential selection effects since our kinematic dataset is more spatially constrained and less numerous. As expected, based on orbital frequencies we can detect stars that have bar-supporting orbits, and their selection results in a bar-like feature in their spatial properties.

\section{Chaoticity and orbital families of RR~Lyrae stars} \label{sec:RetroOrbitsFam}

In this section we examine the orbital frequencies of our dataset, evaluate their chaoticity (using a parameter describing how much the orbital frequencies are conserved), and identify associated orbital families. Orbital frequencies often serve to identify resonances, distinguish chaotic and regular orbits, and classify orbits into orbital families. We focus only on RR~Lyrae variables that satisfy the condition in Eq.~\ref{eq:Condition1} and those which stay on their orbit within $3.5$\,kpc ($4877$ single-mode RR~Lyrae stars). This dataset covers Jacobi integrals, $E_{\rm J}$, in the interval $E_{\rm J} = (-2.6, -1.6) / 10^{5}$\,[km$^{2}$\,s$^{-2}$]. 

To determine whether an orbit is chaotic or regular, we employ two approaches. The first one is based on frequency drift \citep[$\log_{10}\Delta\Omega$, also known as frequency diffusion rate,][]{Laskar1993,Valluri2010}. The frequency drift is defined as the maximum difference in frequencies:
\begin{equation}
\log_{10}\Delta\Omega = \log_{10} \left( \max \left| \frac{\Omega_{i}(t_{2}) - \Omega_{i}(t_{1})}{\Omega_{i}(t_{1})} \right|_{i=1,2,3} \right) \\,
\end{equation}
where $t_{1}$ and $t_{2}$ represent two equal segments of the orbit time interval, and $\Omega_{i}$ represents the three frequencies in cylindrical coordinates. For regular orbits, $\Delta\Omega$ is nearly zero, as regular orbits conserve frequencies, while chaotic orbits exhibit larger $\Delta\Omega$ values ($0.1$ and higher). Our second metric of orbit chaoticity is the Lyapunov exponent \citep[$\Lambda$,][]{Lyapunov1992}. The Lyapunov exponent measures the sensitivity of orbits to initial conditions by quantifying the rate of exponential separation of initially close trajectories. A higher Lyapunov exponent indicates chaotic behavior of orbits, while lower values suggest more regular behavior. We estimated the Lyapunov exponent using the implementation in \texttt{AGAMA} \citep{Vasiliev2013}.

\begin{figure}
\includegraphics[width=\columnwidth]{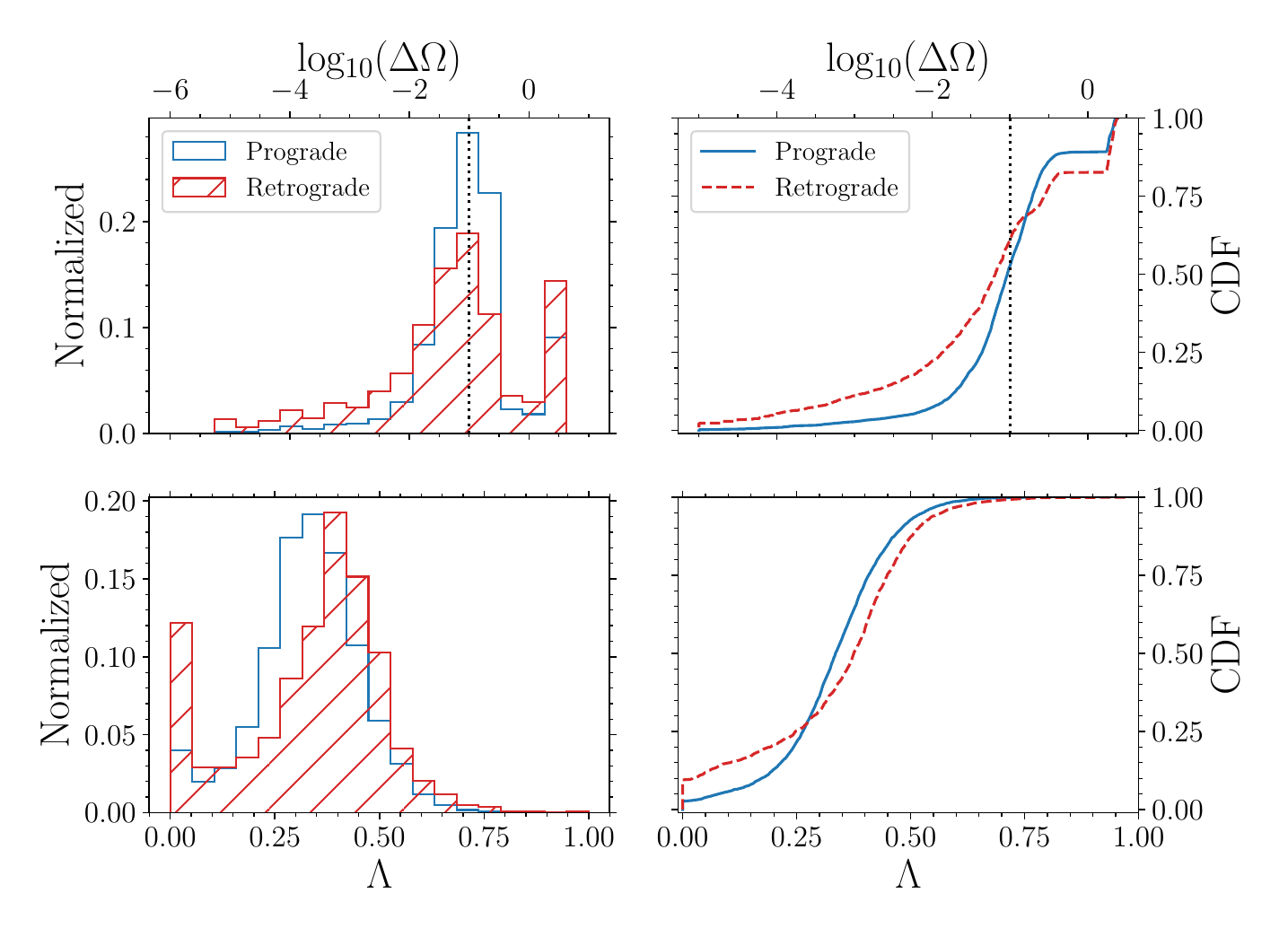}
\caption{The distribution (left-hand panels) and cumulative distribution functions (right-hand panels) of the frequency drift (top panels) and Lyapunov exponent (bottom panels) for retrograde (red lines) and prograde (blue lines) RR~Lyrae stars. Vertical black lines in upper panels denote our separation between chaotic and regular orbits (see Eq.~\ref{eq:ChaoticityCond}).}
\label{fig:ChaoticityOfRR}
\end{figure}

In Figure~\ref{fig:ChaoticityOfRR}, we display the estimated chaoticity and regularity properties of the RR~Lyrae stars. We particularly investigate the differences in periodicity between prograde and retrograde orbits. Using both metrics—frequency drift and the Lyapunov exponent—we find that the distributions of both orbital groups overlap reasonably well. This indicates that orbits in both groups appear to be similarly chaotic or regular and that retrograde orbits seem to form a somewhat stable structure in the central part of the Galactic bulge.

\subsection{Orbital families - Cartesian coordinate system}

Initially, we utilized the combination of fundamental frequencies (time derivatives of the angle variables) in the Cartesian coordinate system (in a bar-rotating frame) to generate frequency maps and automatically identify resonant orbits within our dataset. In frequency maps, resonant orbits are not distributed randomly in the plane, but group together along resonance lines that satisfy the resonant condition $k_{1}\Omega_{1} + k_{2}\Omega_{2} + k_{3}\Omega_{3} = 0$ (applicable to both Cartesian and cylindrical coordinates), where the vector $k=(k_{1}, k_{2}, k_{3})$ consists of integers. By examining frequency maps (or frequency ratios), we can discern the majority of orbital families \citep[e.g.,][]{Portail2015,Valluri2016} associated with various resonances. For simplicity, we classify them into two main groups: tube and box orbits. Tube orbits rotate around the long ($x$, bar is oriented along the x-axis), intermediate ($y$)\footnote{Note that the intermediate, $y$-axis, tubes are unstable and may not exist \citep{Adams2007}.}, or short ($z$) axes. Conversely, box orbits are mostly non-resonant, and do not have a definite sense of rotation.

\begin{figure}
\includegraphics[width=\columnwidth]{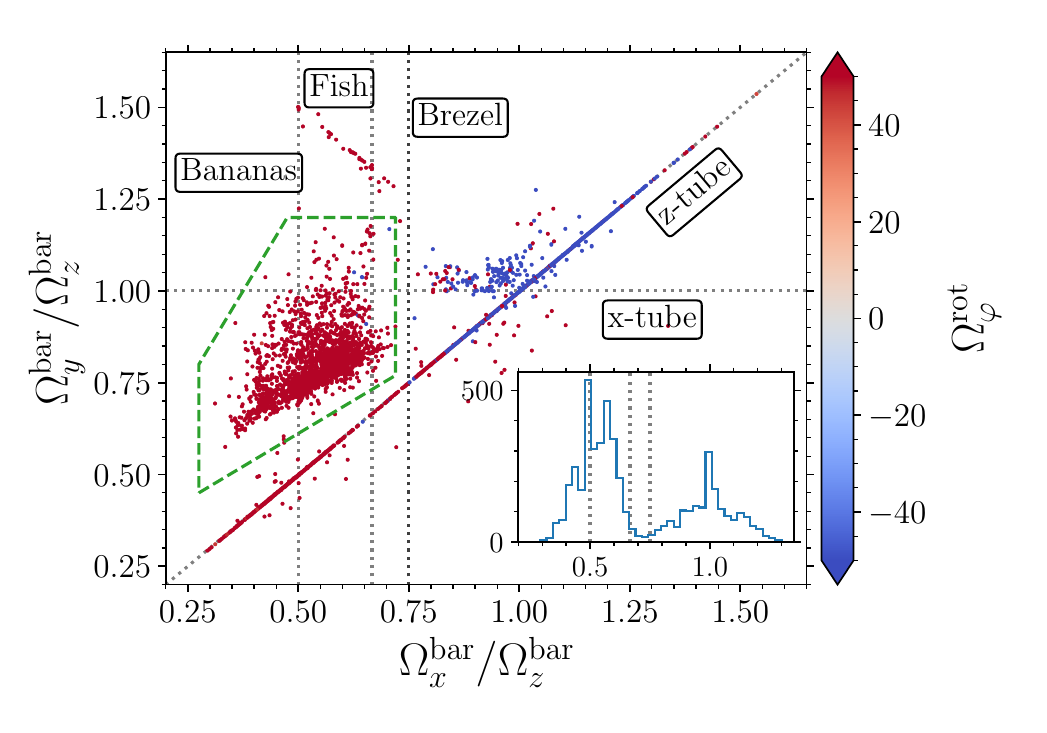}
\caption{Frequency map illustrating fundamental frequencies in the Cartesian coordinate system and bar-rotating reference frame for our RR~Lyrae dataset. Color-coding is used to represent positive and negative values of $\Omega_{\varphi}^{\rm rot}$. Major orbital families are labeled and delineated with dotted lines. Additionally, dashed green lines highlight a region above the $z$-tube resonance. In the inset, we present the distribution of $\Omega_{x}^{\rm bar} / \Omega_{z}^{\rm bar}$ using the same vertical grey dotted lines as in the main panel.}
\label{fig:FreqMapOmegaxyz}
\end{figure}

In Figure~\ref{fig:FreqMapOmegaxyz}, we present a frequency map for our RR~Lyrae bulge dataset. A notable observation is the clustering along the short axis $z$-tubes (diagonal line) and a significant concentration above the $z$-tube resonance. Furthermore, we observe a high density of stars with negative azimuthal frequency, $\Omega_{\varphi}^{\rm rot}$, near frequency ratios close to one ($\Omega_{x}^{\rm bar} / \Omega_{z}^{\rm bar} \approx 1.0$ and $\Omega_{y}^{\rm bar} / \Omega_{z}^{\rm bar} \approx 1.0$). The green area marked in Fig.~\ref{fig:FreqMapOmegaxyz} contains orbits exhibiting mainly chaotic behavior.

Among the most populated groups exhibiting prograde motion in the bar frame are the so-called banana orbits, situated at $\Omega_{x}^{\rm bar} / \Omega_{z}^{\rm bar} = 0.5$. Additionally, the fish-orbits at $\Omega_{x}^{\rm bar} / \Omega_{z}^{\rm bar} = 2/3$, and the brezel-orbits \citep[$\Omega_{x}^{\rm bar} / \Omega_{z}^{\rm bar} = 3/5$, see][and condition in Eq.~\ref{eq:BrezelOrb}]{Portail2015} are also present (although in low numbers), exhibiting characteristics of both box and tube ($z$-tube) orbits.

Conversely, there are few to no orbits at $\Omega_{x}^{\rm bar} / \Omega_{z}^{\rm bar} = 3/4$, as well as other orbits such as $\Omega_{x}^{\rm bar} / \Omega_{z}^{\rm bar} = 4/5$ and $\Omega_{x}^{\rm bar} / \Omega_{z}^{\rm bar} = 5/6$. A small proportion of the remaining prograde variables fall within the region $\Omega_{x}^{\rm bar} / \Omega_{z}^{\rm bar} \simeq 1.0$ and $\Omega_{y}^{\rm bar} / \Omega_{z}^{\rm bar} \simeq 1.0$, where they can be classified as $z$-tubes of the x2 orbital family \citep[e.g.,][]{Contopoulos1980}. This region also hosts the majority of retrograde RR~Lyrae stars, which are centered around $\Omega_{x}^{\rm bar} / \Omega_{z}^{\rm bar} = 1.0$ and $\Omega_{y}^{\rm bar} / \Omega_{z}^{\rm bar} = 1.0$, at the intersection of $z$-tube and $x$-tube resonances. 

\subsection{Orbital families - cylindrical coordinate system}

In this subsection, we further explore the association with orbital families based on the frequency maps. We adopt a framework similar to that of \citet{BeraldoSilva2023} and concentrate on the cylindrical coordinate system in the inertial reference frame. Additionally, we closely examine the orbital shapes associated with various regions of the frequency map, primarily displaying those orbits that exhibit regular behavior (non-chaotic, see condition in Eq.~\ref{eq:ChaoticityCond}). 

\subsubsection{Prograde stars}

\begin{figure*}
\includegraphics[width=2\columnwidth]{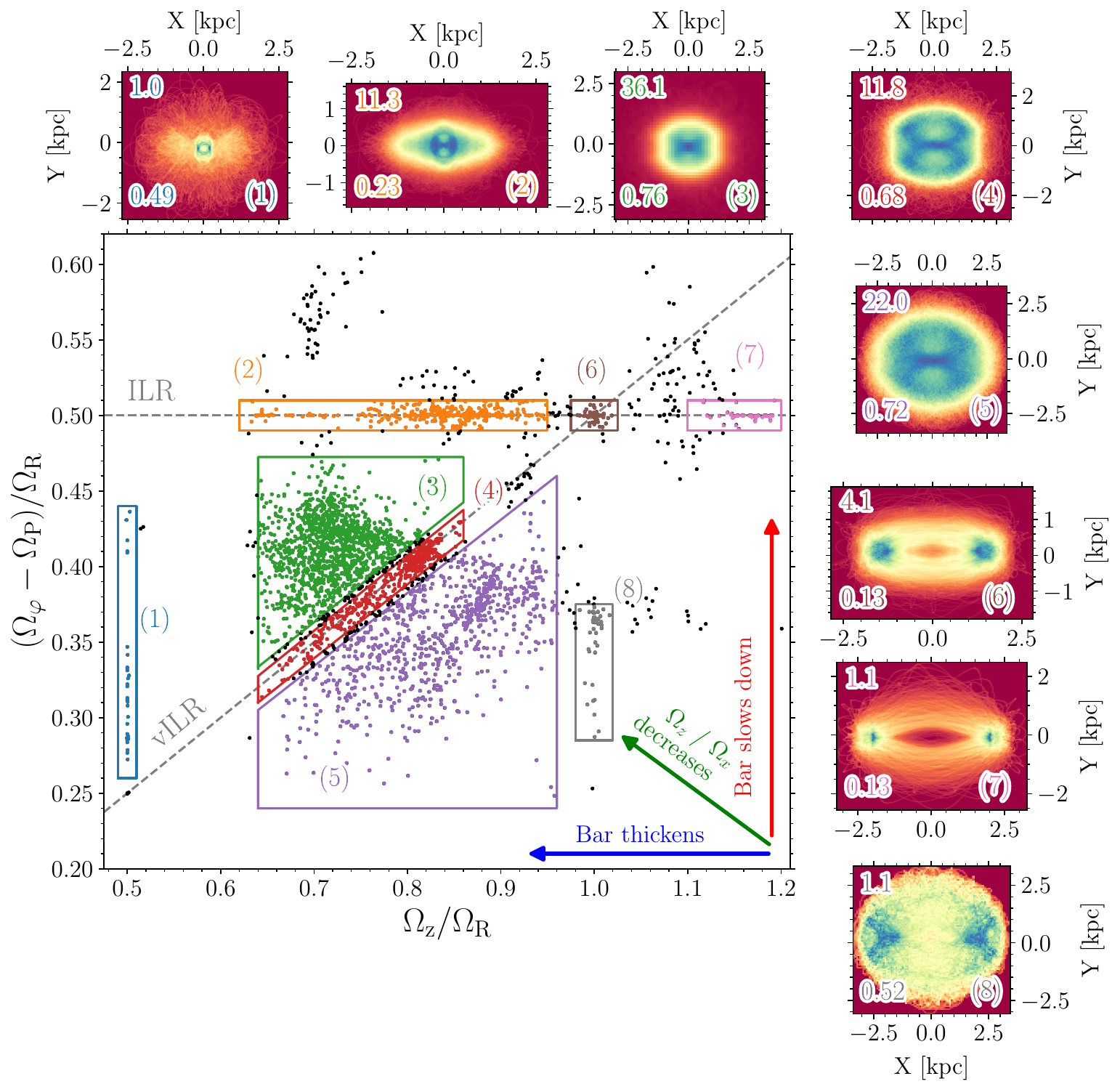}
\caption{Frequency map showing fundamental frequencies in cylindrical coordinates for prograde RR~Lyrae stars in the Galactic bulge dataset. The insets present 2D density maps of the face-on projection (with power-law normalization) of non-chaotic ($\log\Delta\Omega < -1.0$) orbits in a rotating reference frame for eight selected regions, with high density depicted in blue and low density in red. The insets for each region (marked in the right bottom corner) also contain an estimate of circularity for non-chaotic orbits (bottom left corner) of a given region and the percentage of stars (both with chaotic and periodic orbits) in a given region to the total prograde sample (top left corner). Two significant resonances are identified: the inner Lindblad resonance and the vertical inner Lindblad resonance.}
\label{fig:MapCylPROGRADE-x-y}
\end{figure*}

In Figure~\ref{fig:MapCylPROGRADE-x-y}, we show a frequency map for prograde (in the bar reference frame) RR~Lyrae confined to the Galactic bulge in the cylindrical coordinate system and insets depicting orbits, fulfilling the following condition \citep{Valluri2016}:
\begin{equation} \label{eq:ChaoticityCond}
\log\Delta\Omega < -1.0 \\ .
\end{equation}
Orbits that fulfill this condition are considered regular. For interested readers, we include two additional Figures (Figure~\ref{fig:MapCylPROGRADE-x-z} and Figure~\ref{fig:MapCylPROGRADE-y-z}) with insets depicting different orbital planes in the Appendix~\ref{sec:AppFigure}. We divided the frequency map into eight regions and examined the orbits in each selected sector. The level of chaoticity in individual sectors is listed in Table~\ref{table:ChaoticityProret}. For orbits in each sector, we also estimated the flattening of non-chaotic orbits, which we refer to as 'circularity' in this study. The circularity was estimated as a ratio between the minimum and maximum eigenvalue calculated from each region's covariance matrix of the ensemble of orbits coordinates (e.g., in Figure~\ref{fig:MapCylPROGRADE-x-y} for coordinates X and Y). From the circularity we see that orbits in nearly all eight sectors have mostly elongated shapes supporting the bar-like morphology.

In Figure~\ref{fig:MapCylPROGRADE-x-y}, regions 2, 3, and 4 correspond to the inner Lindblad resonance (ILR), the vertical inner Lindblad resonance cloud \citep[vILR cloud,][]{BeraldoSilva2023} and the vILR. These regions are densely populated by RR~Lyrae variables and overlap with the green area indicated in Figure~\ref{fig:FreqMapOmegaxyz}. Sectors 3 and 5, along with regions 2 and 4, have the highest star counts among all regions and contribute to the boxy-peanut shape of the RR~Lyrae variables (see Figure~\ref{fig:MapCylPROGRADE-x-z}). In addition, regions 2 and 3 are occupied by stars with the lowest $E_{\rm J}$ (confined to the central regions) and $L_{z}$ among the prograde stars.

From the insets we notice that regions 4, 5, 6, 7, and 8 do not contribute significantly to the Galaxy's central region, suggesting their association with $z$-tube orbits, which are located in the $x$–$y$ plane and revolve around the $z$-axis. This classification is further confirmed when examining the net angular momentum in the $z$ direction ($L_{z}$), which $z$-tube orbits have (see Figure~\ref{fig:LzAngProg} in the Appendix). In contrast, variables in regions 1 and 2 display lower chaoticity; interestingly region 2 exhibits a bar-like shape in its face-on projection, while a box-like shape is evident in its side-on projection (see Figure~\ref{fig:MapCylPROGRADE-x-z}).

\subsubsection{Retrograde stars}

\begin{figure*}
\includegraphics[width=2\columnwidth]{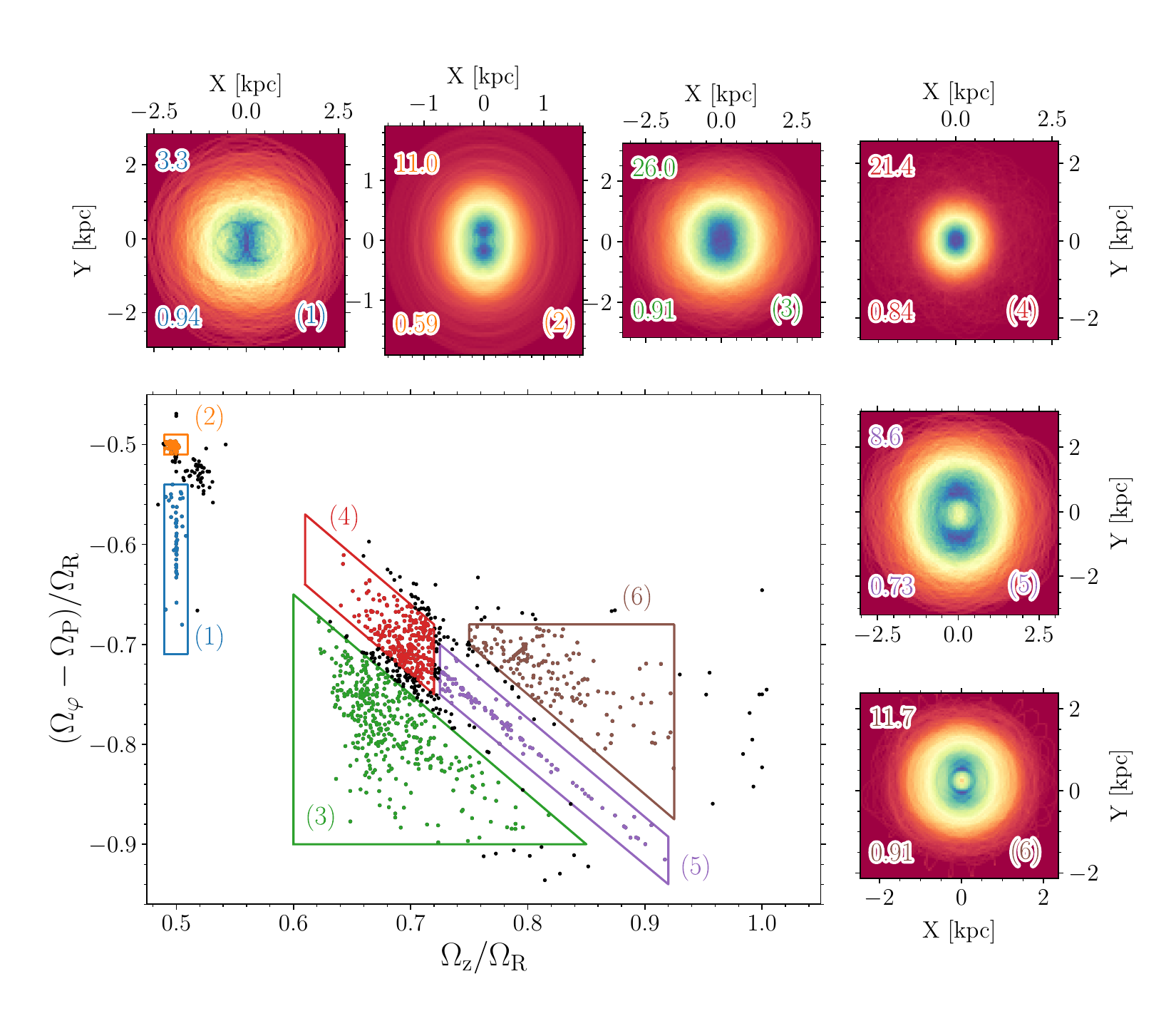}
\caption{Same as Figure~\ref{fig:MapCylPROGRADE-x-y} but for retrograde RR~Lyrae stars with insets (face on projection) showing 2D density maps of mostly stable orbits (with $\textrm{log}(\Delta\Omega) < -1$).} 
\label{fig:MapCylRETROGRADE-x-y}
\end{figure*}

Figure~\ref{fig:MapCylRETROGRADE-x-y} presents the same frequency map as depicted in Figure~\ref{fig:MapCylPROGRADE-x-y}, but for retrograde RR~Lyrae stars. We have partitioned the frequency space into six regions and displayed regular orbits (based on condition in Eq.~\ref{eq:ChaoticityCond}) of each sector in the accompanying insets (in a face-on projection). For side-on and end-on projections of these retrograde RR~Lyrae orbits, see Figures~\ref{fig:MapCylRETROGRADE-x-z} and \ref{fig:MapCylRETROGRADE-y-z}. In general, stars (both variable and non-variable) with retrograde orbits have on average lower $E_{\rm J}$ and $\left| L_{z} \right |$ than their prograde counterparts, meaning that they are more tightly bound to the innermost regions of the Galactic bulge.

A significant difference in the shape of the retrograde orbits compared to the prograde ones is immediately evident. Retrograde RR~Lyrae orbits exhibit a rounder shape, whereas their prograde counterparts predominantly display x-axis elongated orbits (looking at the circularity in subplots of Figures~\ref{fig:MapCylRETROGRADE-x-y} and~\ref{fig:MapCylPROGRADE-x-y}). The same applies if we look at the other orbital projections (side-on and end-on) depicted in the Appendix (see Figures~\ref{fig:MapCylPROGRADE-x-z}, \ref{fig:MapCylPROGRADE-y-z}, \ref{fig:MapCylRETROGRADE-x-z}, and \ref{fig:MapCylRETROGRADE-y-z}). The spatial distribution of retrograde orbits is more spheroidal while prograde orbits show more ellipsoidal shapes.

We again use the angular momenta (see Figure~\ref{fig:LzLxAngRet} in the Appendix) of regular orbits to classify individual sectors into some of the orbital families. Stars in regions 1, 5, and 6 exhibit a median non-zero angular momentum in the $z$ direction, indicating that they predominantly follow $z$-tube orbits. The orbits in the two most numerous and chaotic sectors (3 and 4) are challenging to classify; however, we note that sector 3 shows some net angular momenta in the $z$ direction. Conversely, sectors 1 and 2 display median net angular momenta in the $x$ direction, and coupled with the shape of their orbits, we suggest their classification as $x$-tube orbits (see insets in Figure~\ref{fig:MapCylRETROGRADE-y-z}). 

Expanding on the discussion of chaoticity among retrograde and prograde stars (displayed in Figure~\ref{fig:ChaoticityOfRR}), the degree of regularity among retrograde RR~Lyrae stars appears to be comparable (in median) to that of the prograde moving bulge RR~Lyrae variables. For instance, out of the $3429$ RR~Lyrae stars depicted in the main panel of Figure~\ref{fig:MapCylPROGRADE-x-y}, $52$ percent have $\log\Delta\Omega < -1.0$. Conversely, in the frequency map of retrograde RR~Lyrae variables (Figure~\ref{fig:MapCylRETROGRADE-x-y}), we observed $61$ percent of objects fulfilling the aforementioned condition (out of $1374$ stars). 

As a summary Table~\ref{table:ChaoticityProret} lists the median chaoticity values for each region for RR~Lyrae and non-variable datasets. We note the similar frequency of regular orbits for the prograde red giants as for the prograde RR~Lyrae variables. The same can also be said about median values of frequency drift and Lyapunov exponent. The only exception seems to be region 8 where RR~Lyrae stars appear more chaotic than red giants, but for still the majority of variables in this area show regular orbits. 

\begin{table*}
\caption{Estimated level of chaoticity and circularity for eight prograde sectors, marked in the frequency map depicted in Figure~\ref{fig:MapCylPROGRADE-x-y} and six retrograde regions Figure~\ref{fig:MapCylRETROGRADE-x-y}.}
\setlength{\tabcolsep}{4.5pt}
\begin{tabular}{c|ccccccc|ccccccc} 
& \multicolumn{7}{c}{Prograde RR~Lyrae $\star$} & \multicolumn{7}{|c}{Prograde Red giant $\star$} \\ \hline 
\rule{0pt}{13pt} Reg. & N.~$\star$ & $\widetilde{\textrm{log}_{10}(\Delta\Omega)}$ & $\widetilde{\Lambda}$ & \% of reg. & Circ$_{x}^{y}$ & Circ$_{x}^{z}$ & Circ$_{y}^{z}$ & N.~$\star$ & $\widetilde{\textrm{log}_{10}(\Delta\Omega)}$ & $\widetilde{\Lambda}$ & \% of reg. & Circ$_{x}^{y}$ & Circ$_{x}^{z}$ & Circ$_{y}^{z}$ \\ \hline
1 & 34 & $-1.26$ & 0.16 & 52.9 & 0.487 & 0.486 & 0.237 & 20 & $-0.75$ & 0.22 & 45.0 & 0.565 & 0.460 & 0.260 \\ 
2 & 389 & $-1.07$ & 0.28 & 53.5 & 0.231 & 0.374 & 0.619 & 489 & $-1.20$ & 0.26 & 57.5 & 0.189 & 0.307 & 0.615 \\ 
3 & 1239 & $-1.04$ & 0.35 & 52.6 & 0.761 & 0.455 & 0.598 & 1546 & $-1.06$ & 0.35 & 54.5 & 0.740 & 0.430 & 0.582 \\ 
4 & 406 & $-1.08$ & 0.33 & 56.9 & 0.682 & 0.267 & 0.391 & 587 & $-1.06$ & 0.33 & 56.6 & 0.653 & 0.282 & 0.432 \\ 
5 & 754 & $-0.95$ & 0.40 & 44.3 & 0.720 & 0.261 & 0.362 & 1017 & $-0.96$ & 0.40 & 46.7 & 0.640 & 0.212 & 0.331 \\ 
6 & 139 & $-1.95$ & 0.13 & 73.4 & 0.130 & 0.157 & 0.823 & 290 & $-2.15$ & 0.11 & 74.8 & 0.122 & 0.134 & 0.913 \\ 
7 & 37 & $-1.67$ & 0.19 & 70.3 & 0.129 & 0.086 & 0.668 & 136 & $-1.66$ & 0.19 & 84.6 & 0.125 & 0.086 & 0.685 \\ 
8 & 38 & $-1.06$ & 0.28 & 60.5 & 0.516 & 0.093 & 0.181 & 93 & $-1.23$ & 0.25 & 64.5 & 0.408 & 0.069 & 0.168 \\ \hline
\multicolumn{8}{c}{} \\ 
& \multicolumn{7}{c}{Retrograde RR~Lyrae $\star$} & \multicolumn{7}{|c}{Retrograde Red giant $\star$} \\ \hline
\rule{0pt}{13pt} Reg. & N.~$\star$ & $\widetilde{\textrm{log}_{10}(\Delta\Omega)}$ & $\widetilde{\Lambda}$ & \% of reg. & Circ$_{x}^{y}$ & Circ$_{x}^{z}$ & Circ$_{y}^{z}$ & N.~$\star$ & $\widetilde{\textrm{log}_{10}(\Delta\Omega)}$ & $\widetilde{\Lambda}$ & \% of reg. & Circ$_{x}^{y}$ & Circ$_{x}^{z}$ & Circ$_{y}^{z}$ \\ \hline
1 & 45 & $-1.41$ & 0.34 & 62.2 & 0.936 & 0.492 & 0.494 & 17 & $-0.59$ & 0.37 & 41.2 & 0.976 & 0.483 & 0.494 \\ 
2 & 152 & $-2.62$ & 0.22 & 90.8 & 0.588 & 0.556 & 0.866 & 96 & $-2.55$ & 0.20 & 87.5 & 0.582 & 0.469 & 0.866 \\ 
3 & 358 & $-1.03$ & 0.44 & 52.2 & 0.914 & 0.771 & 0.826 & 214 & $-0.89$ & 0.44 & 44.9 & 0.922 & 0.755 & 0.822 \\ 
4 & 295 & $-1.01$ & 0.41 & 50.2 & 0.845 & 0.583 & 0.648 & 255 & $-0.92$ & 0.41 & 43.5 & 0.830 & 0.538 & 0.649 \\ 
5 & 119 & $-1.26$ & 0.30 & 65.5 & 0.730 & 0.796 & 0.670 & 69 & $-1.13$ & 0.28 & 56.5 & 0.712 & 0.650 & 0.670 \\ 
6 & 162 & $-2.37$ & 0.06 & 75.3 & 0.911 & 0.232 & 0.223 & 154 & $-1.96$ & 0.09 & 77.9 & 0.924 & 0.231 & 0.213 \\  \hline
\end{tabular}
\tablefoot{The first two columns list the sector and number of RR~Lyrae variables in each. The third and fourth columns represent the median values for frequency drift and Lyapunov exponent in a given region. The fifth column contains a fraction of RR~Lyrae pulsators with regular orbits (based on log$_{10}\Delta\Omega<-1$). Columns six, seven, and eight list circularity values for different coordinate projections. Subsequent columns list the same parameters but for red giants.}
\label{table:ChaoticityProret}
\end{table*}

Figure~\ref{fig:MapCylPROGRADE-x-y} and~\ref{fig:MapCylRETROGRADE-x-y} show that RR~Lyrae stars on retrograde orbits are more abundant in the innermost regions of the Galactic bulge and although metal-poor RR~Lyrae stars have a higher tendency to be retrograde, metal-rich stars on retrograde orbits are also present. A population of retrograde stars with high eccentricities can give rise to different features that have been observed previously and attributed to a classical bulge, as discussed more in Section~\ref{sec:Discussion}.

\section{Temporal evolution of orbits in the simulation} \label{sec:simulationEND}

In the previous Sections, we confirmed that the older, metal-poor population lags in rotation compared to the younger, metal-rich counterpart (both for variable and non-variable stars). We also showed that the rotation curve of metal-poor stars can be recovered using orbital frequencies (see Section~\ref{subsec:metPoor}). There seems to be a nearly monotonic progression where, as we move from metal-rich to metal-poor stars, the fraction of stars on retrograde orbits (in the bar reference frame) increases. This trend is also recovered when we use stellar ages for our Red Giant data set. The fraction of retrograde stars increases as we move toward the Galactic center. Similar trends are also observed in the simulation, particularly for the spatial distribution, where we find a good match between RR~Lyrae stars and stellar particles of all ages. In the case of RR~Lyrae stars, the retrograde orbits seem to be as regular (non-chaotic) as the prograde orbits. Therefore, they can be considered as a spheroidal structure within the Galactic bulge. Since this structure with similar trends in spatial properties and metallicity is also observed in the simulation, we believe it has a secular origin. A secular origin in this context means an internal formation since our simulated galaxy evolved without any accretion event.

The emergence of the retrograde population (in the inertial frame) in this simulation has been previously explored by \citet[][their model M3\_nc\_b]{Fiteni2021}. They found that approximately $15$ percent of stellar particles within the inner $4$\,kpc of the simulation have retrograde orbits (based on angular momenta, $L_{z}$), and they associate their origin with the bar. All of the retrograde stellar particles in their study for this simulation are found within $4$\,kpc, and their number increases with time as the bar grows. 

In the following, we study the evolution of orbital frequencies as a function of stellar ages (as a proxy for metallicity) by looking at the individual snapshots of the simulation. We use different strategies to analyze the simulation, going beyond that in Sections~\ref{subsec:Sim} and~\ref{subsec:MetalOmega}. We start by using all stellar particles from a snapshot at $10$\,Gyr and follow the approach described in \citet{GoughKelly2022} to align the simulation with the MW. We adjust its spatial properties by a factor of $1.7$ and velocities by a factor of $0.48$ to match the bar and bulge characteristics. Additionally, we rotate the stellar particles by $27$\,degrees around the $z$-axis to replicate the tilt of the MW bar as estimated by \citet{Wegg2013}. Employing coordinate transformation routines implemented in \texttt{galpy} \citep{Bovy2015}, we position the observer at $X_{\rm GC} = -8.2$\,kpc and transform the coordinates and velocities of the stellar particles to Galactic coordinates, distances, and Galactocentric velocity $v_{\rm GV}$. This scaled simulation enables us to select stellar particles with orbital information that approximately matches the footprint of our observational dataset.

We proceed as in Section~\ref{subsec:Sim} to obtain orbital properties and frequencies of randomly selected stellar particles located within galactocentric radius of $2$\,kpc. The $2$\,kpc is the prescaled value equivalent to $3.5$\,kpc in the MW. The key difference is that now we divide the stellar particles based on their ages into two groups. The first group includes the oldest particles in the simulation born within the first Gyr, while the second group has the younger component born between $4$-$5$\,Gyr later. Each group has $100$k stellar particles and they map distinct stages in the formation history of the simulated galaxy. The first group forms before the formation of a bar (the bar forms between $2$-$4$\,Gyr), and the second group after the bar has formed. For the simulation snapshots at $10, 9, 8, 7, 6$, and $5$\,Gyr we repeated the approach outlined in Section~\ref{subsec:Sim}: we center our simulation using \texttt{PYNBODY}, calculated the pattern speed using the method of \citet{Dehnen2023}. We align the bar in our snapshots with the x-axis. We created an analytic potential from each snapshot using \texttt{AGAMA} and integrated orbits for the same stellar particles in each group across snapshots. Here it is important to emphasize that we used the unrescaled model for creating the analytical potential and for the orbit integration. Using this approach, we obtained kinematic properties for the selected stellar particles across multiple snapshots to examine where particles change from prograde to retrograde motion.

\subsection{Evolution of bulge stellar particles in frequency maps} \label{subsec:EvolutionInFreqMap}

To better understand how stellar particles turn from prograde to retrograde motion in the bar reference frame, we must first examine how they evolve in frequency space. In principle stellar particles can turn retrograde, for example, due to a change in pattern speed (as the bar slows down) or due to chaoticity of their orbit. Groundwork on this front has been done by \citet{BeraldoSilva2023} and is depicted with colored arrows in Figure~\ref{fig:MapCylPROGRADE-x-y}. While \citet{BeraldoSilva2023} investigated the orbital evolution of particles selected in the shoulder region, that is at the ends of the bar, here we focus on the central parts of the simulation. We use the first group of stellar particles that were born in the first Gyr of the simulation, before the bar's formation and do not impose any cut or selection on their distribution. 

\begin{figure} 
\includegraphics[width=\columnwidth]{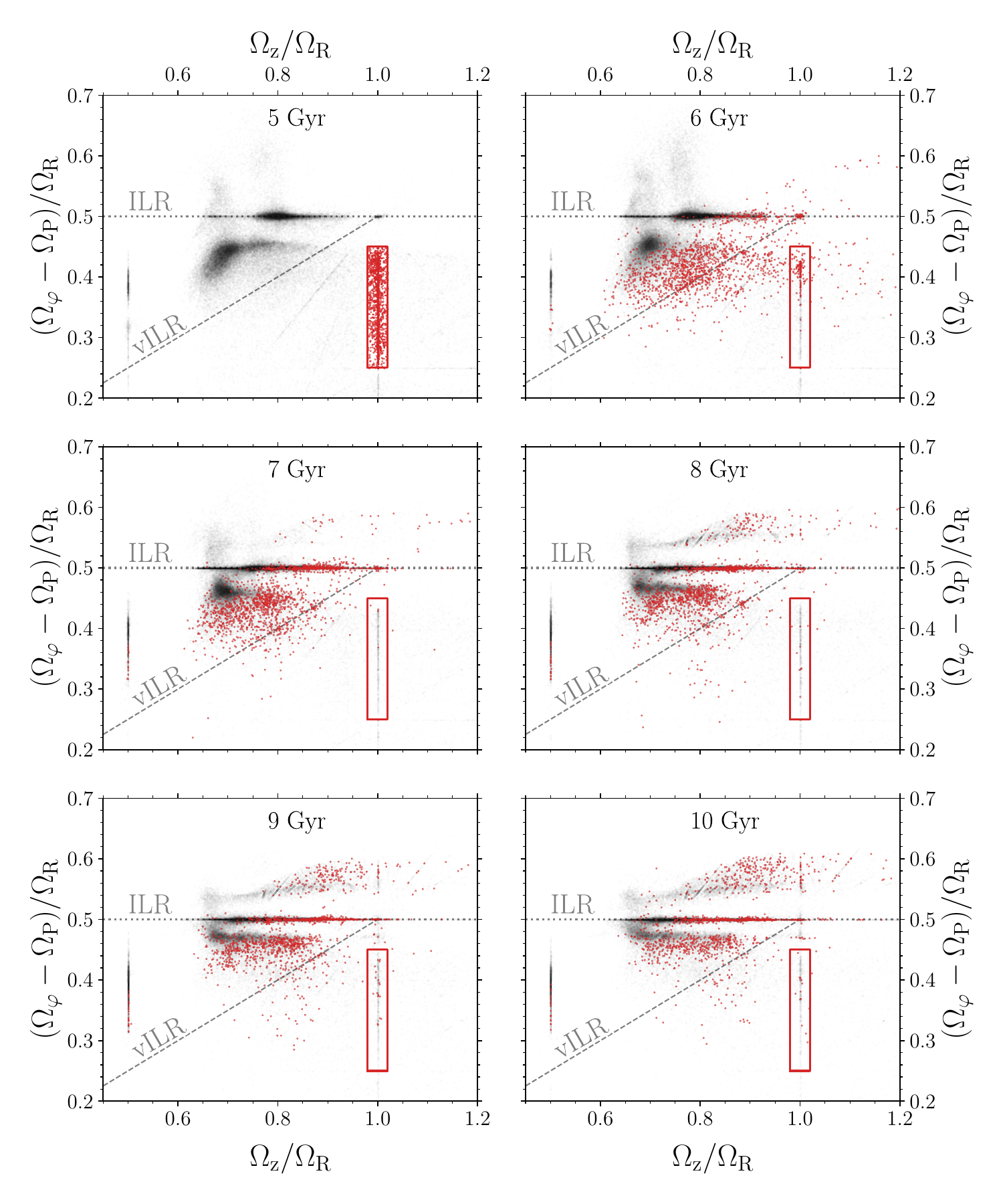}
\caption{The frequency maps for the same prograde stellar particles in six simulation snapshots. As in Figure~\ref{fig:MapCylPROGRADE-x-y} we depict two main resonances with dashed lines. The red dots represent stellar particles at resonance around $\Omega_{z} / \Omega_{\rm R} \approx 1.0$ (at $5$\,Gyr), that we trace to $10$\,Gyr.}
\label{fig:FreqEvolutionSim}
\end{figure}

Figure~\ref{fig:FreqEvolutionSim} shows the evolution of orbital frequencies for stellar particles from $5$ to $10$\,Gyr, for the prograde stellar particles ($(\Omega_{\varphi} - \Omega_{\rm P}) / \Omega_{\rm R} > 0$). We selected a group of prograde stellar particles around the $\Omega_{z} / \Omega_{\rm R} \approx 1.0$ resonance at $5$\,Gyr, and traced them throughout the remaining snapshots. This group was selected due to their clear separation from the rest. From this Figure, we see that as the potential evolves, stellar particles move mainly through the vILR to the vILR cloud, and from there, they progress mainly in two directions. A small portion of stellar particles end up in the resonance at $\Omega_{\rm z} / \Omega_{\rm R} = 0.5$, while the majority clumps at the ILR. The number of stellar particles at the aforementioned resonances doubles between snapshots at $5$ and $10$\,Gyr, going from $1.5$ to $2.9$ percent at $\Omega_{\rm z} / \Omega_{\rm R} = 0.5$, and from $19.7$ to $43.0$ percent at the ILR (using the same boundaries as in Figure~\ref{fig:MapCylPROGRADE-x-y}). In the last snapshot, the vILR cloud is depopulated and the vast majority of stellar particles are at the ILR or above it. Looking at the evolution for stellar particles that remain prograde throughout the six snapshots the evolution in the frequency map is mainly driven by the decrease of $\Omega_{\rm R}$ (preferentially) and $\Omega_{\varphi}$ although we note that $(\Omega_{\varphi} - \Omega_{\rm P})$ together with $\Omega_{\rm z}$ remain nearly constant.

\begin{figure}
\includegraphics[width=\columnwidth]{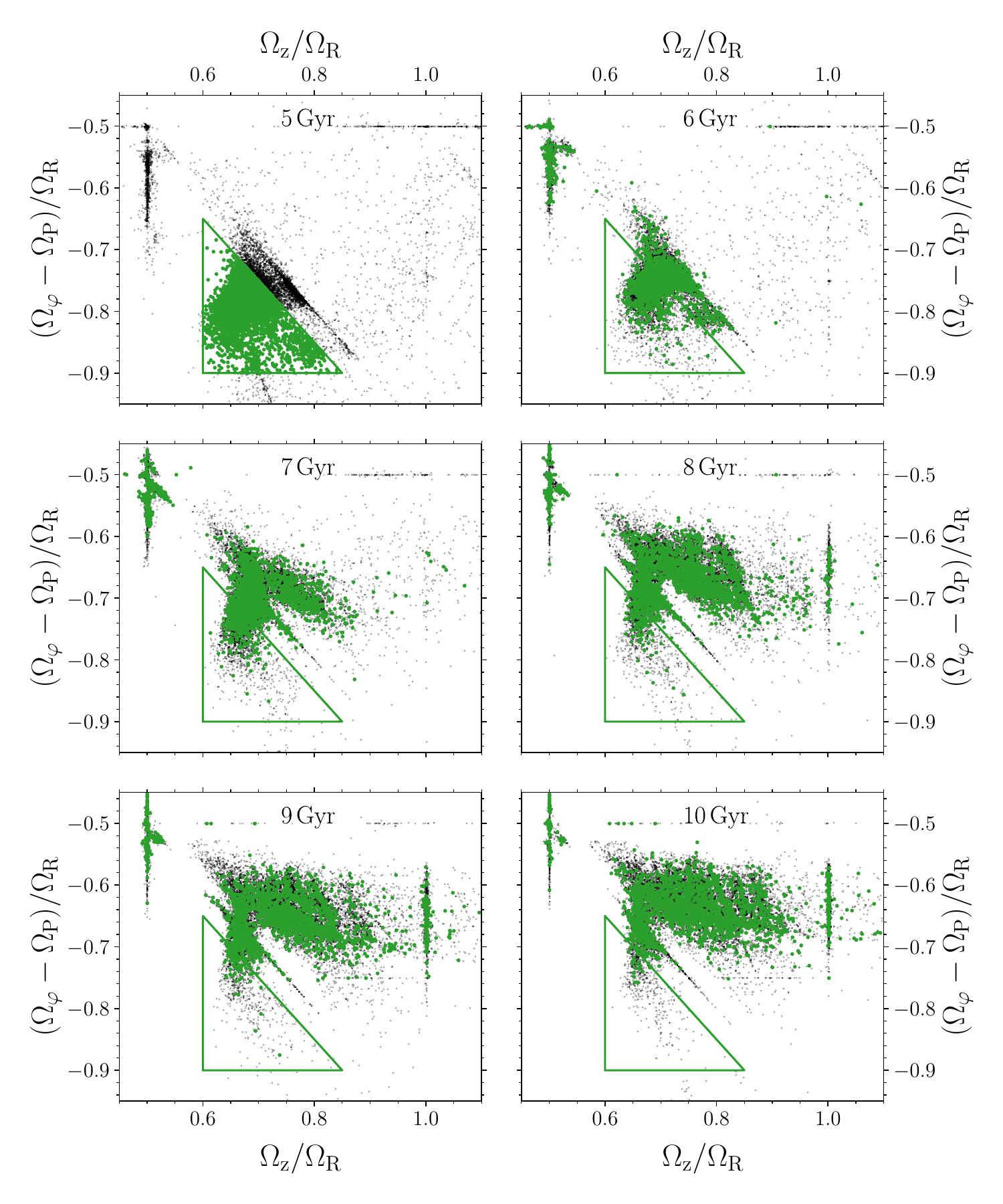}
\caption{The same as Figure~\ref{fig:FreqEvolutionSim} but for retrograde stellar particles (black dots). The green region points to the initial space of the traced particles (green points).}
\label{fig:FreqEvolutionSimNeg}
\end{figure}

The frequency map for the retrograde stellar particles can be seen in Figure~\ref{fig:FreqEvolutionSimNeg}. Looking at the stellar particles that remain retrograde between $5$ and $10$\,Gyr, we see an increase in $\Omega_{\varphi}$ and $\Omega_{\rm z}$ (preferentially). This increase shifts the distribution slightly upwards and toward the right (higher $(\Omega_{\varphi} - \Omega_{\rm P}) / \Omega_{\rm R}$ and $\Omega_{\rm z} / \Omega_{\rm R}$) as the potential evolves. On the other hand $\Omega_{\rm R}$ does not change significantly over the six snapshots. Unlike in the prograde case, $(\Omega_{\varphi} - \Omega_{\rm P})$ still increases between consecutive snapshots. We note that the particles that remain retrograde throughout the six snapshots are a minority among the total sample (around one percent), and retrograde dataset (seven percent).

To better compare simulation frequency space with our observational dataset we tested how the frequency maps change if we impose the following criteria on orbital properties and Galactic coordinates (to mimic the spatial properties of our observational data sets). We perform the following set of cuts at the $10$\,Gyr timestep of the simulation: 
\begin{equation} \label{eq:ConditionOnSim}
\left| \mathcal{l} \right| < 9^{\circ} \hspace{0.2cm} \cap \hspace{0.2cm} r_{\rm apo} < 2\,\text{kpc} \hspace{0.2cm} \cap \hspace{0.2cm} 2^{\circ} < \left| \mathcal{b} \right| < 6^{\circ} \hspace{0.2cm} \cap \hspace{0.2cm} \left| z_{\rm max} \right| > 0.3\,\text{kpc} \\.
\end{equation}
These cuts reduced our dataset from $100000$ to $34682$ stellar particles. The condition on the maximum value of $r_{\rm apo}$ was used to select only those stars that stay within the bulge (similar to our condition on the observational dataset). In addition, the simulation has a massive nuclear stellar disk, and by applying the cut on $z_{\rm max}$ we remove all stellar particles associated with this structure. The two remaining criteria on the Galactic coordinates matched stellar particles to our observational footprint. These cuts on the stellar particle distribution do not change our conclusions from the entire dataset.

\subsection{Stellar particles before becoming retrograde} \label{subsec:RightBefore}

Using the above-mentioned approach we study the kinematic properties of stellar particles before and after they become retrograde. We focus again on the oldest stellar particles\footnote{Since the fraction of the retrograde stellar particles is the highest among this age group.} in our simulation that were formed before the bar. We look at the percentage of retrograde stellar particles across snapshots from $5$ to $10$\,Gyr, and find a fairly stable value of $20 \pm 1$\,percent across six snapshots. If we apply the condition in Eq.~\ref{eq:ConditionOnSim} to this dataset we get $18 \pm 1$\,percent of particles. Only one percent of the total dataset (seven percent of the retrograde particles) remains retrograde throughout six snapshots, while more than $44$ remains prograde. Thus, the remaining $55$ percent of stellar particles during the six snapshots are retrograde at least once. To be specific, in the six analyzed snapshots, $26$ percent of stellar particles turn retrograde once, $13$ percent are retrograde twice, seven percent thrice, six percent four times, and three percent five times. Therefore, we see that the retrograde motion of stellar particles in our six snapshots is mainly temporary, and only a negligible subset remains retrograde long-term. This is expected since $L_{\rm z}$ is not conserved, resulting in particles (and stars) oscillating between retrograde and prograde motion (in bar reference frame).

We examine the stellar particles in the frequency maps, and by looking at how their angular momentum and Jacobi energy change throughout the analyzed snapshots. We focus on how these two parameters varied during the prograde and retrograde parts of their orbits. To discern between individual stellar particles we decided to group them based on their prograde/retrograde preference. We used the following nomenclature to classify prograde/retrograde changes into sets; if a stellar particle was prograde in three consecutive snapshots we marked it as \texttt{PPP} (three positive $\Omega_{\varphi}^{\rm rot}$ values). If during the second snapshot, stellar particles became retrograde and in the third it became prograde again we denoted them as \texttt{PRP}. Correspondingly stellar particles that were retrograde in three successive snapshots were marked as \texttt{RRR}. To expand on this, if a stellar particle oscillated between being prograde and retrograde we used marks \texttt{PRPRPR} and \texttt{RPRPRP} where the particles in odd and even snapshots switched between prograde and retrograde orbits.

\begin{figure}
\includegraphics[width=\columnwidth]{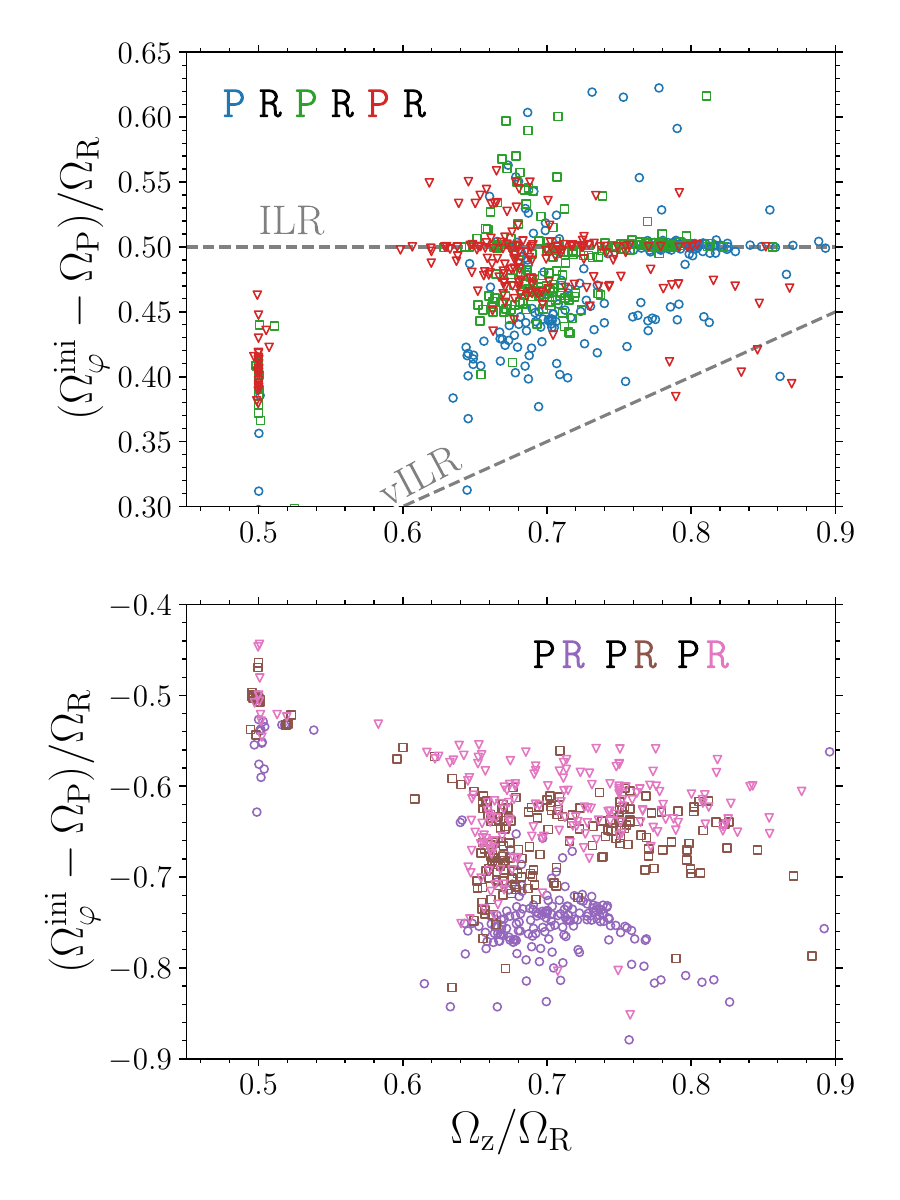}
\caption{Frequency maps for stellar particles that oscillate between prograde and retrograde orbits with respect to the bar during six analyzed snapshots. In the top panel, we see a frequency map similar to the one in Figure~\ref{fig:MapCylPROGRADE-x-y} with blue circles (snapshot at $5$\,Gyr), green squares (snapshot at $7$\,Gyr) and red triangles (snapshot at $9$\,Gyr). The bottom panel depicts frequency maps like the one in Figure~\ref{fig:MapCylRETROGRADE-x-y}, with purple circles (snapshot at $6$\,Gyr), brown squares (snapshot at $8$\,Gyr), and pink triangles (snapshots at $10$\,Gyr).}
\label{fig:PNPNPN-evol}
\end{figure}

Using the frequency maps and locations of important resonances, particularly the ILR (see Figure~\ref{fig:MapCylPROGRADE-x-y} for region 2), we looked at the positions of stellar particles right before they became retrograde. These particles, just like the entire population, follow the progression outlined in the previous subsection (see Section~\ref{subsec:EvolutionInFreqMap}). As the simulation evolves they move from the vILR cloud upward toward the ILR. In Figure~\ref{fig:PNPNPN-evol} we show the evolution in frequency maps for stellar particles that oscillate between prograde and retrograde orbits during six simulation snapshots (marked as \texttt{PRPRPR}). This subgroup represents only a small fraction (below one percent of the total sample) of the total sample but can serve as a good example of the evolution during prograde/retrograde evolution of the same particles. Its evolution in the frequency maps matches well the entire dataset. In Figure~\ref{fig:PNPNPN-evol} we see that the stellar particles do not return to the same frequencies after being prograde/retrograde but they evolve with the rest of the potential toward the ILR. 

\begin{figure}
\includegraphics[width=\columnwidth]{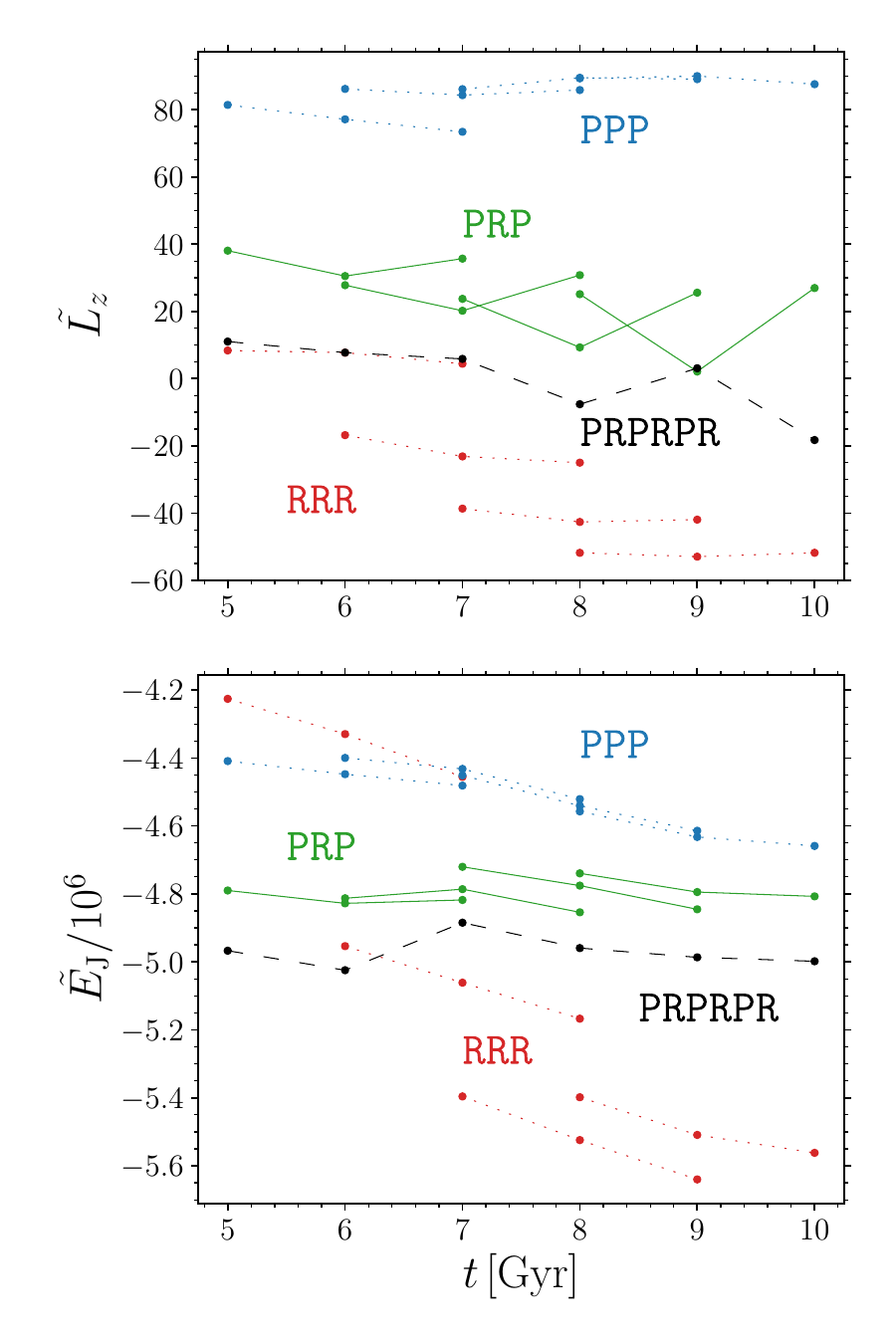}
\caption{Variation in angular momentum (in the inertial frame, top panel) and Jacobi integral (bottom panel) for various sets of prograde/retrograde stellar particles. The individual lines represent variations of median values of quantities as mentioned across six simulation snapshots. Four different sets of stellar particles are depicted here. Particles that keep their prograde or retrograde orbit (marked as \texttt{PPP} and \texttt{RRR} with dotted lines), and those that oscillate between prograde and retrograde direction (denoted as \texttt{PRP} and \texttt{PRPRPR} with solid and dashed lines).}
\label{fig:AngEne-evol}
\end{figure}

As the stellar particles oscillate between prograde and retrograde, their $L_{z}$ and Jacobi energy change. This is depicted in Figure~\ref{fig:AngEne-evol} using median values of both quantities for various particle sets (e.g., \texttt{PPP}, \texttt{PRP}, etc.). As the potential evolves the Jacobi energy of stellar particles predictably decreases as particles sink deeper in the potential. This applies across all particle sets, regardless of their prograde/retrograde preference. Regarding the angular momentum, the situation is quite different. In the case of stellar particles that remain prograde in at least three snapshots (\texttt{PPP}), we detect in the first three snapshots a small decrease in angular momentum, that subsequently starts to increase as stellar particles pass through the ILR (their $(\Omega_{\varphi}^{\rm ini} - \Omega_{\rm P})/ \Omega_{\rm R}$ increases). Stellar particles that oscillate between prograde and retrograde orbits during three snapshots (\texttt{PRP}) experience first the loss (turning from prograde to retrograde) and then an increase (turning from retrograde to prograde) of angular momentum. The same can be said about stellar particles that during the analyzed snapshots constantly oscillate between prograde and retrograde orbits (\texttt{PRPRPR}) as they repeatedly lose and gain angular momentum. The \texttt{RRR} stellar particles that remain retrograde throughout three consecutive snapshots continue losing angular momentum.

Furthermore, stellar particles that remained on retrograde orbits (\texttt{RRRRRR}) in all analyzed snapshots are by definition preferentially located at the negative end of the angular momentum distribution and with nearly the lowest Jacobi energies. The same applies also for snapshots before bar formation, where the \texttt{RRRRRR} set had one of the lowest values for angular momentum and Jacobi energy. We note that the \texttt{RRRRRR} set is only a minority (around one percent of the total sample, and seven percent of the retrograde dataset) among the analyzed stellar particles.

Based on Figures~\ref{fig:FreqEvolutionSim} and~\ref{fig:PNPNPN-evol} stellar particles evolve toward the ILR. Thus, stellar particles start losing angular momentum (in the $z$-direction). As the bar loses angular momentum \citep[to the dark matter halo, e.g.,][]{Debattista1998,Athanassoula2003} it will grow stronger and slow down (its pattern speed will decrease). We see that when a particle turns from prograde to retrograde between two consecutive snapshots, its angular momentum decreases. Inversely, stellar particles that turn from retrograde to prograde orbits gain angular momentum.

\subsection{Younger trapped population} 

Here, we examine the evolution of stellar particles born between $4$ and $5$\,Gyr. These particles were formed after the bar formed in the simulation. To accurately examine this younger dataset, we impose the conditions in Eq.~\ref{eq:ConditionOnSim} on all the analyses listed in this subsection, to exclude the nuclear stellar disk.

\begin{figure}
\includegraphics[width=\columnwidth]{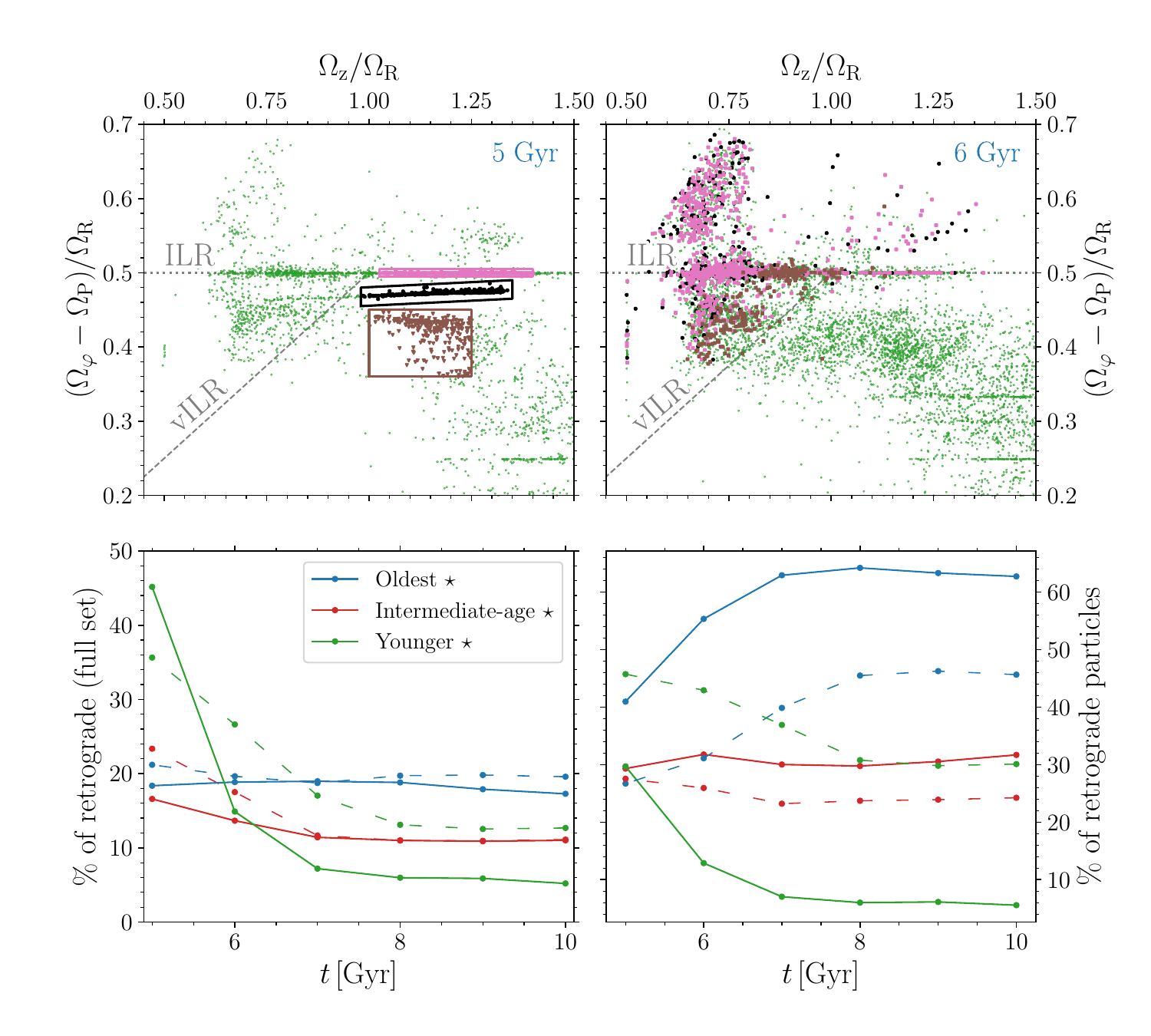}
\caption{Frequency maps (top panels) and the fraction of retrograde stellar particles (bottom panels) of different ages across several snapshots. The top panels display the distribution of younger stellar particles in frequency maps for two different snapshots. Three groups of stellar particles are selected colored brown, pink and black and brown, in the initial snapshot (left-hand panel, at $5$\,Gyr) and traced in the following snapshot (right-hand panel at $6$\,Gyr). Similar to Figure~\ref{fig:FreqEvolutionSim}, the two main resonances are marked with dashed lines. The bottom panels show the fraction of retrograde stars within the total sample (left bottom panel) and the fraction split by formation age within the retrograde category (for a given snapshot) alone (right bottom panel). The solid lines represent analyses conducted on the dataset where the condition in Eq.~\ref{eq:ConditionOnSim} was applied, while the dotted lines represent the entire dataset. The blue, red, and green lines represent the oldest (age $9-10$\,Gyr), intermediate (age $8-9$\,Gyr), and youngest (age $5-6$\,Gyr) stellar particles, respectively.}
\label{fig:YoungerStePar}
\end{figure}

In Section~\ref{subsec:EvolutionInFreqMap}, we examined how the oldest stellar particles evolved in frequency space. We found that they move mainly through the vILR to the vILR cloud and from there upward to the ILR. Stellar particles born after the formation of the bar move slightly differently. Unlike the oldest stellar particles, the younger objects populate the warm ILR \citep[$1.0 < \Omega_{\rm z} / \Omega_{\rm R} < 1.5$,][]{BeraldoSilva2023}, and from there they move leftward with decreasing $\Omega_{\rm z} / \Omega_{\rm R}$ (see top panels of Figure~\ref{fig:YoungerStePar}).

The bottom panels of Figure~\ref{fig:YoungerStePar} illustrate the evolution of retrograde stellar particles across six snapshots. In these panels, we present the entire datasets with dotted lines (without applying the condition in Eq.~\ref{eq:ConditionOnSim}) and a subsample resembling our observational dataset with solid lines (applying the spatial condition in Eq.~\ref{eq:ConditionOnSim}). Initially, the younger population (green color) exhibits a significant retrograde fraction ($\Omega_{\varphi}^{\rm ini} < \Omega_{\rm P}$) and then gets trapped (with $\Omega_{\varphi}^{\rm ini} > \Omega_{\rm P}$) and aligns with the bar. Comparing the solid and dotted lines highlights the substantial impact of applying the condition from Eq.~\ref{eq:ConditionOnSim}. Conversely, the overall trend remains consistent: after a few gigayears, the ratio of retrograde younger stellar particles stabilizes with the average percentage of retrograde stellar particles across snapshots from $7$ to $10$ gigayears equal to $6\pm1$\,percent. We note that this number is related to the sample where the condition of Eq.~\ref{eq:ConditionOnSim} was applied. The older population (blue color) maintains a relatively constant number of retrograde stars, unaffected by whether conditions replicating our observational footprint are applied or not. Considering both age groups collectively, the ratio of retrograde stars stabilizes around $7$\,Gyr (approximately $3$\,Gyr after the bar formed) and remains nearly constant throughout the remainder of the simulation. 

In addition to the previously mentioned two populations we also included stellar particles that formed between $1$ to $2$\,Gyr, i.e., before the bar but after the oldest population formed a disk. This intermediate-age population shows a similar trend to the oldest stellar particles with a rather stable retrograde fraction in the bulge. On the other hand, the ratio of retrograde vs. prograde stellar particles is lower than for the oldest population, and percentage-wise, the oldest stellar population still dominates the retrograde distribution.

\section{Discussion} \label{sec:Discussion}

The analysis of the $N$-body+SPH simulation suggests that the origin of retrograde stars in our observational dataset could be a natural progression of the MW bar evolution, particularly the bar's slow-down and growth. A fraction of retrograde stars in our dataset likely had retrograde orbits (in the inertial frame) even before the bar formed \citep[similar to the finding by][]{Fiteni2021}. Based on their energies and angular momentum, these stars have been part of the central regions (or the bulge) from the very beginning. This may reflect the fact that more centrally concentrated stars, formed from lower angular momentum material, require smaller torques to reverse the sense of their angular momentum vector. After the bar forms, the majority of retrograde stars oscillate between prograde and retrograde orbits.

In the simulation, the percentage of retrograde stars of a given age varies strongly only after the bar forms for the initial $3$\,Gyr, and then stabilizes. Assuming that the MW bar is at least $6$\,Gyr old \citep[e.g.,][]{Sanders2024} then Figure~\ref{fig:OmegaPhiChemistry} reflects a trend over the past few gigayears. The metal-rich RR~Lyrae variables ([Fe/H]$_{\rm phot} > -1.0$\,dex) are probably slightly younger (by $1$ - $2$\,Gyr) and likely initially had more angular momentum compared to the metal-poor RR~Lyrae stars (commonly assumed age $\approx 12$\,Gyr). This is in contrast with \citet{Savino2020} and highlights the need for a full understanding of the RR~Lyrae formation channels. To further strengthen our assumption we need spectroscopically measured [$\alpha$/Fe] abundances. In the case, where the metal-rich RR~Lyrae stars are $\alpha$-poor, they are likely slightly younger than their metal-poor counterparts. If they turn out to be $\alpha$-rich, then they would be likely very old (low-mass) RR~Lyrae stars.

The results presented here, particularly those in Sections~\ref{sec:OrbitalParam}, \ref{sec:RotMetalPoor}, and \ref{sec:RetroOrbitsFam}, depend on the selected analytical potential for the MW and its properties. The chosen MW potential affects the number of identified interlopers (e.g., depending on selection criteria and total mass of the potential), but we would not expect this to change our results significantly. However, a significant impact on the results would be selecting an unbarred potential for the MW \citep[e.g., \texttt{MWPotential2014} or \texttt{McMillan} potentials][as used in some previous studies]{Bovy2015,McMillan2017}. By using an axisymmetric analytical potential, we would not observe the bulk structure in $z_{\rm max}$ vs. $r_{\rm apo}$ plane (see Figure~\ref{fig:ApocenZmaxRot}). In our case, with a rotating barred potential, the bar properties significantly affect the orbits and estimated frequencies. In Appendix~\ref{secAp:Pattern}, we address the dependence of the results presented here on the different values of the bar's pattern speed.  

The effect of uncertainties in the orbital parameters and frequencies examined here has a negligible impact on the presented results. The orbital frequencies are generally significant, and since we focus exclusively on bulge stars, our condition on the apocentric distance ($r_{\rm apo} < 3.5$\,kpc) primarily excludes stars with small $\Omega_{\varphi}^{\rm rot} \approx 0$. This results in only a minimal chance of misclassifying a star as prograde or retrograde (in the bar reference frame). The bulk structure in the $z_{\rm max}$ vs. $r_{\rm apo}$ plane remains clearly visible even when we select only stars with high significance in the aforementioned orbital parameters.

We note that, in the following discussion, we do not address recent studies of the central Milky Way and some of its spheroidal features, as discussed in \citet{Rix2022, Rix2024, Belokurov2022}. The connection between our dataset and the findings of these studies will be explored in a forthcoming paper.

\subsection{Secular spherical bulge} \label{subsec:ImperatorSomnium}

Recent spatial and kinematical studies focusing on the Galactic bulge's stellar content reveal a prevailing dichotomy between the metal-poor ([Fe/H]~$\lesssim-0.5$\,dex in metallicity) and metal-rich (approximately $>-0.5$\,dex in metallicity) populations. Metal-poor bulge stars exhibit a more spheroidal spatial distribution with slower rotation, while metal-rich stars predominantly support a barred boxy-peanut bulge spatial distribution and rotation curve \citep[e.g.,][]{Zoccali2017,Rojas-Arriagada2020,Kunder2020,Arentsen2020PIGSI,Queiroz2021,Lim2021,Arentsen2023}. The aforementioned studies sometimes invoked the composite morphology of the MW bulge with both boxy-peanut and classical (product of mergers) components. 

Starting from the metal-poor end, Figure~\ref{fig:OmegaPhiChemistry} shows that the most metal-poor ([Fe/H]$<-1.0$\,dex) stars residing in the Galactic bulge have a high percentage of counter-rotating orbits ($\approx 30$\,percent). These objects mostly have boxy orbits (see, e.g., Figure~\ref{fig:MapCylRETROGRADE-x-y}) and neither spatially nor kinematically follow the bar as traced from metal-rich red clump stars. The large number of these stars leads to our view of the bulge as more spherical with lagging rotation. If we remove the counter-rotating metal-poor stars, we obtain bulge properties characteristic of their metal-rich counterparts, such as the rotation curve and double-peaked distance distribution (considering only banana-orbits) and a barred spatial distribution (see Figures~\ref{fig:MetalPoorRot} and~\ref{fig:SpatBar}).

The stars with intermediate metallicities ($-1.0 < \textrm{[Fe/H]} < 0.0$\,dex) have a lower fraction of retrograde stars (between $15$ to $25$ percent). Kinematically, they exhibit a rotation curve similar to that of metal-rich stars, and spatially, they show a nearly barred distribution \citep[see, e.g.,][]{Lim2021,Johnson2022}. This intermediate metallicity region is likely sensitive to how close the analyzed dataset is to the Galactic center\footnote{As shown in Figure~\ref{fig:OmegaPhiRgc}, where counter-rotating stars are more often found closer to the center than on the outskirts of the bulge.}, and the metalicity distribution function (MDF) is skewed toward the metal-rich or metal-poor end. 

The super metal-rich bulge stars ($\textrm{[Fe/H]}>0.0$\,dex) are notable for tracing the boxy/peanut shape both spatially and kinematically \citep[e.g.,][]{Zoccali2017}. The percentage of counter-rotating stars in this group is low, around $10$-$15$ percent in our dataset. Despite being the smallest group in terms of percentage, they constitute nearly $40$\,percent of retrograde non-variable giants in absolute numbers for our dataset.

A significant portion of retrograde (in the bar reference frame) stars in the Galactic bulge likely oscillates between prograde and retrograde orbits. According to our simulation analysis, their density compared to generally prograde stars does not change significantly over several gigayears. The orbits of retrograde stars exhibit similar regularity as prograde orbits. Almost all retrograde orbits form a nearly axisymmetric distribution (in the aggregate), contributing to an overall spheroidal distribution in the central parts of the bulge. 

There are several reasons why we do not necessarily need to invoke a significant classical bulge in the MW. The currently observed features of the MW bulge can be explained using an $N$-body+SPH simulation without external influences (e.g., merger events). The difference in the rotation curve between metal-rich and metal-poor stars can be explained, in addition to kinematic fractionation \citep{Debattista2017,Fragkoudi2018,GoughKelly2022}, by the varying numbers of retrograde stars, which appear to increase linearly with metallicity (see Figure~\ref{fig:OmegaPhiChemistry}). A similar trend can be observed in the increase of retrograde stars towards the central parts of the MW (shown in Figure~\ref{fig:OmegaPhiRgc}). Lastly, chemical abundances of bulge non-variable giants (with metallicities $> -1.0$\,dex), particularly in the [Mg/Mn] vs. [Al/Fe] plane, do not exhibit an over-enhancement of accreted stars (i.e., [Al/Fe]-poor and [Mg/Mn]-rich abundances). Although we note that on the metal-poor side of the metallicity distribution, accreted stars identified using aforementioned abundances could become relevant as discussed in \citet{Horta2021}.

Therefore, the observed transition from a metal-rich, barred, cylindrically rotating Galactic bulge to a less barred, slowly rotating bulge can be largely explained by the secular evolution of the MW's central regions. While we do not completely exclude the contribution of mergers to the Galactic bulge, our analysis points to their potential minimal relevance to the current morphology of the Galactic bulge. Furthermore, the higher fraction of retrograde orbits among metal-poor giants (e.g., RR~Lyrae stars) may be a consequence of angular momentum exchange, potentially facilitated by the Galactic bar. Nonetheless, it is also possible that these stars had intrinsically lower angular momentum before the bar was formed. 

It is important to emphasize that our results, along with comparisons to $N$-body+SPH simulations, provide a specific perspective on the formation history of the MW bulge. Further investigation using additional $N$-body (+SPH) and cosmological simulations would be useful to fully understand the evolution and origin of the retrograde stellar population in the Galactic bulge. On the other hand, we show that a secularly evolved galaxy can form a substructure that, in some of its properties, resembles a classical bulge. The spatial and kinematical characteristics of this substructure can be reasonably well matched with a similar component in the Milky Way bulge.

\subsection{Note on the metal-rich RR~Lyrae stars} \label{subsec:MetalRichKinAge}

Metal-rich RR~Lyrae stars ($\textrm{[Fe/H]} > -1.0$\,dex), as shown in Figures~\ref{fig:VGVRotDisp-FEH}, \ref{fig:RRlbanana}, and~\ref{fig:OmegaPhiChemistry}, present a different perspective on the Galactic bulge compared to their metal-poor counterparts. This observation aligns with our previous findings \citep{Prudil20253D} and corroborates the results of other studies \citep[e.g.,][]{Kunder2016,Du2020}. The spatial and kinematic properties of metal-rich RR~Lyrae stars closely resemble those of red giants and red clump stars of the same metallicity. Potentially a key to understanding the differences between metal-rich and metal-poor RR~Lyrae stars likely lies in their distinct evolutionary pathways, as proposed by \citet{Iorio2021}, \citet{Bobrick2024} and \citet{Zhang2025}. These studies suggest that metal-rich RR~Lyrae stars arise from binary evolution, in contrast to their single-star metal-poor counterparts. Binary interactions may allow these stars to pass through the instability strip at a significantly younger age (compared to single-star evolution), which could explain their bar-like kinematics and spatial distributions.

We examine the red giants in Figure~\ref{fig:OmegaPhiChemistry} and focus on those corresponding to the most metal-rich bin for RR~Lyrae stars. The age distribution of red giants within these metallicity bins, based on the APOGEE value-added catalog \citep{Mackereth2019}, is centered around $8.5$\,Gyr\footnote{It is important to note that the median error on age in our APOGEE dataset is $\sim3$\,Gyr.}, with a significant spread and a non-Gaussian distribution. Notably, approximately $10$ percent of stars in this metallicity bin are younger than $5$\,Gyr. This finding contrasts with the presumed age distribution of the Galactic bulge, which is predominantly composed of intermediate-age and old stars, with most ages exceeding $5$\,Gyr \citep{Ortolani1995,Clarkson2008,Bensby2013,Renzini2018,Joyce2023}. The kinematics of metal-rich RR~Lyrae variables suggest they are slightly younger than their metal-poor counterparts, although ages below $8$\,Gyr appear, in light of our analysis, unlikely.

{Lastly, it is important to note the Galactic bulge does contain old \citep[$12$-$13$\,Gyr, see][]{Massari2023}, metal-rich RR~Lyrae stars within globular clusters NGC~6441 and NGC~6388 \citep{Clementini2005}. These cluster metal-rich RR~Lyrae stars have long pulsation periods ($\approx 0.7$\,day for RRab variables) unlike their thin-disk associated metal-rich counterparts \citep[$0.43$\,day for RRab variables, e.g.,][]{Prudil2020Disk}. Then perhaps there must be a formation channel for metal-rich RR~Lyrae that does not require binary evolution, for example, helium enrichment as suggested in \citet{Savino2020}.

\section{Summary} \label{sec:Summary}

In this work, we derived systemic velocities for $8456$ RR~Lyrae stars, representing the largest dataset of these variables toward the Galactic bulge assembled to date. We obtained their orbits by integrating their spatial and kinematic properties in an analytical gravitational potential analogous to that of the MW. We separated the presence of RR~Lyrae interlopers ($22$ percent of our RR~Lyrae dataset), belonging to other MW substructures such as the disk and halo, by imposing a condition on the apocentric distance ($r_{\rm apo} < 3.5$\,kpc), thus removing the variables that are merely passing through the bulge. We note that the population of interlopers within the Galactic bulge depends significantly on the selected analytical potential and their location in the MW. There is a steep increase in the fraction of interlopers with increasing Galactocentric radius and Galactic latitude. Approximately $75$ percent of RR~Lyrae interlopers can be associated with the Galactic halo, while the remaining $25$ percent spend most of their orbit close to the Galactic plane ($z_{\rm max} < 1$\,kpc).

The identified interlopers cannot explain the lag observed in the bulge rotation curve (as observed in non-variable giants) across the entire RR~Lyrae dataset. Removing interlopers from our dataset primarily affects the velocity dispersion. Dividing our RR~Lyrae sample based on their photometric metallicities reveals a variation in the rotation pattern. Metal-rich RR~Lyrae stars ($\textrm{[Fe/H]}_{\textrm{phot}} = (0, -1)$\,dex) exhibit rotation patterns expected from non-variable giants and confirm results from our spatial and transverse velocity analysis \citep{Prudil20253D}, that this subgroup of RR~Lyrae pulsators follows the MW bar. As we transition to metal-poor RR~Lyrae ($-2.0 \leq \textrm{[Fe/H]}_{\textrm{phot}} \leq -1.0$\,dex), we observe a slower rotation trend that almost entirely disappears for very metal-poor RR~Lyrae variables ($-2.6 \leq \textrm{[Fe/H]}_{\textrm{phot}} \leq -2.0$\,dex).

Examining the distribution of RR~Lyrae pulsators in orbital parameter space ($z_{\rm max}$ vs. $r_{\rm apo}$), we observed a bifurcation for variables associated with the Galactic bulge (difference in vertical extent). This bifurcation arises due to the inclusion of the Galactic bar in the analytical potential, segregating stars that rotate prograde with the bar (located in the bulk) from those on retrograde orbits (found along the identity line in the $z_{\rm max}$ vs. $r_{\rm apo}$ plane). A similar structure is also noted for the non-variable giants. Stars likely get boosted to these orbits when they hit a vertical resonance, probably the vertical inner Lindblad resonance (vILR).

To distinguish stars in both structures, we employ the angular frequency in the bar frame, where negative values align with the identity line and positive values fall within the observed bulk (see Figure~\ref{fig:ApocenZmaxRot}). RR~Lyrae stars in orbits that are prograde in the bar frame exhibit a clear rotation and dispersion pattern consistent with predictions from simulations and observations of non-variable giants. Conversely, stars with retrograde orbits in the bar reference frame show counter-rotation and a flat velocity dispersion. Utilizing this criterion on positive and negative $\Omega_{\varphi}^{\rm rot}$, we also discern a rotation trend for metal-poor RR~Lyrae stars ([Fe/H] $< -2.0$\,dex), where $38$ percent are interlopers, $19$ percent exhibit retrograde motion, and only $43$ percent are prograde RR~Lyrae stars.

Analyzing the binned metallicity distribution for RR~Lyrae and non-variable giant stars as a function of the percentage of negative $\Omega_{\varphi}^{\rm rot}$ objects, we find a linearly increasing trend in retrograde fraction moving from metal-rich to metal-poor stars. This linear increase in the fraction of stars with negative $\Omega_{\varphi}^{\rm rot}$ is linked to the lag observed in the rotation curve for RR~Lyrae stars, and for metal-poor non-variable giants. Here, the younger (metal-rich) giant population is less affected by objects on retrograde orbits compared to the older (metal-poor) population and exhibits a clear rotation trend. The analysis of prograde and retrograde orbits showed, as expected, that stars moving on prograde orbits have more elongated paths than those on retrograde orbits. In the frequency analysis described in Section~\ref{sec:RetroOrbitsFam}, we identified several orbital families, and we showed that retrograde stars form nearly a spherical shape in the central parts of the MW (mostly within $2.0$\,kpc). To further support our conclusions, previous studies using RRab variables have demonstrated that the central regions of the Galactic bulge exhibit slower rotation compared to the outer Bulge regions. The latter appears to be more kinematically aligned with the Galactic bar \citep{Kunder2020,Han2024}. Lastly, the level of orbit regularity is almost the same for retrograde and prograde orbits, leading to the conclusion that the spherical structure supported by the retrograde orbits is a stable substructure within the Galactic bulge. 

Further analysis using the $N$-body+SPH simulation showed that retrograde motion is long-term only for a small subset of stellar particles. The majority of retrograde stellar particles oscillate between prograde and retrograde motion. In the frequency maps, these variations in orbits occur close to the ILR. Examining stellar particles that were born before or after the formation of the bar, we find that the percentage of retrograde stars does not change significantly for a given age group after the bar forms. This suggests that the trend observed in Figure~\ref{fig:OmegaPhiChemistry} has likely remained the same for the past few gigayears \citep[assuming the Galactic bar is older than $6$\,Gyr,][]{Sanders2024}.

Our analysis leads us to conclude that the previously reported spheroidal feature in the Galactic bulge is a real secondary component. Based on the examined simulation, the spheroidal element can be related to the secular evolution of the Galactic bulge. The evolution is mainly driven by angular momentum exchange and is likely facilitated by the Galactic bar. The flow of angular momentum induces the oscillation between retrograde and prograde orbits. The RR~Lyrae stars and some of the non-variable giants in our study have been part of the inner MW even before the bar was formed. Due to their low angular momentum, a large fraction of them now oscillate between prograde and retrograde configurations. The stability (i.e., non-chaotic nature) of retrograde orbits, along with their high prevalence, particularly at the metal-poor end of the bulge's metallicity distribution function (MDF), can explain certain spatial and kinematic features of the MW's bulge that are often attributed to a classical bulge morphology. 

One should bear in mind that our conclusions listed above are partially based on secularly evolved simulation; additional investigations with a diverse set of $N$-body and cosmological simulations would be useful. That said, we showed that a secular evolution can produce a substructure resembling a classical bulge, with spatial and kinematic properties matching those observed in the Milky Way.

\section{Data availability}
\href{https://zenodo.org/records/15724090}{https://zenodo.org/records/15724090}

\begin{acknowledgements}
Z.P. thanks Christian Johnson for providing the BDBS catalog. A.M.K. acknowledges support from grant AST-2009836 from the National Science Foundation.

This work has made use of data from the European Space Agency (ESA) mission {\it Gaia} (\url{https://www.cosmos.esa.int/gaia}), processed by the {\it Gaia} Data Processing and Analysis Consortium (DPAC, \url{https://www.cosmos.esa.int/web/gaia/dpac/consortium}). Funding for the DPAC has been provided by national institutions, in particular the institutions participating in the {\it Gaia} Multilateral Agreement.

Funding for the Sloan Digital Sky Survey IV has been provided by the Alfred P. Sloan Foundation, the U.S. Department of Energy Office of Science, and the Participating Institutions. SDSS acknowledges support and resources from the Center for High-Performance Computing at the University of Utah. The SDSS web site is \url{www.sdss4.org}.

SDSS is managed by the Astrophysical Research Consortium for the Participating Institutions of the SDSS Collaboration including the Brazilian Participation Group, the Carnegie Institution for Science, Carnegie Mellon University, Center for Astrophysics | Harvard \& Smithsonian (CfA), the Chilean Participation Group, the French Participation Group, Instituto de Astrofísica de Canarias, The Johns Hopkins University, Kavli Institute for the Physics and Mathematics of the Universe (IPMU) / University of Tokyo, the Korean Participation Group, Lawrence Berkeley National Laboratory, Leibniz Institut für Astrophysik Potsdam (AIP), Max-Planck-Institut für Astronomie (MPIA Heidelberg), Max-Planck-Institut für Astrophysik (MPA Garching), Max-Planck-Institut für Extraterrestrische Physik (MPE), National Astronomical Observatories of China, New Mexico State University, New York University, University of Notre Dame, Observatório Nacional / MCTI, The Ohio State University, Pennsylvania State University, Shanghai Astronomical Observatory, United Kingdom Participation Group, Universidad Nacional Autónoma de México, University of Arizona, University of Colorado Boulder, University of Oxford, University of Portsmouth, University of Utah, University of Virginia, University of Washington, University of Wisconsin, Vanderbilt University, and Yale University.
 
This research made use of the following Python packages: \texttt{Astropy} \citep{astropy2013,astropy2018}, \texttt{PYNBODY} \citep{Pontzen2013}, \texttt{emcee} \citep{Foreman-Mackey2013}, \texttt{AGAMA} \citep{Vasiliev2019Agama}, \texttt{IPython} \citep{ipython}, \texttt{Matplotlib} \citep{matplotlib}, \texttt{NumPy} \citep{numpy}, and \texttt{SciPy} \citep{scipy}.

\end{acknowledgements}

\bibliographystyle{aa}
\bibliography{biby} 

\begin{appendix} 

\section{Estimation of line-of-sight and systemic velocities from available spectra} \label{sec:App:VlosVsys}

To estimate systemic velocities, $v_{\rm sys}$, from available spectra, we proceeded in the same way for all three spectral sources (BRAVA-RR, APOGEE, and program ID 093.B-0473) with only minor adjustments due to different spectral ranges. We used photometric information in the OGLE and \textit{Gaia} survey for all targeted RR~Lyrae stars (pulsation periods, time of brightness maxima, and peak-to-peak amplitudes in $I$ and $G$-band). 

To derive $v_{\rm sys}$ we first measure the line-of-sight velocities, $v_{\rm los}$, for available spectra. We use the \texttt{iSpec} package \citet{Blanco2014,Blanco2019iSpec} for this procedure (spectral synthesis and line-of-sight velocity determination). We proceed in a similar way as in \citet{Prudil2024GBEXII}; we first synthesize a grid of spectra with physical parameters covering the typical ranges for RR~Lyrae stars \citep[based on stellar parameters found in the literature, e.g.,][]{Sneden2017,Crestani2021Alpha,Crestani2021}, with the following stellar parameters:
\begin{itemize}  \setlength\itemsep{0.3em}
\item $T_{\rm eff} = (6000, 6500, 7000, 7500)$\,K
\item log\,$g = (2.0, 2.5, 3.0)$\,dex
\item {[Fe/H]~$= (-2.5, -2.0, -1.5, -1.0, -0.5, 0.0)$\,dex}
\item Microturbulence velocity $\xi_{\rm turb} = 3.5$\,km\,s$^{-1}$ \hspace{1cm} .
\end{itemize} 
We also needed to synthesize different spectral ranges based on three spectral sources used in our study. We used $8450 - 8720$\,\AA~for BRAVA-RR to mimic spectral ranges from \textit{Gaia} \citep{Katz2023} that were used for creating line-of-sight velocity templates and scaling relations. For the APOGEE survey, we used the $15000 - 17000$\,\AA~range, and for data from the FLAMES GIRAFFE spectrograph, we used the $5200 - 5700$\,\AA~range. All of the aforementioned synthetic spectra were created using \texttt{iSpec}, and its wrapper for radiative transfer code, MOOG \cite[February 2017 version,][]{Sneden1973}, in combination with the ATLAS9 model atmosphere \citep{Castelli2003}, lines from the VALD database\footnote{\url{http://vald.astro.uu.se/}}, and solar scale \citep{Asplund2009}. 

In estimating $v_{\rm los}$, we proceeded in the following way. In the first step, we estimated $v_{\rm los}$ using all synthesized spectra for a given observed spectrum. The synthesized spectrum that resulted in the lowest scatter in the $v_{\rm los}$ was selected for the final $v_{\rm los}$ calculation, where we varied the fluxes of observed spectra within their uncertainties (assuming they follow a Gaussian distribution) and in each iteration recorded the $v_{\rm los}$ ($200$ interactions in total). Akin to \citet{Prudil2024GBEXII}, we used the bootstrap method to estimate the average and standard deviation of $v_{\rm los}$ and its variation, $\sigma_{v_{\rm los}}$. In Figure~\ref{fig:Dist-vlos}, we show the distribution of uncertainties on estimated line-of-sight velocities for fundamental and first overtone RR~Lyrae stars. Using the observation time and coordinates on the sky, we performed Heliocentric correction for each measured $v_{\rm los}$ and estimated the Heliocentric Julian date of the observation, $T_{\rm HJD}$.

\begin{figure}
\includegraphics[width=\columnwidth]{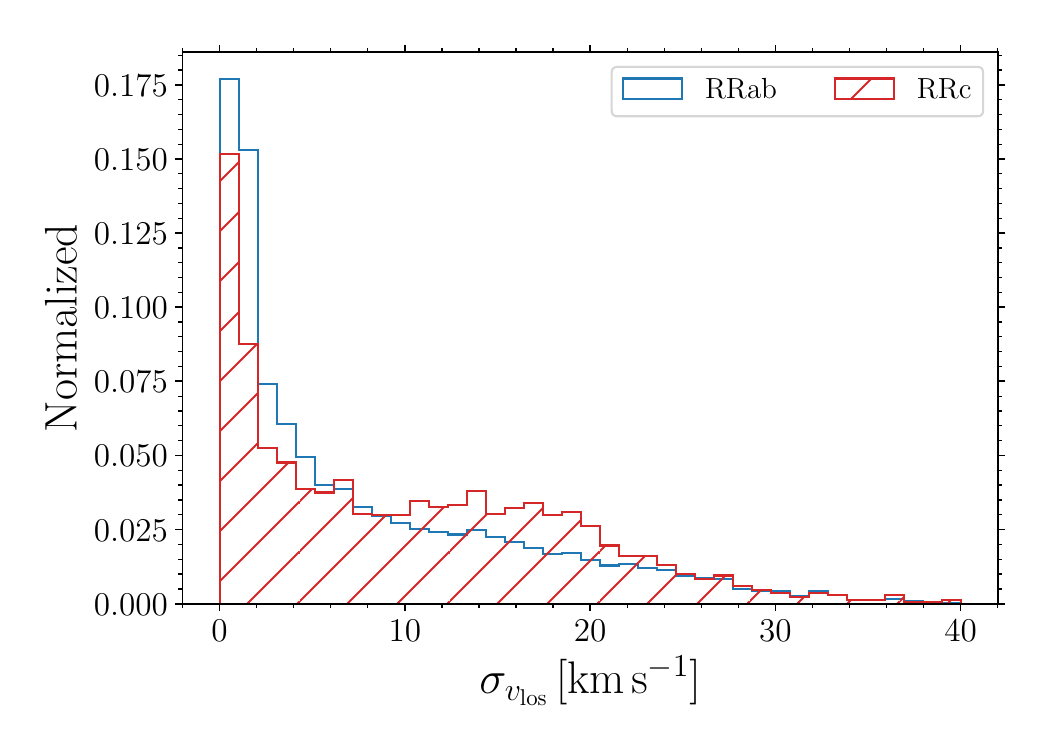}
\caption{The distribution of $\sigma_{v_{\rm los}}$ for our sample of RR~Lyrae stars, the blue and red histograms represent uncertainties for RRab and RRc variables.}
\label{fig:Dist-vlos}
\end{figure}

\subsection{Systemic velocities} \label{subsec:AppVsysFin}

We divided our procedure based on the spectra source to determine $v_{\rm sys}$ of individual stars. In the case of BRAVA-RR and APOGEE data, we used our newly derived line-of-sight velocity templates and scaling relations for RRab and RRc variables \citep{Prudil2024GBEXII}. For spectra from the VLT/FLAMES program 093.B-0473 we used templates and scaling relations from \citet{Braga2021}. Their templates for Fe lines include the majority of features present in these spectra. We note that there is a partial overlap in RR~Lyrae variables between the three spectral sources (four objects in common). We did not combine $v_{\rm los}$ to estimate $v_{\rm sys}$ since our data set covers different spectral ranges that require different treatments to estimate $v_{\rm sys}$. Variables that have data from more than one source were used to verify derived systemic velocities and their uncertainties. The estimated line-of-sight velocities for some of the analyzed spectra exhibited unreliable measurements. Thus, we decided to employ the following conditions on the significance of $v_{\rm los}$:
\begin{equation} \label{eq:condONbravavlos}
\left| v_{\rm los} / \sigma_{v_{\rm los}} \right| > 10 \hspace{1cm} \text{OR} \hspace{1cm} \sigma_{v_{\rm los}} < 20 \hspace{0.2cm}\text{km\,s}^{-1}
\end{equation}
These two conditions preserved stars with $v_{\rm los}$ around zero that still have low $\sigma_{v_{\rm los}}$. If a given $v_{\rm los}$ and $\sigma_{v_{\rm los}}$ did not fulfill one of the conditions above, it was not used in subsequent systemic velocity determination.

For the determination of $v_{\rm sys}$ of individual RR~Lyrae stars, we draw inspiration from our previous work \citep{Prudil2024GBEXII}. The calculation of $v_{\rm sys}$ is an optimization problem with a set of data and model parameters. Therefore, we used the Bayesian approach \citep{Bayes1763} that the following equation can summarize: 
\begin{equation} \label{eq:BayPosMain}
p(\boldsymbol{\theta}\,|\,\mathbf{D}) \propto p(\mathbf{D}\,|\,\boldsymbol{\theta}) \, p(\boldsymbol{\theta}) \\,
\end{equation}
where $p(\boldsymbol{\theta}\,|\,\mathbf{D})$ represents a posterior probability for a model $\boldsymbol{\theta}$ given data $\mathbf{D}$, variable $p(\mathbf{D}\,|\,\boldsymbol{\theta})$ is the likelihood function and $p(\boldsymbol{\theta})$ is the prior on model parameters. Data for a given star in our spectroscopic sample can be described as:
\begin{equation}
\mathbf{D} =\left\{ P, M_{\rm 0}, \text{Amp}_{\rm los}, \sigma_{\text{Amp}_{\rm los}}, v_{\rm los}, \sigma_{v_{\rm los}}, T_{\rm HJD} \right \} \\.
\end{equation}
The amplitudes for line-of-sight velocity curves and their associated errors ($\text{Amp}_{\rm los}$ and $\sigma_{\text{Amp}_{\rm los}}$) were determined in the following way. For stars in the \textit{Gaia} RR~Lyrae catalogue \citep{Clementini2023} we used their Amp$_{G}$, while for variables that did not have Amp$_{G}$, we used OGLE peak-to-peak amplitudes ($\text{Amp}_{I}$), and converted to Amp$_{G}$ using relations from \citet[][see Eqs.~10 and~12]{Prudil2024GBEXII}.

To convert the amplitudes from $G$-passband to $\text{Amp}_{\rm los}$ for BRAVA-RR and APOGEE spectra, we also used equations from \citet{Prudil2024GBEXII}. In the case of spectra from the 093.B-0473 program, we used relations from \citet[][see Eqs.~11 and~13]{Prudil2024GBEXII} to convert our $G$-band magnitudes to Amp$_{V}$ and subsequently to $\text{Amp}_{\rm los}$ through relations from \citet{Braga2021} for Fe lines. We propagated uncertainties, using covariance matrices from \citet{Prudil2024GBEXII}, in case of relations from \citet{Braga2021}, we refitted the relations to estimate the correlation ($\rho_{\rm RRab} = 0.99$ and $\rho_{\rm RRc} = 0.97$) between parameters to construct the covariance matrix. Here, it is important to emphasize that \citet{Braga2021} uses as a time reference, $M_{\rm 0}$, the time of the mean magnitude along the rising branch for 093.B-0473 spectra. We used the same reference point instead of the $M_{\rm 0}$ determined by the OGLE team.

$\boldsymbol{\theta}$ for a given pulsator is represented by the $v_{\rm sys}$ and a phase shift $\Delta_{\rm phase}$. The phase shift parameter was implemented to mitigate the possible effects of imprecision in $M_{\rm 0}$. These can be caused by the data analysis (in case of a low number of observations) or related to a given RR~Lyrae star, e.g., modulations of light curves \citep{Netzel2018,Prudil2017a,Prudil2017b,Skarka2020} or binarity \citep{Hajdu2015,Prudil2019Binary,Hajdu2021}. For the phase shift, we selected the following prior:
\begin{equation} \label{eq:VsysPriors}
p(\boldsymbol{\theta}) = \mathcal{N} \left(\Delta_{\rm phase} \,|\, 0.0, 0.1 \right)\,\,\,\cap\,\,\,\left| \Delta_{\rm phase} \right| < 0.3 \\.
\end{equation}
Here $\mathcal{N}$ represents the normal distribution. The central value at $0.0$ and a small spread equal to $0.1$ were motivated by our initial expectation that our calculated phases would not shift significantly due to the above-mentioned effects. Lastly, the likelihood $p(\mathbf{D}^{k}\,|\,\boldsymbol{\theta}^{k})$ for given $k$-star we defined as:
\begin{equation} \label{eq:vsyslike}
p(\mathbf{D}\,|\,v_{\rm sys}) = \mathcal{N}\left( v^{\rm model}_{\rm los} \,|\,v_{\rm los}, \sigma^{\rm model}_{\rm data} \right) \\,
\end{equation}
where $v^{\rm model}_{\rm los}$ represents predicted line-of-sight velocities estimated based on the line-of-sight velocity templates and spectro-photometric data. We defined the $v^{\rm model}_{\rm los}$ in the same way as in \citet{Prudil2024GBEXII} for BRAVA-RR and APOGEE data, therefore using both the template $\texttt{Temp}^{\rm RR}$ and the scatter along the template $\texttt{Temp}_{\text{err}}^{\rm RR}$. For the 093.B-0473 program spectra, we used only the template from \citet{Braga2021}.

For each star, we maximalized the posterior log-probability defined in the Eq.~\ref{eq:BayPosMain} with priors and likelihood described in Equations~\ref{eq:VsysPriors} and \ref{eq:vsyslike} using the \textsc{emcee} package \citep{Foreman-Mackey2013}. For each star, we used $50$ walkers for $4000$ steps. Subsequently, we thinned the samples by $\tau=10$ and discarded the initial $3900$ samples. The resulting posterior distribution of parameters $\boldsymbol{\theta}^{k}$ was then used to estimate the $v_{\rm sys}$ its uncertainty, $\sigma_{v_{\rm sys}}$, and shift in phase, $\Delta_{\rm phase}$.

\subsection{Comparison of systemic velocities} \label{subsec:AppVsysComp}

Here, we compare our determine systemic velocities for BRAVA-RR and APOGEE spectra with the systemic velocities estimated by \citet{Kunder2020} in the original BRAVA-RR Data Release 2. We note that \citet{Butler2024} conducted a similar comparison between BRAVA-RR and APOGEE velocities, finding good agreement between APOGEE line-of-sight velocities and individual line-of-sight measurements from the Ca Triplet.

In Figure~\ref{fig:CompBRAVA_MY}, we measured our determined systemic velocities and the systemic velocities estimated by BRAVA-RR ($v_{\rm sys}^{\rm BRAVA-RR}$). For the comparison, we used only variables marked with $\text{Flag}=0$ from \citet{Kunder2020}. In total, $2213$ systemic velocities for single-mode RR~Lyrae stars were compared. We found a mild scatter (assessed based on the root-mean-square error, RMSE) between our values and values determined in \citet{Kunder2020} that was most likely a result of an updated approach in the $v_{\rm sys}$ determination.

Figure~\ref{fig:CompBRAVA_APO} shows a comparison of systemic velocities determined here based on BRAVA-RR data \citet[][again, using only those with $\text{Flag}=0$]{Kunder2020} and $v_{\rm sys}$ estimated from the APOGEE spectra. We see a larger scatter than in the previous comparison (BRAVA-RR velocities only), most likely caused by several different sources. First, APOGEE spectra have, in general, lower SNRs than BRAVA-RR spectra, resulting in increased uncertainties of individual $v_{\rm los}$. Second, APOGEE in most cases provides only a single epoch measurement, which in the case of low-SNR spectra can lead to inaccurate $v_{\rm sys}$ determination. Third, the line-of-sight velocity templates and amplitude scaling relations for estimating $v_{\rm sys}$ from APOGEE spectra are based only on a few RR~Lyrae variables, thus contributing to the large scatter in the comparison. 

In this Figure, we also see that some of the analyzed RR~Lyrae stars have extremely different estimated systemic velocities between APOGEE and BRAVA-RR data sets (marked with red points). Particularly, we noticed that $34$ RR~Lyrae stars in the comparison (seven percent of the dataset for comparison) exhibit more than $50$\,km\,s$^{-1}$ difference in the determined systemic velocities. From these $34$ RR~Lyrae stars, nine do not pass the criteria in Eq.~\ref{eq:Condition1} and their discrepant systemic velocity can be explained by a blended signal with a nearby star. Two stars from the remaining $25$ objects have different \textit{Gaia} \texttt{source\_id}'s in our dataset and the APOGEE catalog. We note that the detected outliers do not show significantly lower SNR than other APOGEE RR~Lyrae stars and the APOGEE flags also do not provide additional information that would shed light on the clear reason behind their discrepacy. We note that we prioritized BRAVA-RR RR~Lyrae systemic velocities over those derived based on APOGEE data. Lastly, the average difference between both estimated systemic velocity data sets equals $0.8$\,km\,s$^{-1}$. Thus, we do not detect any significant offset between both surveys.

In conclusion, our derived systemic velocities match reasonably well with the literature values. The discrepancies in the individual $v_{\rm sys}$ are partially covered by the uncertainties that are on average $4$\,km\,s$^{-1}$ and $8$\,km\,s$^{-1}$ for BRAVA-RR and APOGEE $v_{\rm sys}$, respectively.

\begin{figure}
\includegraphics[width=\columnwidth]{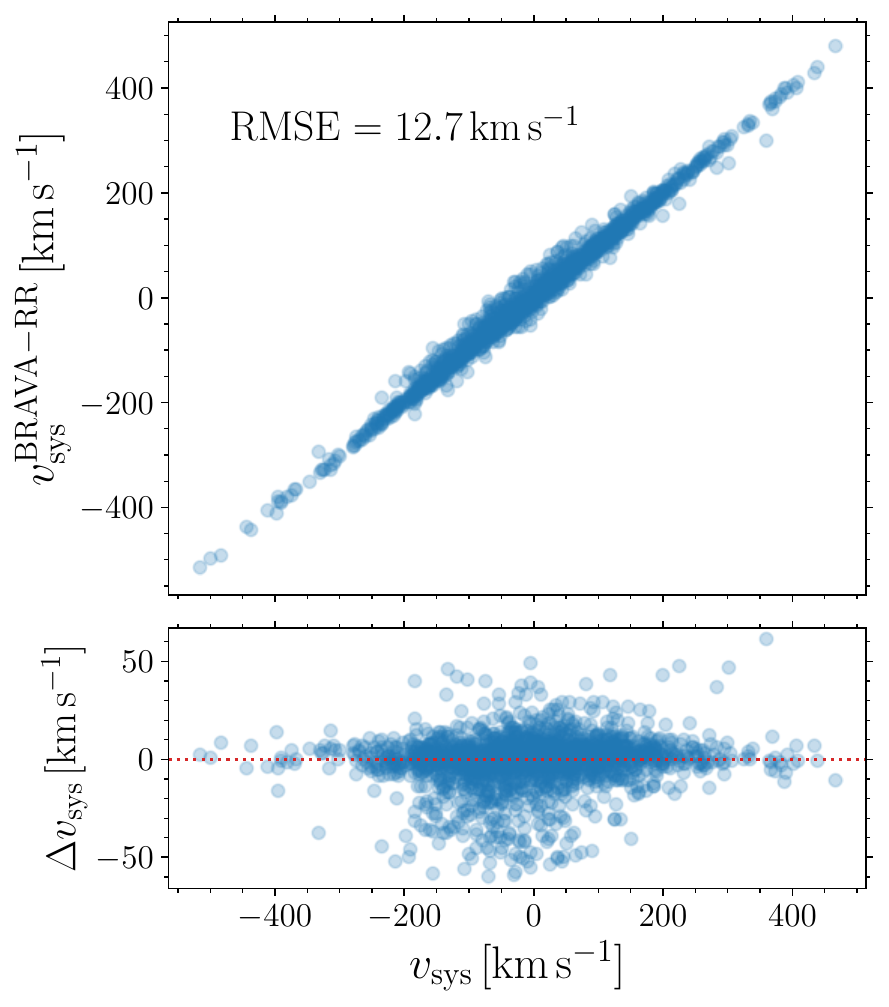}
\caption{The comparison of determined systemic velocities with literature values \citep[BRAVA-RR systemic velocities,][]{Kunder2020} and using our approach for BRAVA-RR spectra.}
\label{fig:CompBRAVA_MY}
\end{figure}

\begin{figure}
\includegraphics[width=\columnwidth]{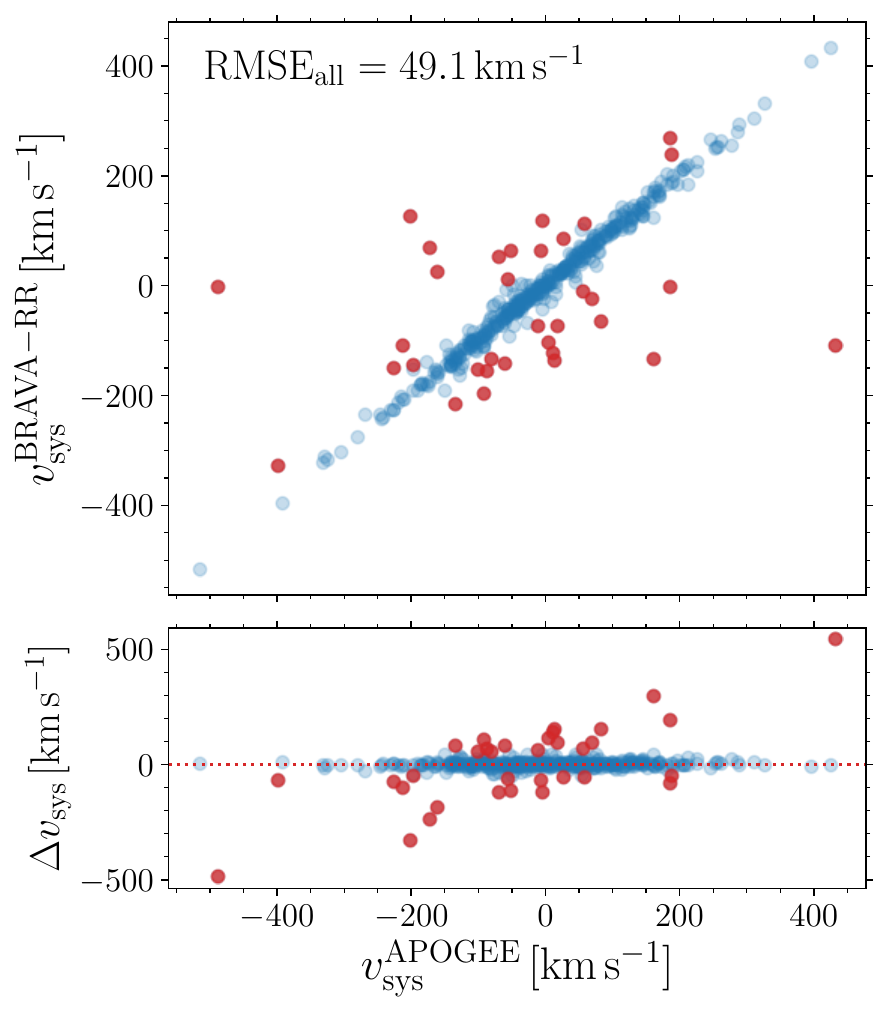}
\caption{The same as Figure~\ref{fig:CompBRAVA_MY} but comparing with APOGEE spectra and values determined by BRAVA-RR \citep{Kunder2020}. The red points mark stars with extremely different systemic velocities estimated based on APOGEE and BRAVA-RR spectra and do not pass the condition in Equation~\ref{eq:Condition1}.}
\label{fig:CompBRAVA_APO}
\end{figure}

\section{Foreground and background of the Galactic bulge} \label{sec:ForeBack}

Here, we focus on the population in the foreground and background (based on their orbital properties) of the Galactic bulge. The orbital properties of non-variable giants are compared with our RR~Lyrae kinematical data set in Figure~\ref{fig:StarsInForeground}. This image focuses on the orbital properties in $z_{\rm max}$ and $r_{\rm apo}$ space for both data sets. In particular, we explore their approximate orbital association with the MW's substructures, disk, and halo (roughly separated by a boundary using $z_{\rm max}$ values). We noticed that most bulge RR~Lyrae interlopers come from the Galactic halo, while for non-variable giants, most of the interlopers come from the Galactic disk or the outer bulge. We see that many interloper stars still have mostly positive $\Omega_{\varphi}^{\rm rot}$, for example prograde motion with respect to the Galactic bar, for $83$\,\% of RR~Lyrae stars and $96$\,\% of non-variable giants (with the apocentric distances between $3.5$\,kpc and $7.0$\,kpc). 

This suggests that when examining the distribution of line-of-sight velocities for stars toward the Galactic bulge, we will see more younger stars on prograde orbits (disk and outer bulge origin) with respect to the bar than older ones. Thus, this will affect spectroscopic surveys and studies predominantly targeting younger/metal-rich populations toward the Galactic bulge. Without taking into account whether these objects are actually located in the Galactic bulge, they will find a preferential rotation trend with the Galactic bar. This is different for the metal-poor population, which is mostly associated with the Galactic halo and contributes less to the rotation pattern of stars located toward the Galactic bulge. We note that this effect contributes only marginally and cannot explain the lag in the rotation for the older metal-poor population.

\begin{figure}
\includegraphics[width=\columnwidth]{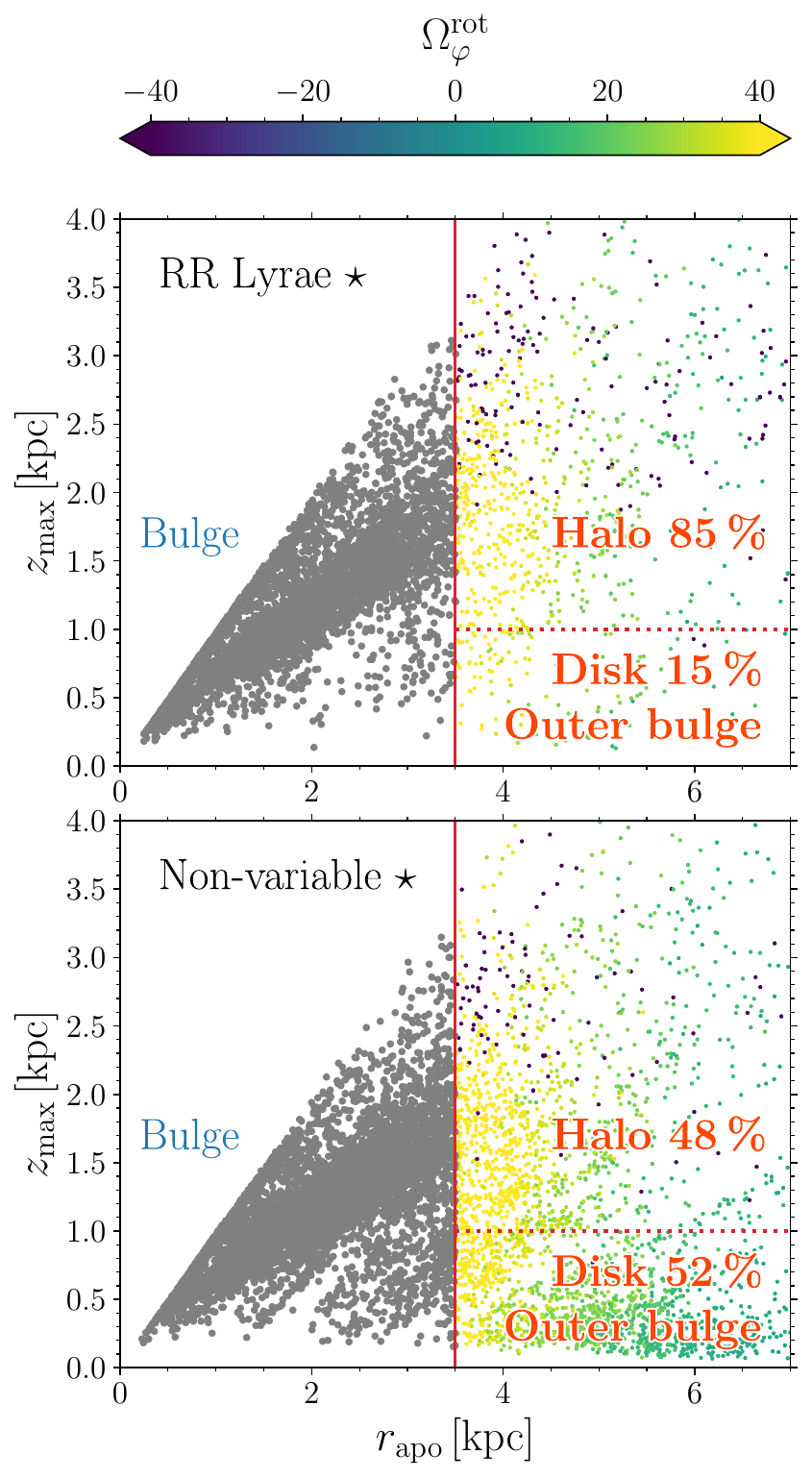}
\caption{The distribution of orbital properties $z_{\rm max}$ vs. $r_{\rm apo}$ for RR~Lyrae (top panel) and non-variable giants stars (bottom panel). The color coding is based on values of $\Omega_{\varphi}^{\rm ine}- \Omega_{\rm P}$. The solid vertical line represents our separation between the Galactic bulge and other MW substructures (Galactic disk, outer bulge and halo). The dotted horizontal line approximately separates the Galactic disk and halo based on values for $z_{\rm max}$.}
\label{fig:StarsInForeground}
\end{figure}

\section{Dependence of orbital properties on pattern speed} \label{secAp:Pattern}

Here we explore if and how our results change if we apply a different value for the $\Omega_{\rm P}$. We focus on the change in the pattern speed since we expect it to correlate with the bar length. The value for the pattern speed used in this study and for the applied model in \texttt{AGAMA} was $\Omega_{\rm P} = 37.5$\,km\,s$^{-1}$\,kpc$^{-1}$. We note that other values for the $\Omega_{\rm P}$ appear in the literature, ranging between $30$-$60$\,km\,s$^{-1}$\,kpc$^{-1}$, meaning that MW either has long-slow or fast-short bar \citep[see, e.g.,][]{Debattista2002,Antoja2014,Sormani2015,Portail2017Pattern,Clarke2022}. In addition, an even smaller estimate for the bar pattern speed was recently presented by \citet[][$\Omega_{\rm P} = 24$\,km\,s$^{-1}$\,kpc$^{-1}$]{Horta2024Knot} using the APOGEE survey data.

We explore the effect of different pattern speeds on the results presented here by recalculating the orbital properties and frequencies using a set of values for $\Omega_{\text{P}} = 24, 30, 37.5, 40, 50$\,km\,s$^{-1}$\,kpc$^{-1}$. We only tested these values for our RR~Lyrae kinematical dataset, but we do not expect any significant differences for red giants and red clump stars analyzed here. For all five values of $\Omega_{\text{P}}$ we looked at $z_{\rm max}$ vs. $r_{\rm apo}$ space and found a similar bulk structure (similar to the one found in the middle panels of Figure~\ref{fig:ApocenZmaxRot}) and nearly the same rotation pattern (using angular frequency in the bar frame, top panels of Figure~\ref{fig:ApocenZmaxRot}). The percentage of halo and disk interlopers remains the same (within one to two percent) across all values of $\Omega_{\text{P}}$ as for our fiducial value. The same can be said about the fraction of retrograde and prograde RR~Lyrae stars, which does not change, more than one to two percent, from values reported in Section~\ref{sec:OrbitalParam}. The trend in circularity for non-chaotic orbits also matches our results in Section~\ref{sec:RetroOrbitsFam}, the prograde moving RR~Lyrae stars exhibit a more bar-like shape while retrograde variables show more circular orbits.

\begin{figure}
\includegraphics[width=\columnwidth]{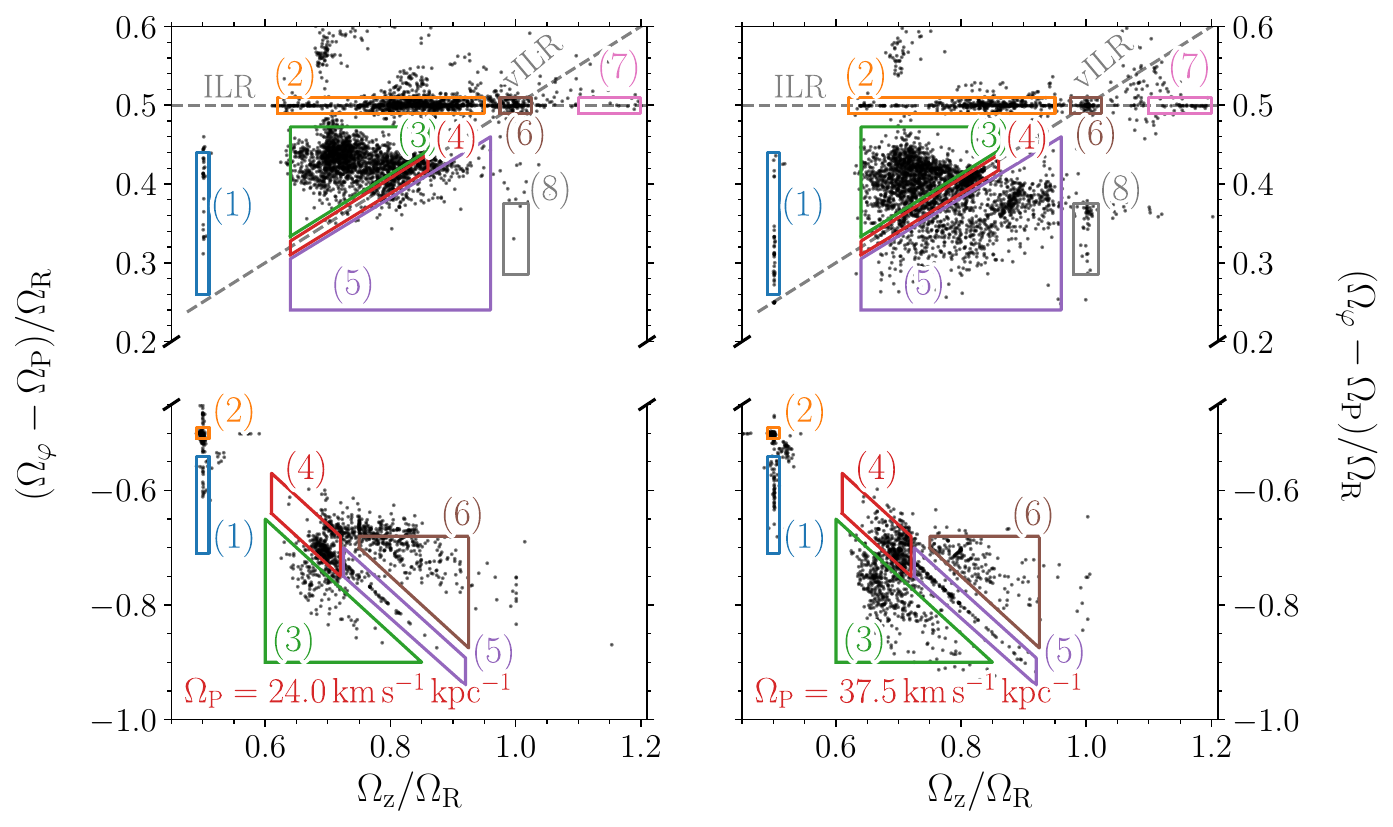}
\caption{Two frequency maps showing fundamental frequencies in cylindrical coordinates for two pattern speeds, $\Omega_{\text{P}} = 24$\,km\,s$^{-1}$\,kpc$^{-1}$ (left-hand panels) and $\Omega_{\text{P}} = 37.5$\,km\,s$^{-1}$\,kpc$^{-1}$ (right-hand panels). Only bulge RR~Lyrae variables fulfilling conditions in Eq.~\ref{eq:Condition1} are depicted. Similar to Figures~\ref{fig:MapCylPROGRADE-x-y} and~\ref{fig:MapCylRETROGRADE-x-y} we depict analysed regions.}
\label{fig:MapDiffPattern}
\end{figure}

In Figure~\ref{fig:MapDiffPattern} we observe differences in the overall distribution of stars in the frequency maps (both for prograde and retrograde). The prograde frequency maps constructed using lower values for $\Omega_{\text{P}}$ appear more evolved, in the sense that regions 5, 7, and 8 are more depopulated, and significantly more stars are located in regions 2 and 3 (ILR and vILR cloud). This results in a higher number of prograde stars with chaotic orbits ($\log\Delta\Omega < -1.0$) for lower values of $\Omega_{\text{P}}$. The retrograde side of the frequency map also shows differences between lower and higher values of $\Omega_{\text{P}}$. Regions 2 and 6 (depicted, e.g., in Figure~\ref{fig:MapCylRETROGRADE-x-y}) are more populated for lower $\Omega_{\text{P}}$ than for higher values of the pattern speed, while other regions exhibit the opposite trend. The exception here is region 4 which shows a similar fraction or number of stars throughout the various values for pattern speed.

Since for lower $\Omega_{\text{P}}$ we see an increased number of stars in the regions of the ILR and the vILR cloud and for regions 2 and 6 on the retrograde side, this results in a larger difference in overall regularity of orbits between the prograde and retrograde part of our dataset. For lower values of the pattern speed, the retrograde stars have on average more regular orbits (in both frequency drift and the Lyapunov coefficient), while for larger values of $\Omega_{\text{P}}$ the difference in chaoticity between prograde and retrograde orbits decreases. Our test showed that values in Table~\ref{table:ChaoticityProret} strongly depend on the selected pattern speed, particularly the number of stars in each region. The overall trend however remains the same, orbits of prograde moving RR~Lyrae stars exhibit ellipsoidal distribution while retrograde orbits have spherical shapes.

\section{Additional images} \label{sec:AppFigure}

\begin{figure}[h!]
\includegraphics[width=\columnwidth]{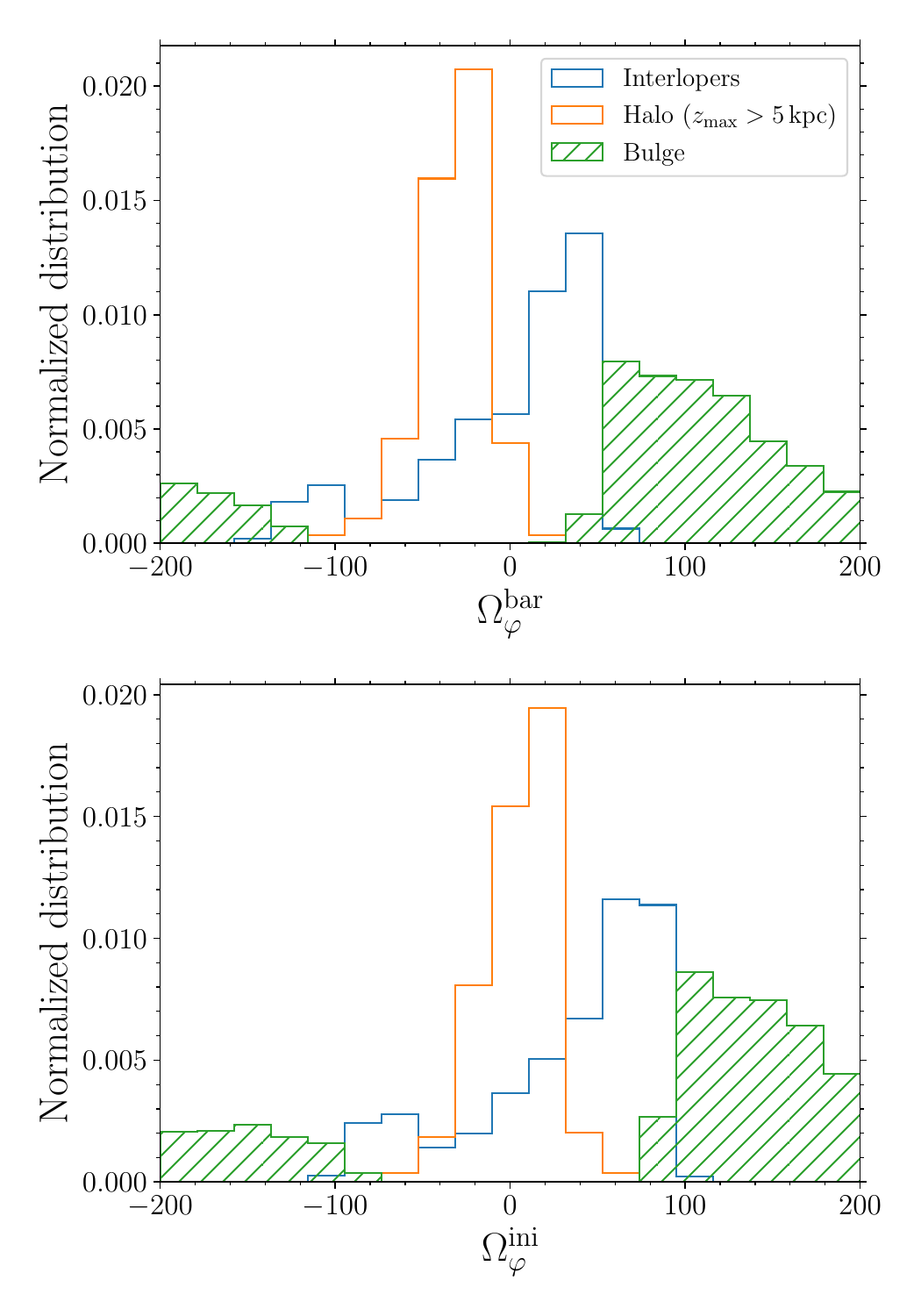}
\caption{The distribution of interloping (blue histogram) and bulge (green histogram) RR~Lyrae stars. In addition, we include likely halo RR~Lyrae stars selected based on their $z_{\rm max}$.}
\label{fig:HaloFreq}
\end{figure}

\begin{figure}[h!]
\includegraphics[width=\columnwidth]{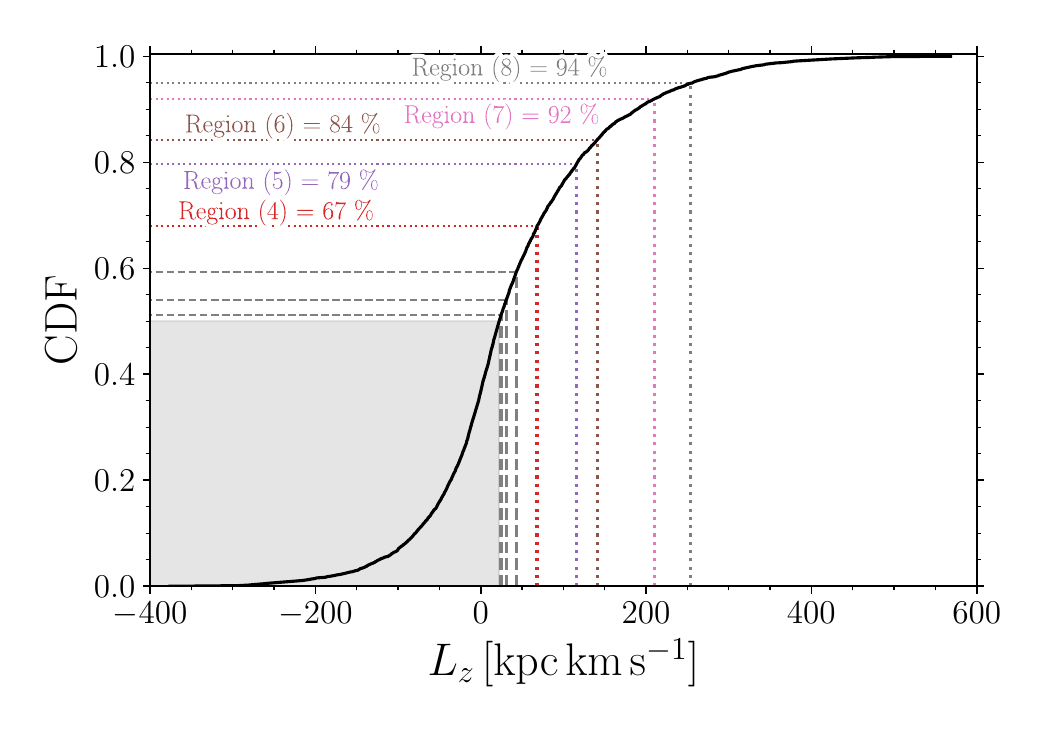}
\caption{Cumulative distribution of $L_{z}$ for bulge RR~Lyrae stars (black solid line). The dotted color lines (same as depicted in Figure~\ref{fig:MapCylPROGRADE-x-y}) represent median angular momentum values for the regular orbits in five regions (4, 5, 6, 7, and 8). Dashed lines and grey lines depict median $L_{z}$ for regions 1, 2, and 3. The grey-filled rectangle outlines percentile $50$ and lower for $L_{z}$.}
\label{fig:LzAngProg}
\end{figure}

\begin{figure}[h!]
\includegraphics[width=\columnwidth]{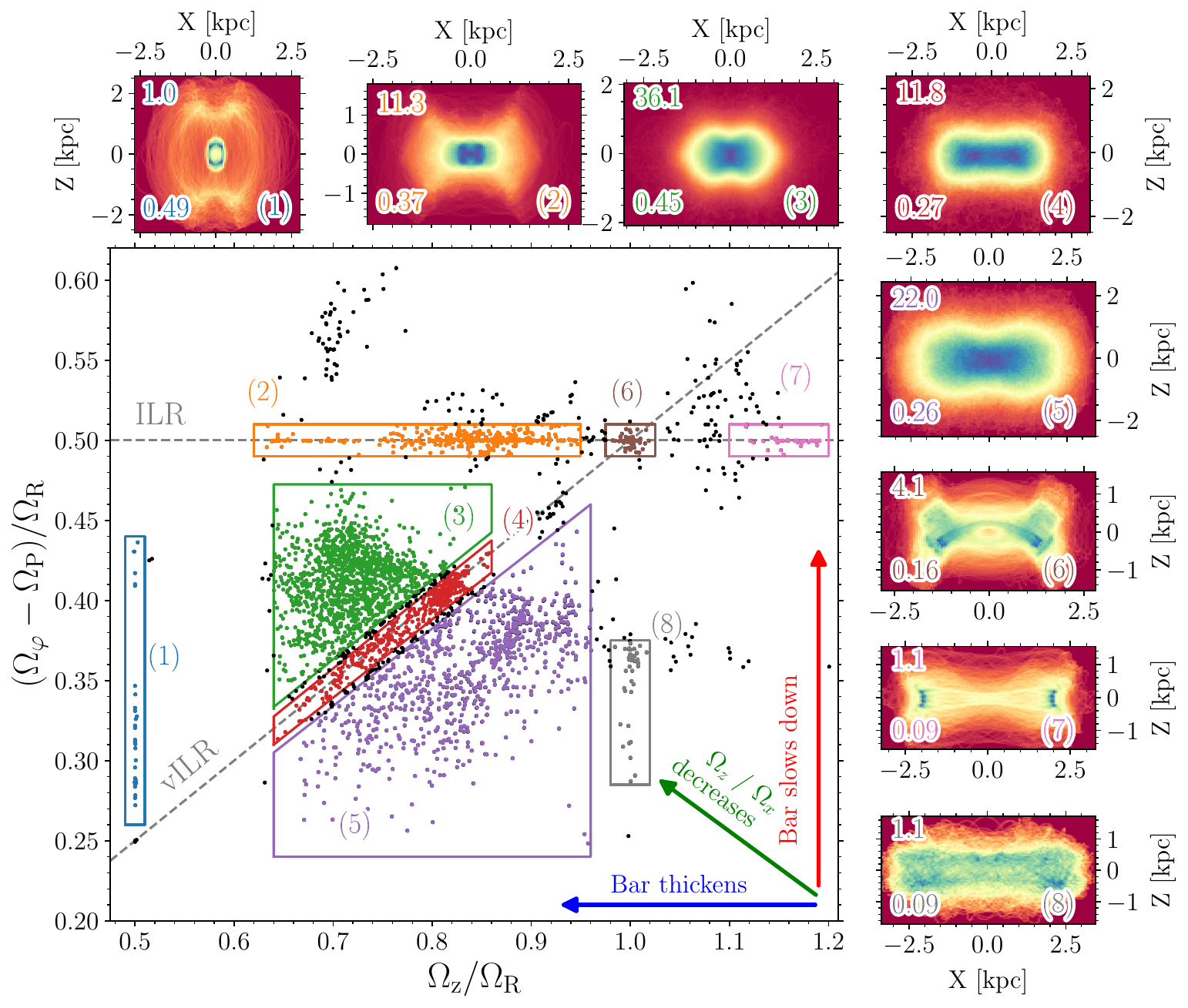}
\caption{Same as Figure~\ref{fig:MapCylPROGRADE-x-y} for prograde RR~Lyrae stars but in the insets we display orbits in the $x$ vs. $z$ plane (side on projection).}
\label{fig:MapCylPROGRADE-x-z}
\end{figure}

\begin{figure}
\includegraphics[width=\columnwidth]{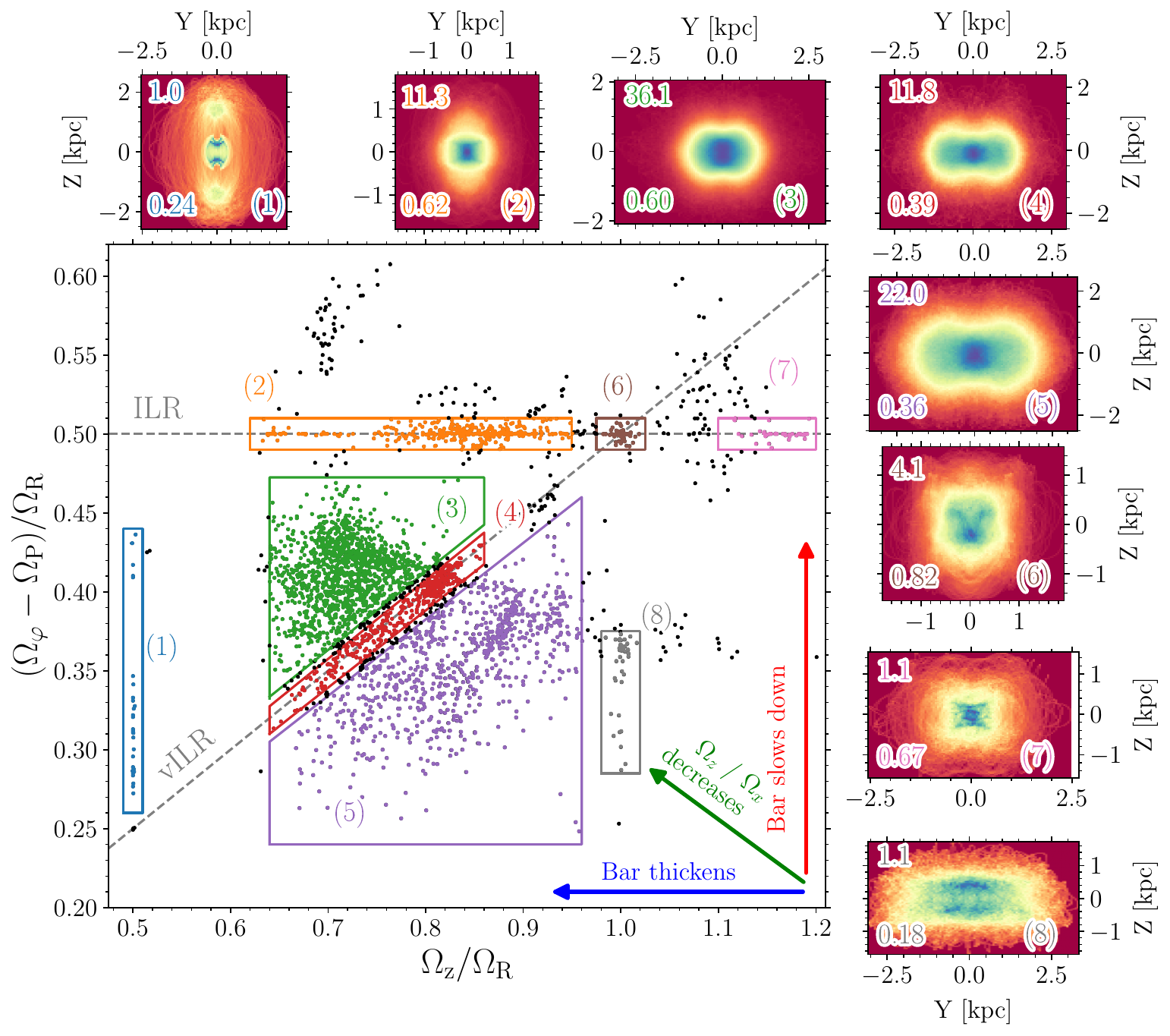}
\caption{Same as Figure~\ref{fig:MapCylPROGRADE-x-y} for prograde RR~Lyrae stars but in the insets we display orbits in the $y$ vs. $z$ plane (end on projection).}
\label{fig:MapCylPROGRADE-y-z}
\end{figure}

\begin{figure}
\includegraphics[width=\columnwidth]{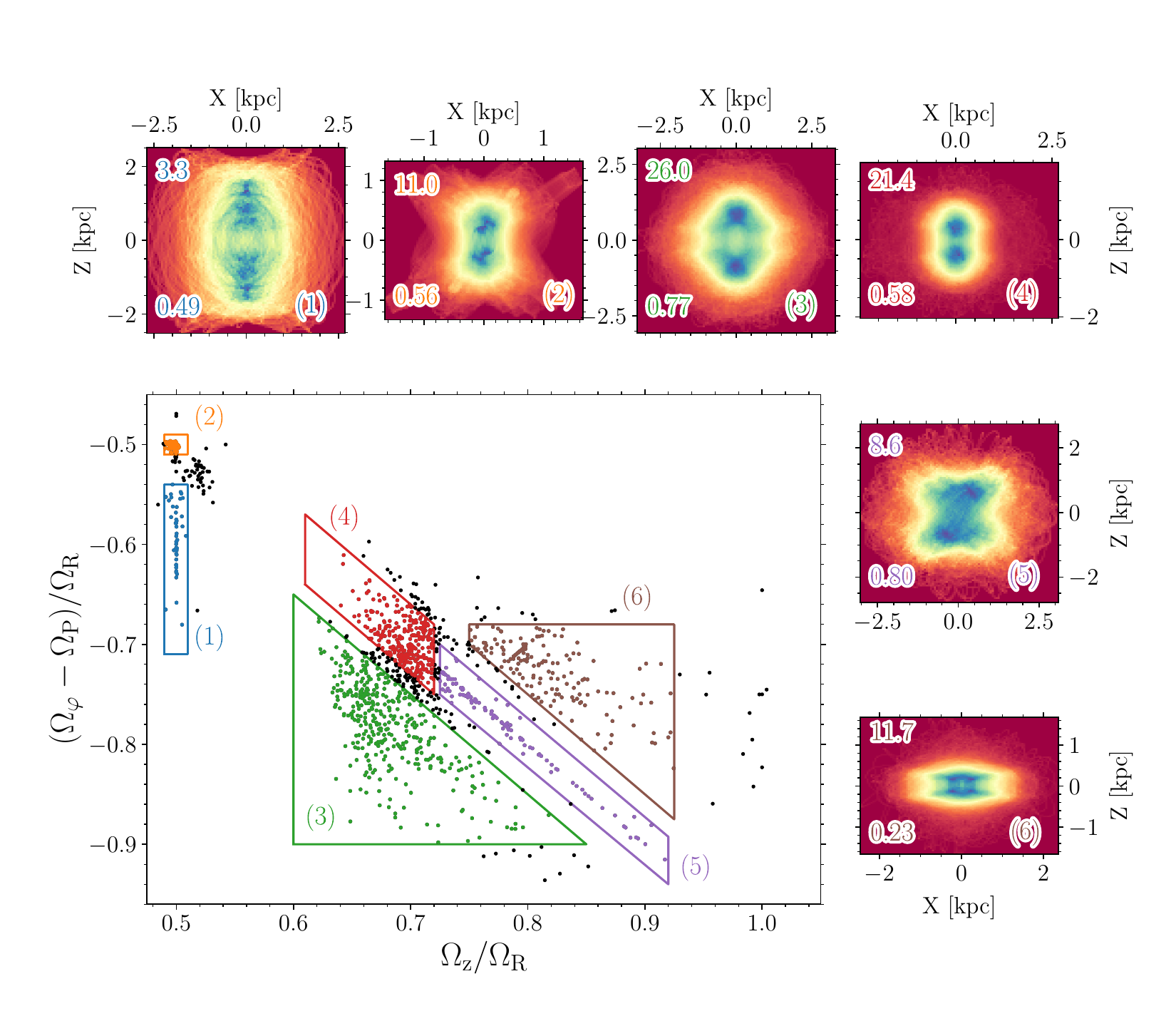}
\caption{Same as Figure~\ref{fig:MapCylRETROGRADE-x-y} for retrograde RR~Lyrae stars but in the insets we display orbits in the $x$ vs. $z$ plane (side on projection).}
\label{fig:MapCylRETROGRADE-x-z}
\end{figure}

\begin{figure}
\includegraphics[width=\columnwidth]{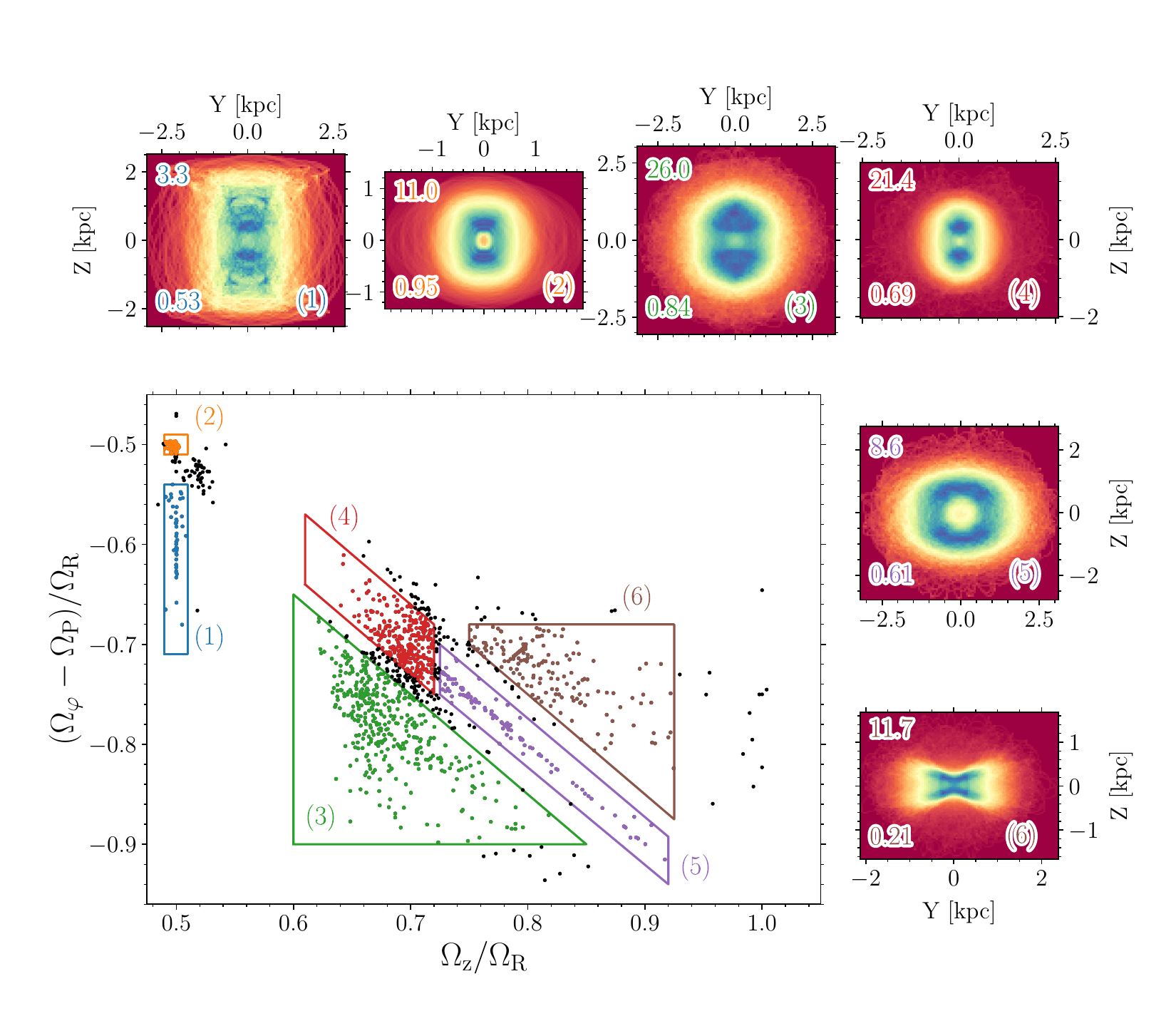}
\caption{Same as Figure~\ref{fig:MapCylRETROGRADE-x-y} for retrograde RR~Lyrae stars but in the insets we display orbits in the $y$ vs. $z$ plane (end on projection).}
\label{fig:MapCylRETROGRADE-y-z}
\end{figure}

\begin{figure}
\includegraphics[width=\columnwidth]{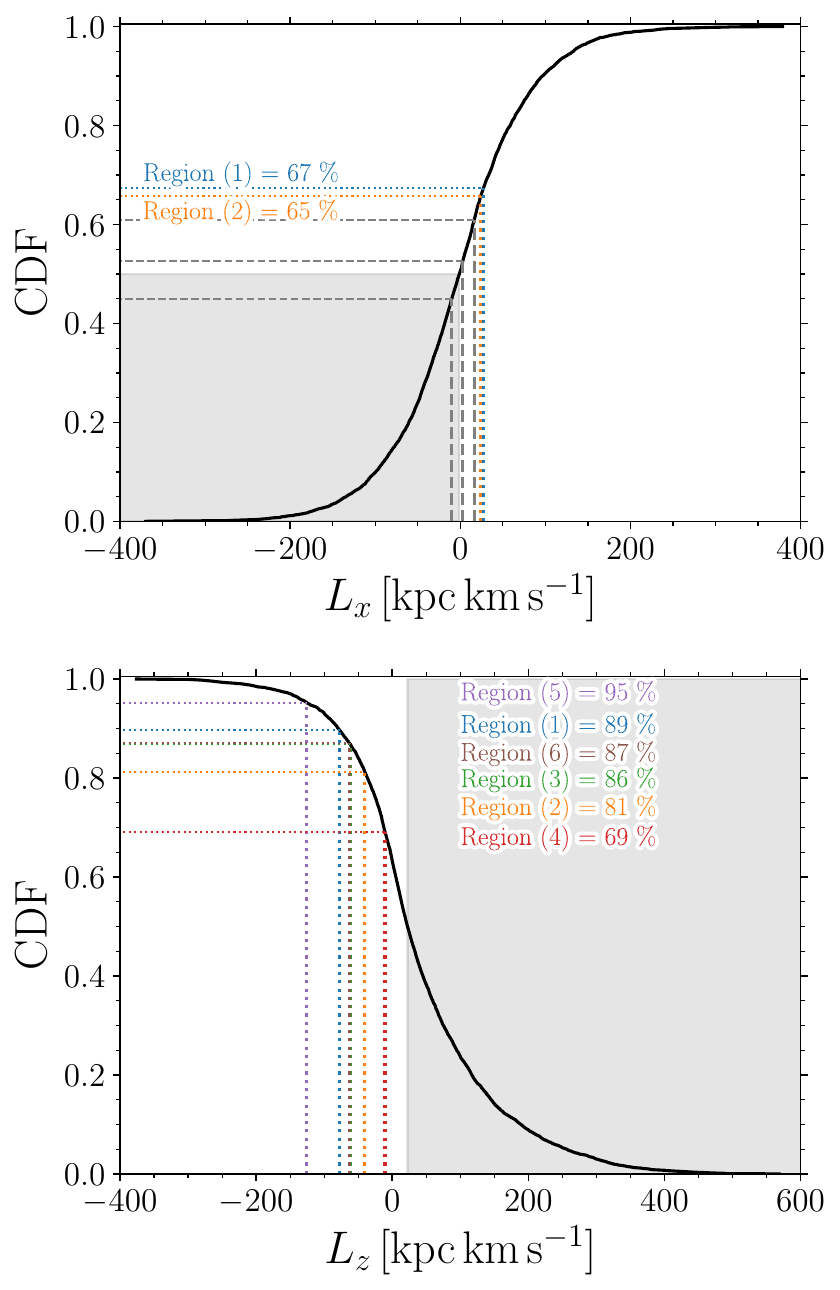}
\caption{Similar to Figure~\ref{fig:LzAngProg} we display cumulative distribution for $L_{x}$ (top panel) and $L_{z}$ (bottom panel) of our retrograde RR~Lyrae dataset. With color-coded dotted lines, we highlighted the median angular momenta values for selected regions from the retrograde frequency map (Figure~\ref{fig:MapCylRETROGRADE-x-y}). The grey-filled rectangles outline percentile $50$ and lower for $L_{x}$ and $50$ and higher for $L_{z}$.}
\label{fig:LzLxAngRet}
\end{figure}

\end{appendix}
\end{document}